%% file: article.tex
\documentclass[sigconf]{acmart}

\usepackage{xspace}
\usepackage{rotating}
\usepackage{makecell}
\usepackage{graphicx}
\usepackage{mathtools}
\usepackage{booktabs}
\usepackage{array}

\DeclarePairedDelimiter{\ceil}{\lceil}{\rceil}

\usepackage{tikz}
\usetikzlibrary{matrix,arrows,patterns,positioning,trees,calc,fit,decorations}
\usepackage{adjustbox}

\usepackage{pgfplots}
\pgfplotsset{every axis/.append style={
		scaled y ticks = false,
		scaled x ticks = false,
		y tick label style={/pgf/number format/.cd, fixed, fixed zerofill,
			int detect,1000 sep={\;},precision=3, /tikz/.cd},
		x tick label style={/pgf/number format/.cd, fixed, fixed zerofill,
			int detect, 1000 sep={},precision=3, /tikz/.cd}
	}
}
\pgfplotsset{compat=newest}

\newcommand{\NoShow}[1]{}
\input GemmPartStateless

\newcommand{\dgemm}{{\sc dgemm}\xspace}

\newboolean{showtikz}
\setboolean{showtikz}{true}

\title{The MOMMS Family of Matrix Multiplication Algorithms}
\author{Tyler M. Smith and Robert A. van de Geijn}

\begin{abstract}
\input 00abstract  
\end{abstract}

\begin{document}
\keywords{matrix multiplication, dense linear algebra, performance, caches}
\copyrightyear{2019}
\acmYear{2019}
\setcopyright{acmlicensed}
\acmConference[SC19]{The International Conference for High Performance Computing, Networking, Storage, and Analysis}{November 17--27, 2019}{Denver, CO}

\maketitle

\input body

\newpage

\bibliographystyle{ACM-Reference-Format}
\bibliography{biblio}


\end{document}

%% file: GemmPartStateless.tex
\newcommand{\GPSBaseBody}[4]{
    \node (C#4) [fit={(-#2/2, #1/2) (#2/2, -#1/2)}, inner sep=0pt, draw=black, thick] {}; \pgfmatrixnextcell 
    \node {$\mathrel{+}=$}; \pgfmatrixnextcell
    \node (A#4) [fit={(-#3/2, #1/2) (#3/2, -#1/2)}, inner sep=0pt, draw=black, thick] {}; \pgfmatrixnextcell 
    \node (B#4) [fit={(-#2/2, #3/2) (#2/2, -#3/2)}, inner sep=0pt, draw=black, thick] {}; \pgfmatrixnextcell
    \\
}

\newcommand{\GPSBase}[5]{
    \node (#4) [matrix, matrix of nodes, nodes={anchor=center},
                     row sep=0.1cm, column sep=0.1cm,#5] 
                     {\GPSBaseBody{#1}{#2}{#3}{#4}};
}

\newcommand{\GPSPartM}[5]{
    \foreach \a in {1,...,#4}{
        \draw let \p1 = (C#5) in (\x1 - #2/2, \y1 - #1/2 + \a * #1/#4 ) -- ( \x1 + #2/2 , \y1 - #1/2 + \a * #1 / #4); 
    } 
    \foreach \a in {1,...,#4}{
        \draw let \p1 = (A#5) in (\x1 - #3/2, \y1 - #1/2 + \a * #1/#4 ) -- ( \x1 + #3/2 , \y1 - #1/2 + \a * #1 / #4); 
    } 
}

\newcommand{\GPSPartK}[5]{
    \foreach \a in {1,...,#4}{
        \draw let \p1 = (A#5) in (\x1 - #3/2 + \a * #3 / #4, \y1 - #1/2 ) -- ( \x1 - #3/2 + \a * #3 / #4, \y1 + #1/2); 
    } 
    \foreach \a in {1,...,#4}{
        \draw let \p1 = (B#5) in (\x1 - #2/2, \y1 - #3/2 + \a * #3 / #4 ) -- ( \x1 + #2/2, \y1 - #3/2 + \a * #3 / #4 ); 
    }
}

\newcommand{\GPSPartN}[5]{
    \foreach \a in {1,...,#4}{
        \draw let \p1 = (C#5) in (\x1 - #2/2 + \a * #2 / #4, \y1 - #1/2 ) -- ( \x1 - #2/2 + \a * #2 / #4, \y1 + #1/2); 
    } 
    \foreach \a in {1,...,#4}{
        \draw let \p1 = (B#5) in (\x1 - #2/2 + \a * #2 / #4, \y1 - #3/2 ) -- ( \x1 - #2/2 + \a * #2 / #4, \y1 + #3/2); 
    }
}

\newcommand{\GPSShade}[8]{
    \draw let \p1 = (#1) in [#8] (\x1-#3/2+#6*#5, \y1+#2/2-#7*#4) rectangle (\x1-#3/2+#6*#5+#5, \y1+#2/2-#7*#4-#4);
}

%% file: 00abstract.tex
As the ratio between the rate of computation and rate with which data can be retrieved 
from various layers of memory continues to deteriorate,
a question arises: Will the current best algorithms for computing matrix-matrix multiplication on future CPUs continue to be (near) optimal?
This paper provides compelling analytical and empirical evidence that the answer is ``no''.
The analytical results guide us to a new family of algorithms of which the current state-of-the-art
``Goto's algorithm" is but one member.
The empirical results, on architectures that were custom built to reduce the amount of bandwidth to main memory,
show that under different circumstances, different and particular members of the family become more superior.   Thus, this family will likely start playing a prominent role going forward.

%% file: body.tex
\section{Introduction}
\input 01intro

\section{Theory and Fundamental Shapes}
\label{sec:single_cache}
\input 02theory_and_shapes

\section{Multiple Levels of Cache}
\label{sec:multi_cache}
\input 03three_levels

\section{Multilevel Cache Tradeoffs}
\input 04tradeoffs

\section{Experiments}
\label{sec:experiments}

\input 05experiments

\section{Summary}
\input 06conclusion

%% file: 01intro.tex
For almost two decades, the so-called Goto's algorithm for matrix-matrix multiplication (MMM) has guided practical implementations on current CPUs~\cite{GotoTR,GotoBLAS}.  
The algorithm orchestrates computation so as to keep a packed copy of a roughly square submatrix (block) of $ A $ in the L2 cache and a packed copy of a row panel of $ B $ in the L3 cache.
Major innovations of Goto's algorithm include 
staging a block of $A$ in the L2 cache rather than the L1 cache to reduce the amount of data movement by allowing the block of $A$ to be larger,
packing the block of $A$ and panel of $B$ into specially-formatted contiguous buffers for better spatial locality,
and showing that translation-lookaside buffer (TLB) misses can be a performance impediment that is alleviated by reducing the footprint of the block of $A$.

\begin{figure}
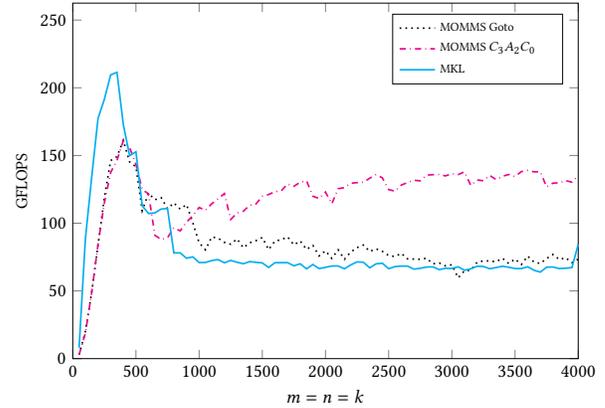

    \begin{center}
    \resizebox{.95\linewidth}{!}{
	\ifthenelse{\boolean{showtikz}}{
	\input g_l3_packing_800_intro 
    }{}}
    \end{center}
    \caption{Comparison of various MMM implementations  for an architecture with very low bandwidth to main memory.}
    \label{fig:intro}
    \end{figure}

For years, the rate of peak computation and the memory movement have been diverging~\cite{mccalpinmemory},
and MMM has been predicted to soon become a memory-bound operation based on such hardware trends~\cite{czechowski2011balance},
taking into the hardware requirements for this computation to be balanced~\cite{kung1985memory}.
While Goto's algorithm is well-suited to the relative speeds of caches in current memory hierarchies, we show, through analysis and empirical studies,
that this will not continue to be the case as bandwidths between various memory layers continue to deteriorate relative to the rate of computation.
Figure~\ref{fig:intro}
reports the attained performance of various implementations 
on a custom-built computer that allows the 
bandwidth to different memory layers to be artificially reduced%
\footnote{Details of the experiment are given in Section~\ref{sec:experiments}.}.
The curve labeled MOMMS Goto uses Goto's algorithm and we believe MKL uses an algorithm similar to it, and the curve labeled
MOMMS $C_3 A_2 C_0$ uses an algorithm that more effectively utilizes the $L_3$ cache.
The performance degradation of Goto's algorithm on such an architecture is significant.
\textbf{As MMM becomes near memory bound, it is essential that algorithms for this extremely widely-used operation to utilize the memory hierarchy as effectively as possible.}

This paper first reviews recent theoretical results by Smith et al.~\cite{smith2019tight} that establish an (essentially) tight lower bound on the memory traffic incurred by a MMM under a simple model.
It then makes a number of new contributions:
\begin{itemize}
	\item
	With these theoretical results as a foundation, it proposes the Multilevel Optimized 
	Matrix-matrix Multiplication Sandbox (MOMMS) family of practical algorithms of which Goto's algorithm is a member.
	\item
    It analyzes tradeoffs in the number of transfers between layers of the memory hierarchy
    that arise when simultaneously optimizing for multiple levels of cache.
    These tradeoffs are not explained by current state-of-the-art theoretical results.
	\item
	It analytically exposes different scenarios under which different algorithms in the family exhibit beneficial characteristics.
	\item
	It empirically demonstrates the benefits of different algorithms on custom-built hardware that allows memory bandwidths to be varied.
\end{itemize}
Together, these lay the foundation for practical solutions if and when the balance between computation and memory bandwidth changes in the future.

%% file: g_l3_packing_800_intro.tex
\begin{tikzpicture}
  \begin{axis}[width=4.0in, height=3in,
               solid,
               xlabel={$m=n=k$},
               ylabel={\small GFLOPS},
               xmin=0,xmax=4000,ymin=0,ymax=262.5,
               legend pos=north east,
               clip=false]
\addplot[color=black,dotted,thick] coordinates {
    (   50  ,   2.83682 )
    (   100 ,   20.13531    )
    (   150 ,   49.18821    )
    (   200 ,   83.87854    )
    (   250 ,   118.81949   )
    (   300 ,   146.25149   )
    (   350 ,   149.15923   )
    (   400 ,   161.11471   )
    (   450 ,   145.23752   )
    (   500 ,   141.65012   )
    (   550 ,   109.24184   )
    (   600 ,   122.14355   )
    (   650 ,   117.34928   )
    (   700 ,   118.82014   )
    (   750 ,   110.67545   )
    (   800 ,   114.98699   )
    (   850 ,   110.41671   )
    (   900 ,   113.78317   )
    (   950 ,   100.35809   )
    (   1000    ,   85.60086    )
    (   1050    ,   80.32817    )
    (   1100    ,   89.82297    )
    (   1150    ,   88.95691    )
    (   1200    ,   85.83049    )
    (   1250    ,   84.51274    )
    (   1300    ,   88.41581    )
    (   1350    ,   81.84037    )
    (   1400    ,   85.53972    )
    (   1450    ,   87.12534    )
    (   1500    ,   89.44569    )
    (   1550    ,   80.4537 )
    (   1600    ,   86.85733    )
    (   1650    ,   87.78379    )
    (   1700    ,   89.91487    )
    (   1750    ,   83.98151    )
    (   1800    ,   86.88975    )
    (   1850    ,   80.63893    )
    (   1900    ,   83.61943    )
    (   1950    ,   75.8469 )
    (   2000    ,   79.62408    )
    (   2050    ,   74.06139    )
    (   2100    ,   80.35416    )
    (   2150    ,   73.38735    )
    (   2200    ,   78.70814    )
    (   2250    ,   81.95239    )
    (   2300    ,   83.91953    )
    (   2350    ,   79.14343    )
    (   2400    ,   81.07894    )
    (   2450    ,   79.22854    )
    (   2500    ,   76.14337    )
    (   2550    ,   75.37591    )
    (   2600    ,   77.74674    )
    (   2650    ,   73.62772    )
    (   2700    ,   73.5072 )
    (   2750    ,   73.34238    )
    (   2800    ,   74.1486 )
    (   2850    ,   70.03286    )
    (   2900    ,   70.62576    )
    (   2950    ,   68.45839    )
    (   3000    ,   69.32473    )
    (   3050    ,   59.65569    )
    (   3100    ,   64.90067    )
    (   3150    ,   65.25927    )
    (   3200    ,   71.16858    )
    (   3250    ,   72.16361    )
    (   3300    ,   72.10177    )
    (   3350    ,   71.52895    )
    (   3400    ,   74.13336    )
    (   3450    ,   70.26765    )
    (   3500    ,   73.55419    )
    (   3550    ,   69.54425    )
    (   3600    ,   75.69977    )
    (   3650    ,   71.3189 )
    (   3700    ,   70.24562    )
    (   3750    ,   73.93207    )
    (   3800    ,   76.91303    )
    (   3850    ,   73.20249    )
    (   3900    ,   74.52726    )
    (   3950    ,   70.88363    )
    (   4000    ,   73.43104    )
};
\addplot[color=magenta,dashdotted,thick] coordinates {
    (   50  ,   2.81126 )
    (   100 ,   18.95932    )
    (   150 ,   48.85923    )
    (   200 ,   84.33525    )
    (   250 ,   114.89604   )
    (   300 ,   137.48924   )
    (   350 ,   146.38352   )
    (   400 ,   160.87152   )
    (   450 ,   155.90248   )
    (   500 ,   143.39393   )
    (   550 ,   124.81629   )
    (   600 ,   121.66413   )
    (   650 ,   91.09056    )
    (   700 ,   87.94721    )
    (   750 ,   89.27853    )
    (   800 ,   96.6449 )
    (   850 ,   94.32151    )
    (   900 ,   100.74818   )
    (   950 ,   105.5658    )
    (   1000    ,   111.61634   )
    (   1050    ,   109.48877   )
    (   1100    ,   114.37474   )
    (   1150    ,   117.97742   )
    (   1200    ,   121.90699   )
    (   1250    ,   102.71797   )
    (   1300    ,   107.51754   )
    (   1350    ,   108.81248   )
    (   1400    ,   113.26724   )
    (   1450    ,   114.23816   )
    (   1500    ,   119.83996   )
    (   1550    ,   121.17103   )
    (   1600    ,   123.24675   )
    (   1650    ,   123.89226   )
    (   1700    ,   128.96735   )
    (   1750    ,   127.27331   )
    (   1800    ,   130.21759   )
    (   1850    ,   131.37796   )
    (   1900    ,   119.91797   )
    (   1950    ,   118.22849   )
    (   2000    ,   123.02504   )
    (   2050    ,   114.30017   )
    (   2100    ,   125.48821   )
    (   2150    ,   126.24432   )
    (   2200    ,   129.48698   )
    (   2250    ,   129.97615   )
    (   2300    ,   130.23891   )
    (   2350    ,   133.1952    )
    (   2400    ,   135.9252    )
    (   2450    ,   133.50273   )
    (   2500    ,   124.83729   )
    (   2550    ,   123.62059   )
    (   2600    ,   128.13712   )
    (   2650    ,   130.04355   )
    (   2700    ,   131.72211   )
    (   2750    ,   131.28016   )
    (   2800    ,   135.45842   )
    (   2850    ,   135.83183   )
    (   2900    ,   135.97493   )
    (   2950    ,   134.92405   )
    (   3000    ,   136.58883   )
    (   3050    ,   135.88058   )
    (   3100    ,   137.93898   )
    (   3150    ,   127.87591   )
    (   3200    ,   131.91504   )
    (   3250    ,   131.35532   )
    (   3300    ,   135.27636   )
    (   3350    ,   132.24293   )
    (   3400    ,   134.61407   )
    (   3450    ,   135.97737   )
    (   3500    ,   134.44903   )
    (   3550    ,   138.10834   )
    (   3600    ,   138.94944   )
    (   3650    ,   137.92626   )
    (   3700    ,   137.97515   )
    (   3750    ,   126.78893   )
    (   3800    ,   129.5792    )
    (   3850    ,   129.83044   )
    (   3900    ,   131.49421   )
    (   3950    ,   130.29623   )
    (   4000    ,   135.173     )
};
\addplot[color=cyan,solid,thick] coordinates {
    (   50  ,   7.9811  )
    (   100 ,   88.5975 )
    (   150 ,   134.2669    )
    (   200 ,   177.38949   )
    (   250 ,   191.49106   )
    (   300 ,   209.54761   )
    (   350 ,   211.43083   )
    (   400 ,   172.83961   )
    (   450 ,   149.91573   )
    (   500 ,   152.74069   )
    (   550 ,   112.78617   )
    (   600 ,   107.16295   )
    (   650 ,   107.60308   )
    (   700 ,   110.4846    )
    (   750 ,   110.82045   )
    (   800 ,   78.05201    )
    (   850 ,   78.12012    )
    (   900 ,   74.12821    )
    (   950 ,   75.16478    )
    (   1000    ,   70.94078    )
    (   1050    ,   71.01156    )
    (   1100    ,   72.4355 )
    (   1150    ,   73.11571    )
    (   1200    ,   70.77899    )
    (   1250    ,   72.52593    )
    (   1300    ,   71.1777 )
    (   1350    ,   70.08392    )
    (   1400    ,   71.64088    )
    (   1450    ,   71.10924    )
    (   1500    ,   70.70008    )
    (   1550    ,   67.30399    )
    (   1600    ,   70.83734    )
    (   1650    ,   70.8535 )
    (   1700    ,   70.88697    )
    (   1750    ,   68.5682 )
    (   1800    ,   70.03763    )
    (   1850    ,   66.28213    )
    (   1900    ,   69.4871 )
    (   1950    ,   66.45789    )
    (   2000    ,   67.41955    )
    (   2050    ,   68.39863    )
    (   2100    ,   68.41497    )
    (   2150    ,   66.33427    )
    (   2200    ,   69.2262 )
    (   2250    ,   71.44408    )
    (   2300    ,   71.1745 )
    (   2350    ,   67.04272    )
    (   2400    ,   69.97845    )
    (   2450    ,   70.48666    )
    (   2500    ,   66.38939    )
    (   2550    ,   67.97345    )
    (   2600    ,   68.33573    )
    (   2650    ,   68.30701    )
    (   2700    ,   66.08441    )
    (   2750    ,   66.71696    )
    (   2800    ,   67.6397 )
    (   2850    ,   67.74736    )
    (   2900    ,   65.73489    )
    (   2950    ,   66.73166    )
    (   3000    ,   66.59019    )
    (   3050    ,   67.82709    )
    (   3100    ,   65.25482    )
    (   3150    ,   66.30686    )
    (   3200    ,   68.24104    )
    (   3250    ,   68.01815    )
    (   3300    ,   66.36779    )
    (   3350    ,   67.23742    )
    (   3400    ,   68.24857    )
    (   3450    ,   67.47568    )
    (   3500    ,   66.52977    )
    (   3550    ,   66.47899    )
    (   3600    ,   67.84733    )
    (   3650    ,   65.28293    )
    (   3700    ,   63.88355    )
    (   3750    ,   67.50226    )
    (   3800    ,   67.68627    )
    (   3850    ,   66.4968 )
    (   3900    ,   66.76543    )
    (   3950    ,   67.2221 )
    (   4000    ,   84.31261    )
};
\legend{
	\begin{minipage}{.75in}{\scriptsize  MOMMS Goto}\end{minipage},
	\begin{minipage}{.75in}{\scriptsize MOMMS $C_3 A_2 C_0$}\end{minipage},
	\begin{minipage}{.75in}{\scriptsize MKL}\end{minipage},
}
\end{axis}
\end{tikzpicture}

%% file: 02theory_and_shapes.tex
We briefly review the state-of-the-art theoretical lower bounds for the I/O complexity for MMM,
and describe algorithms that attain those bounds and thus are optimal for a simple model of a two-level memory hierarchy.
These algorithms become our fundamental components from which we compose practical algorithms for multiple levels of cache, in Section~\ref{sec:multi_cache}.

\subsection{An I/O lower bound for MMM}

Smith et al.~\cite{smith2019tight} starts with a simple model of memory with two layers of memory: a small, fast memory with capacity of $ M $ elements and a large, slow memory with unlimited capacity.  It shows that any algorithm for ordinary MMM%
\footnote{We only consider algorithms that compute the $ i,j $ element of $ m \times n $ matrix $ C $ as $ \gamma_{i,j} := \sum_{p=0}^{k-1} \alpha_{i,p} \beta_{p,j} $ where $ \alpha_{i,p} $ and $ \beta_{p,j} $ are the $ i,p $ and $ p,j $ elements of $ m \times k $ matrix $ A $ and $ k \times n $ matrix $ B $, respectively.} 
must read at least ${2mnk}/{\sqrt{M}} - 2M$ elements from slow memory and additionally write at least $mn - M$ elements to slow memory.
Adding these two lower bounds gives a lower bound on the number of transfers between slow and fast memory,
called the I/O lower bound, of approximately $ {2mnk}/{\sqrt{M}} $.  
Importantly, this lower bound is tight, modulo lower order terms.  It improves upon previous work~\cite{redblue,dongarra2008masterworker,irony2004communication}.

\subsection{Resident algorithms for MMM}
In~\cite{smith2019tight}, it is shown that 
three algorithms, named {\em Resident~A}, {\em Resident~B}, and {\em Resident~C},
attain the lower bound on the number of reads from slow memory%
\footnote{Modulo lower order terms.}.  
Additionally, Resident~C attains the lower bound on the number of writes to slow memory\footnotemark[3].  
In each algorithm, the elements of one of the operand matrices are read from slow memory only once,
and each element of the other two operand matrices is reused approximately $\sqrt{M}$ times each time it is brought into fast memory.
While the Resident~A algorithm was described as early as 1991~\cite{lam1991cache}, and all three appear in~\cite{ITXGEMM}, 
their optimality was first noted in~\cite{smith2019tight}.

\input fig_three_shapes

\subsubsection{Resident C}
The operation MMM $ Z \mathrel{+}= X Y $ can be computed by the sequence of rank-1 updates 
$ Z \mathrel{+}= x_0 y_0^T + x_1 y_1^T + \cdots $,
where $ x_i $ and $ y_i^T $ are a row and column of $ X $ and $ Y $, respectively. 
This is illustrated in Figure~\ref{fig:three_shapes} (left),
where $Z$ is the square block on the left, $X$ is the middle operand, and $Y$ is the operand on the right.
The vectors $x_i$ and $y_i^T$ are represented by the thin partitions of $X$ and $Y$.

Suppose we have (larger) matrices $C$, $A$, and $B$. 
We compute $C \mathrel{+}= AB$ in the following way. Partition:
\setlength{\arraycolsep}{2pt}
$$
C \!\rightarrow\! 
\left( \begin{array}{c | c | c}
C_{0,0} & \mbox{\tiny $\cdots$} & C_{0,n-1} \\ \hline
\mbox{\tiny $\vdots $} & & \mbox{\tiny $\vdots$} \\  \hline
C_{m-1,0} & \mbox{\tiny$\cdots$} & C_{m-1,n-1} 
\end{array}
\right), 
A \!\rightarrow\!
\left( \begin{array}{c}
A_0 \\ \hline
\mbox{\tiny $\vdots $} \\ \hline
A_{m-1}
\end{array}
\right),
B \!\rightarrow\! 
\left( \begin{array}{c | c | c}
B_0 & \mbox{\tiny $\cdots$} & B_{n-1} 
\end{array}
\right),
$$
where $ C_{i,j} $ is $ m_c \times n_c $, $ A_i $ is $ m_c \times k$, and $ B_j $ is $ k \times n_c $, except at the margins.
Then we compute the suboperation $C_{i,j} \mathrel{+}= A_i B_j$ using the described algorithm for $ Z \mathrel{+}= X Y $. Now, $C_{i,j}$ is read from slow memory once
at the beginning of the suboperation, and resides in fast memory during the rest of the duration, and 
$A_i$ and $B_j$ are streamed one row and column at a time from slow memory.
Each element of each operand is read once for each $ i,j $, and there are $\ceil{\frac{m}{m_c}} \ceil{\frac{n}{n_c}}$ such suboperations.
Overall, this algorithm incurs
$m n$ reads and $m n$ writes for matrix $C$,
$\ceil{\frac{m n k}{n_c}}$ reads for matrix $A$, and 
$\ceil{\frac{m n k}{m_c}}$ reads for matrix $B$.
When $m_c \approx n_c \approx \sqrt{M}$~\footnote{$m_c$ and $n_c$ must be slightly less than $\sqrt{M}$ to make room for a row of $A_i$ and a column of $B_j$ in fast memory.\label{foot:sqrtm}},
the I/O cost is ${2mnk}/{\sqrt{M}} + 2 mn$.
The highest ordered term in the I/O cost of the Resident~C algorithm is the same as the I/O lower bound for MMM.
Thus the algorithm is essentially optimal.

\subsubsection{Resident A and B}
Similarly, in the MMM $ Z \mathrel{+}= X Y $, each column of $Z$ can be computed by the matrix-vector multiplication$ z_i \mathrel{+}= X y_i$, where $z_i$ and $y_i$ are columns of $Z$ and $Y$, respectively.
This is illustrated in Figure~\ref{fig:three_shapes} (middle), $Z$, $X$, and $Y$ are the left, middle, and right operands, respectively.

Consider $C\mathrel{+}=AB$.
Partition:
$$
C \rightarrow 
\left( \begin{array}{c}
C_0 \\ \hline
\vdots \\ \hline
C_{n-1} \\
\end{array}
\right),
A \rightarrow 
\left( \begin{array}{c | c | c}
A_{0,0} & \cdots & A_{0,n-1} \\ \hline
\vdots & & \vdots \\  \hline
A_{m-1,0} & \cdots & A_{m-1,n-1} 
\end{array}
\right),
B \rightarrow 
\left( \begin{array}{c}
B_0 \\ \hline
\vdots \\ \hline
B_{n-1} \\
\end{array}
\right),
$$
where $C_i$ is $m_c \times n$, $A_{i,p}$ is $ m_c \times  k_c $, and $ B_{p} $ is $ k_c \times n $, except at the margins.
Then we compute the suboperation $C_i \mathrel{+}= A_{i,p} B_{p}$ using the described MVM-based algorithm
for $Z \mathrel{+}= X Y$. 
In Resident~A, $A_{i,p}$ is read from slow memory once at the beginning of the suboperation,
and $C_j$ and $B_p$ are streamed from slow memory one column at a time.
The total I/O costs associated with each matrix are:
$\ceil{\frac{mnk}{k_c}}$ reads and $\ceil{\frac{mnk}{k_c}}$ writes of elements of $ C $;
$m k$ reads of elements of $ A $; and
$\ceil{\frac{mnk}{m_c}}$ reads of elements of $ B $.
If $k_c \approx n_c \approx \sqrt{M}$,
the input cost is approximately ${2mnk}/{\sqrt{M}} + nk$,
and the output cost is approximately ${mnk}/{\sqrt{M}}$.
The input cost attains near the lower bound on reads from slow memory.

The Resident B algorithm is the obvious symmetric equivalent to the Resident A algorithm, 
built upon the suboperation in Figure~\ref{fig:three_shapes} (right).  Its I/O costs mirror that of the Resident A algorithm.

The above descriptions ``stream'' rows and/or columns of two matrices while keeping a block of the third matrix resident in fast memory.  Notice that one can instead stream row panels instead of rows and/or column panels instead of columns as long as the ``small'' dimension of the panel is small relative to the sizes of the block that is resident in fast memory.
This still achieves the I/O lower bound modulo a lower order term.  This insight becomes crucial when we discuss blocking for multiple levels of memory.

\subsection{Algorithms for different shapes of MMM}
\label{sec:single_different_shapes}

The number of reads and writes from slow memory
for the Resident~A, B, and C algorithms
depend on the shape of the input matrices:
There are cases where one of the algorithms is more {\textit{efficient}} than the other two,
where we define efficiency by flops per memop (I/O operations).
There are $2mnk$ flops performed during MMM, and the I/O lower bound is ${2mnk}/{\sqrt{M}}$.
Thus our goal for efficiency is $\sqrt{M}$ flops per memop.
We examine the cases for which algorithms are efficient,
assuming that $m$, $n$, and $k$ are at least $\sqrt{M}$.

Resident C is efficient if and only if $k$ is large.
It reads $\ceil{\frac{mnk}{n_c}} + \ceil{\frac{mnk}{m_c}} + mn$ elements from slow memory during MMM.
If $m_c = n_c = \sqrt{M}$, this is approximately ${2mnk}/{\sqrt{M}} + mn$.
This gives an efficiency of $\left( \frac{1}{\sqrt{M}} + \frac{mn}{2k} \right)^{-1}$.
When $k$ is large, this is approximately $\sqrt{M}$.
We can analyze Resident~A and Resident~B similarly. 
Here we ignore the I/O cost for writes.
If the sizes of the resident blocks are chosen to be equal to $\sqrt{M}$,
Resident~B has an efficiency of $\left( \frac{1}{\sqrt{M}} + \frac{nk}{2m} \right)^{-1}$, which is approximately $\sqrt{M}$ when $m$ is large.
Resident~A has an efficiency of $\left( \frac{1}{\sqrt{M}} + \frac{mk}{2n} \right)^{-1}$, which is approximately $\sqrt{M}$ when $n$ is large.

This shows that one must choose the right algorithm depending on the shape of the problem.
For each of the Resident~A,B, and C algorithms, there is a minimal shape that can be implemented efficiently.
For Resident~C this occurs when $m \approx n \approx \sqrt{M}$, and $k$ is large.
For Resident~B this occurs when $k \approx n \approx \sqrt{M}$, and $m$ is large.
For Resident~A this occurs when $m \approx k \approx \sqrt{M}$, and $n$ is large.
In each case, the resident matrix fits into fast memory,
and the dimension shared by the other two operands should be large
so that the cost of moving the resident matrix into fast memory can be amortized.

The fact that one must choose a different algorithm for MMM depending on problem shape and size
was previously noted for distributed memory MMM~\cite{schatz2012scalable,li1996poly},
and for hierarchical memory MMM~\cite{ITXGEMM,ATLAS}.

\subsection{A balancing act}
\label{sec:balancing}
So far in this section, we have assumed that the costs associated with accessing an element is the same, 
no matter if it is an element of $A$, $B$, or $C$.
In doing so, we arrived at the following strategy:
Place a \textbf{square} block of the resident matrix in fast memory, streaming the other two from slow memory.
This amortizes the I/O costs associated with the resident matrix, 
and \textbf{equalizes} the number of accesses of the two streamed matrices.

We now re-analyze the Resident~A, B, and~C algorithms in the case that the costs associated with accessing elements of the different operands are unequal.
This can happen if, e.g., if we add a third layer of memory of intermediate size and access cost. 
In this case, at the start of a multiplication with submatrices one operand may reside in slow memory while another resides in some intermediate layer.
Furthermore, in many cases reads and writes cannot be overlapped (e.g. main memory is often not dual-ported), and hence it is more expensive to access elements of $C$ since $C$ must be both read and written.
One way to address this  is to select the algorithm where blocks of the operand that is most expensive to access are kept in fast memory as much as possible.
Another is to adjust the sizes used for the the resident block in fast memory.

We now walk through an example of the second solution.
Suppose we are employing the Resident~A algorithm, with an $m_c \times k_c$ block of $A$ in fast memory.
If the cost of accessing an element of $B$ costs $\beta_B$, and accessing an element of $C$ costs $\beta_C$,
when $m$, $n$, and $k$ are large, the efficiency in terms of flops per memop is
$\frac{2 m_c}{\beta_B} + \frac{2 k_c}{\beta_C}$.
This is maximized when 
$m_c = \sqrt{\frac{\beta_B }{\beta_C}M}$ and
$k_c = \sqrt{\frac{\beta_C }{\beta_B}M}$.
With this, the total cost of I/O (rather than the number of accesses) associated with accessing the streamed matrices are equalized {\bf and thus the cost minimized} (modulo lower order terms).

\subsection{Summary} 

The ingredients to an efficient algorithm are:
(1) Fill fast memory with a submatrix of one of the operands (the resident matrix),
(2) Amortize the I/O cost associated with (1) over enough computation,
(3) Choose dimensions for the resident block that equalize the I/O costs (rather then the number of accesses) associated with the two streamed matrices.

%% file: fig_three_shapes.tex
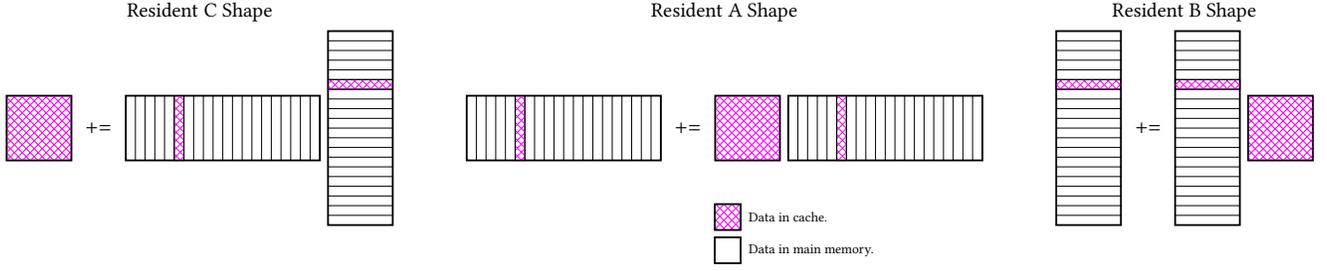
\begin{figure*}

\resizebox{1.0\textwidth}{!}{
\begin{tikzpicture}

\newcommand{\nb}{1cm}
\newcommand{\longdim}{3cm}
\newcommand{\nparts}{20}
\GPSBase{\nb}{\nb}{\longdim}{Cres}{}
\GPSBase{\nb}{\longdim}{\nb}{Ares}{right=1.0cm of BCres}
\GPSBase{\longdim}{\nb}{\nb}{Bres}{right=1.0cm of BAres}

\GPSPartK{\nb}{\nb}{\longdim}{\nparts}{Cres}{}
\GPSPartN{\nb}{\longdim}{\nb}{\nparts}{Ares}{}
\GPSPartM{\longdim}{\nb}{\nb}{\nparts}{Bres}{}

\GPSShade{CCres}{\nb}{\nb}{\nb}{\nb}{0}{0}{pattern=crosshatch, pattern color=magenta}
\GPSShade{AAres}{\nb}{\nb}{\nb}{\nb}{0}{0}{pattern=crosshatch, pattern color=magenta}
\GPSShade{BBres}{\nb}{\nb}{\nb}{\nb}{0}{0}{pattern=crosshatch, pattern color=magenta}

\GPSShade{ACres}{\nb}{\longdim}{\nb}{\longdim/\nparts}{5}{0}{pattern=crosshatch, pattern color=magenta}
\GPSShade{BCres}{\longdim}{\nb}{\longdim/\nparts}{\nb}{0}{5}{pattern=crosshatch, pattern color=magenta}

\GPSShade{CAres}{\nb}{\longdim}{\nb}{\longdim/\nparts}{5}{0}{pattern=crosshatch, pattern color=magenta}
\GPSShade{BAres}{\nb}{\longdim}{\nb}{\longdim/\nparts}{5}{0}{pattern=crosshatch, pattern color=magenta}

\GPSShade{ABres}{\longdim}{\nb}{\longdim/\nparts}{\nb}{0}{5}{pattern=crosshatch, pattern color=magenta}
\GPSShade{CBres}{\longdim}{\nb}{\longdim/\nparts}{\nb}{0}{5}{pattern=crosshatch, pattern color=magenta}

\path (CCres) +(-\nb/2, 0) coordinate (CCleft);
\path (BCres) +(\nb/2, 0) coordinate (BCright);
\draw[draw=none] (CCleft) -- (BCright) coordinate [midway] (Cmid);
\path (CBres) +(-\nb/2, 0) coordinate (CBleft);
\path (BBres) +(\nb/2, 0) coordinate (BBright);
\draw[draw=none] (CBleft) -- (BBright) coordinate [midway] (Bmid);
\path (CAres) +(-\longdim/2, 0) coordinate (CAleft);
\path (BAres) +(\longdim/2, 0) coordinate (BAright);
\draw[draw=none] (CAleft) -- (BAright) coordinate [midway] (Amid);

\node (label1) [above=1.8cm of Cmid, anchor=center] {Resident C Shape};
\node (label2) [above=1.8cm of Amid, anchor=center] {Resident A Shape};
\node (label3) [above=1.8cm of Bmid, anchor=center] {Resident B Shape};

\node (cache) [pattern=crosshatch, pattern color=magenta, fit={(-0.4cm/2, -0.4cm/2) (0.4cm/2, 0.4cm/2)}, inner sep=0pt, draw=black, thick, yshift=-.75cm] at (Ares.south) {};
\node (mem) [fill=none,  fit={(-0.4cm/2, -0.4cm/2) (0.4cm/2, 0.4cm/2)}, inner sep=0pt, draw=black, thick, yshift=-.3cm] at (cache.south) {};
\node [right] at (cache.east) { {\scriptsize Data in cache.} };  
\node [right] at (mem.east) { {\scriptsize Data in main memory.} };  
\end{tikzpicture}
}
\caption{The three shapes of MMM exposed by the algorithms Resident~C (left), Resident~A (middle), and Resident~B (right).
Each algorithm can be implemented as two loops around its corresponding shape.}

\label{fig:three_shapes}
\end{figure*}

%% file: 03three_levels.tex
We now extend the ideas from Section~\ref{sec:single_cache} to  MMM for computers with multiple layers of fast memory.
We carefully build up a particular algorithm for a computer with three layers of memory:
a \textit{slow} main memory, a \textit{medium} cache, and a \textit{fast} cache,
with each level of cache faster and smaller in capacity than the one before it.
We name the fast and medium caches $L_f$ and $L_m$, and name their capacities $M_f$ and $M_m$, respectively.

We start by partitioning the matrices so that the computation is orchestrated as a double loop over a particular subproblem,
one of the shapes in Figure~\ref{fig:three_shapes}, that that effectively utilizes the $L_m$ cache.
The question then becomes how to implement the subproblem in a way that effectively utilizes $ L_f $.

\subsection{A motivating example}

For our motivating example, we choose the Resident~A algorithm when blocking for $ L_m $.

\paragraph{Effectively utilizing $ L_m $}
To effectively utilize $ L_m $,
we select the partitioning that casts computation in terms of a double loop around the middle shape in Figure~\ref{fig:three_shapes}.
We call this subproblem the $L_m$ block-panel multiply.

\paragraph{Effectively utilizing $ L_f $}

The question now becomes how to orchestrate the $ L_m $ block-panel multiply in a way that effectively utilizes $ L_f $.
This suggests again a double loop around one of the shapes in Figure~\ref{fig:three_shapes}.  
It is not hard to see that creating a double loop that again implements a Resident~A algorithm is problematic:
Partitioning $A$ for $L_f$ would expose panels of $B$ and $C$ that by design are too large to fit into  $L_m$,
and these panels are used in multiple $L_f$ subproblems.
Either these panels of $B$ or $C$ would need to be brought into $L_m$ multiple times
or the sizes of the various matrix partitions would need to be reduced.
Either way the effect would be that the operation would no longer be near-optimal with respect to the number of transfers between 
$L_m$ and slower levels of memory.
We conclude that the block-panel multiply should be implemented in terms of a Resident~C or Resident~B algorithm. 
Which of these depends on the choice of the outer and inner loop, which we discuss next.


\paragraph{Choosing the outer loop for the $L_f$ cache}
%
In order to attain near the lower bound, each element in the two long panels of $B$ and $C$
must be used $\approx\sqrt{M}$ times each time it is brought into $L_m$.
This leads us to first partition along the $n$ dimension with blocksize $n_c$,
yielding partitions of $B$ and $C$ that are small enough to fit into the $L_m$ cache along with the block of $A$.
%

\paragraph{The inner loop for the $L_f$ cache}
The next step is to further partition the matrices to optimize for $L_f$.
The subproblem exposed by each iteration of the $L_f$ outer loop is a block of $A$ times a skinny panel of $B$ updating a skinny panel of $C$.
The $L_f$ inner loop will partition this subproblem along one of the two dimensions that the $L_f$ outer loop did not.
We can choose either of these.
For this example, we will choose the $k$ dimension (with blocksize $k_c$).
This $L_f$ inner loop exposes a new subproblem that we will call the $L_f$ subproblem.
In this case, the $L_f$ subproblem is a tall and skinny panel of $A$ times a $k_c \times n_c$ block of $B$,
updating a tall and skinny panel of $C$.
If $k_c \approx n_c \approx \sqrt{M_f}$, then the $L_f$ subproblem corresponds to the furthest left shape seen in Figure~\ref{fig:three_shapes}.
Then, the block of $B$ will reside in the $L_f$ cache, and the panels of $A$ and $C$ will be streamed in from lower levels of cache
for the duration of this subproblem.

\input fig_one_step

\subsection{Building a Family of Algorithms}
In the the motivating example, we started with a problem resembling the middle shape from Figure~\ref{fig:three_shapes},
and used two loops to partition the problem, resulting in a problem resembling one of the other two problem shapes.
This suggests the following methodology to optimize for any number of levels of cache:
We begin with one of the three shapes in Figure~\ref{fig:three_shapes},
optimizing for the I/O cost for the $L_h$ cache.
Then, to optimize for the next smaller and faster level of cache, the $L_{h-1}$ cache,
we first partition the problem along the long dimension,
and then partition along along one of the other two dimensions.
The result is one of the other two shapes shown in Figure~\ref{fig:three_shapes}.
We name the outermost of these two loops the \textit{$L_{h-1}$ outer loop}
and the innermost the \textit{$L_{h-1}$ inner loop}. 
This process is shown in Figure~\ref{fig:onestep}.

We note that~\cite{ITXGEMM} claimed that it was locally optimal to encounter
a subproblem that corresponds to one of the three optimal subproblems at every level of the memory hierarchy.
However that paper did not give details on how this could be accomplished,
nor did it analyze the claim in terms of any I/O lower bounds.

\subsection{Classifying matrix operands and algorithms}
\label{sec:classifying}
The two loops for $L_{h-1}$ have exposed partitions of matrices
that differ in terms of access frequency and size.
From these properties, we can classify these different matrix partitions.

The \textit{$L_h$ resident block} is the block that is designed to remain and reside in the $L_h$ cache during the duration of the $L_h$ subproblem.
The other two operands of an $L_h$ subproblem are called the \textit{$L_h$ streamed panels},
as small partitions of the streamed panels are brought into $L_h$ during an iteration of the $L_{h-1}$ outer loop,
used for computation, and then not used again during the $L_h$ subproblem.

The $L_{h-1}$ inner loop partitions the $L_h$ resident block and one of the $L_h$ streamed panels.
The remaining $L_h$ streamed panel is left unpartitioned.
The matrix partition not partitioned by the $L_{h-1}$ inner loop is used during every iteration of the $L_{h-1}$ inner loop.
Guided by the principle that each element of the $L_h$ subproblem
should only by read into $L_h$ once, 
it must remain in cache during the entire inner $L_{h-1}$ loop.
We name this matrix partition the \textit{$L_h$ guest panel}.
Compare this to the resident block of the $L_h$ cache. 
The elements of the $L_h$ guest matrix, like the elements of the $L_h$ resident block,
are reused from $L_h$  across iterations of a loop.
The difference is that the $L_h$ resident block is reused across every iteration of the outer $L_{h-1}$ loop,
and the $L_h$ guest matrix is reused across the iterations of the inner $L_{h-1}$ loop.

After the two $L_{h-1}$ loops, we have exposed one of the three shapes associated with our algorithms
Resident~A, Resident~B, and Resident~C.
The small block that will then reside in $L_{h-1}$ will be known as the \textit{$L_{h-1}$ resident block}.


The algorithms that arise from our methodology can be identified by the operand that the resident block is from in each level of cache.
We introduce a naming convention for the algorithms that states the level of cache and the operand that resides in it.
For instance if an algorithm has $B$ as the resident block of the $L_2$ cache, $A$ as the resident block of the $L_1$ cache,
and $C$ as the resident block in registers,
it is called $B_2 A_1 C_0$.

%

\subsection{Optimizing for registers}
In our family of algorithms, we think of the register file as $L_0$: the smallest and fastest level of cache.
For practical reasons, it should be treated as a special case.
In many implementations of MMM,
the innermost kernel implements the Resident~C algorithm~\cite{GotoBLAS,BLIS1,wang2013augem,heinecke2016libxsmm}.
There are good reasons for this.
The latency of the computation instructions dictates that there is a minimum number of registers that must be used to store elements of $C$
to avoid the instruction latency becoming a bottleneck.
The number of elements of $C$ that are stored in registers must be at least
the product of the instruction latency and the number of elementary additions that can be performed per cycle~\cite{yotov2005,BLIS4}.
Often this means that a significant portion of the registers must be dedicated to storing elements of $C$,
making it unnatural to use the Resident~A or Resident~B algorithms for the registers.
Therefore, it is often the case that there is no choice but to use Resident~C for the registers.

%% file: fig_one_step.tex
\newcommand{\bigdim}{1.8cm}
\newcommand{\middim}{.6cm}
\newcommand{\smalldim}{.20cm}
\newcommand{\lhres}{pattern=crosshatch dots,pattern color=magenta}
\newcommand{\lhguest}{pattern=crosshatch,pattern color=green}
\newcommand{\lhstream}{fill=cyan}

\begin{figure*}[tb!]

\resizebox{\textwidth}{!}{
\begin{tikzpicture}[
    >=latex,
    every node/.style={align=center},
    grow=right,
    level 1/.style={sibling distance=0.0cm, level distance=6.0cm},
    level 2/.style={sibling distance=1.0cm, level distance=5.0cm},
    level 3/.style={sibling distance=0.0cm, level distance=4.0cm},
    edge from parent/.style={->,draw},
    edge from parent path={(\tikzparentnode.east) -- (\tikzchildnode.west)} 
]

\matrix (BLOCKPANEL) [nodes={anchor=center},
            row sep=0.1cm,column sep=0.1cm,]
            {\GPSBaseBody{\middim}{\bigdim}{\middim}{Lh}}
    child{
        node [matrix, matrix of nodes, nodes={anchor=center},
              row sep=0.1cm, column sep=0.1cm,]
              {\GPSBaseBody{\middim}{\bigdim}{\middim}{outerloop}}
        child{
            node [matrix, matrix of nodes, nodes={anchor=center},
                  row sep=0.1cm, column sep=0.1cm,]
                  {\GPSBaseBody{\middim}{\smalldim}{\middim}{innerloopa}}
            child{
                node [matrix, matrix of nodes, nodes={anchor=center},
                      row sep=0.1cm, column sep=0.1cm,]
                  {\GPSBaseBody{\middim}{\smalldim}{\smalldim}{Lh1a}}
            }
        }
        child{
            node [matrix, matrix of nodes, nodes={anchor=center},
                  row sep=0.1cm, column sep=0.1cm,]
                  {\GPSBaseBody{\middim}{\smalldim}{\middim}{innerloopb}}
            child{
                node [matrix, matrix of nodes, nodes={anchor=center},
                      row sep=0.1cm, column sep=0.1cm,]
                  {\GPSBaseBody{\smalldim}{\smalldim}{\middim}{Lh1b}}
            }
        }
    };
\GPSShade{ALh}{\middim}{\middim}{\middim}{\middim}{0}{0}{pattern=crosshatch dots, pattern color=magenta};

\GPSPartN{\middim}{\bigdim}{\middim}{9}{outerloop};
\GPSShade{Aouterloop}{\middim}{\middim}{\middim}{\middim}{0}{0}{ pattern=crosshatch dots,pattern color=magenta };

\GPSPartK{\middim}{\smalldim}{\middim}{3}{innerloopa};
\GPSShade{Ainnerloopb}{\middim}{\middim}{\middim}{\middim}{0}{0}{pattern=crosshatch dots,pattern color=magenta};
\GPSShade{ALh1b}{\smalldim}{\middim}{\smalldim}{\middim}{0}{0}{pattern=crosshatch dots,pattern color=magenta};
\GPSShade{Cinnerloopb}{\middim}{\smalldim}{\smalldim}{\smalldim}{0}{0}{fill=cyan};
\GPSShade{CLh1b}{\smalldim}{\smalldim}{\smalldim}{\smalldim}{0}{0}{fill=cyan};
\GPSShade{Binnerloopb}{\middim}{\smalldim}{\middim}{\smalldim}{0}{0}{pattern=crosshatch, pattern color=green};
\GPSShade{BLh1b}{\middim}{\smalldim}{\middim}{\smalldim}{0}{0}{pattern=crosshatch, pattern color=green};

\GPSPartM{\middim}{\smalldim}{\middim}{3}{innerloopb};
\GPSShade{Ainnerloopa}{\middim}{\middim}{\middim}{\middim}{0}{0}{pattern=crosshatch dots,pattern color=magenta};
\GPSShade{ALh1a}{\middim}{\smalldim}{\middim}{\smalldim}{0}{0}{pattern=crosshatch dots,pattern color=magenta};
\GPSShade{Binnerloopa}{\middim}{\smalldim}{\smalldim}{\smalldim}{0}{0}{fill=cyan};
\GPSShade{BLh1a}{\smalldim}{\smalldim}{\smalldim}{\smalldim}{0}{0}{fill=cyan};
\GPSShade{Cinnerloopa}{\middim}{\smalldim}{\middim}{\smalldim}{0}{0}{pattern=crosshatch, pattern color=green};
\GPSShade{CLh1a}{\middim}{\smalldim}{\middim}{\smalldim}{0}{0}{pattern=crosshatch, pattern color=green};

\matrix (PANELBLOCK) [nodes={anchor=center},
            row sep=0.1cm,column sep=0.1cm,below=1.0cm of BLOCKPANEL]
            {\GPSBaseBody{\bigdim}{\middim}{\middim}{Lh}}
    child{
        node [matrix, matrix of nodes, nodes={anchor=center},
              row sep=0.1cm, column sep=0.1cm,]
              {\GPSBaseBody{\bigdim}{\middim}{\middim}{outerloop}}
        child{
            node [matrix, matrix of nodes, nodes={anchor=center},
                  row sep=0.1cm, column sep=0.1cm,]
                  {\GPSBaseBody{\smalldim}{\middim}{\middim}{innerloopa}}
            child{
                node [matrix, matrix of nodes, nodes={anchor=center},
                      row sep=0.1cm, column sep=0.1cm,]
                  {\GPSBaseBody{\smalldim}{\middim}{\smalldim}{Lh1a}}
            }   
        }   
        child{
            node [matrix, matrix of nodes, nodes={anchor=center},
                  row sep=0.1cm, column sep=0.1cm,]
                  {\GPSBaseBody{\smalldim}{\middim}{\middim}{innerloopb}}
            child{
                node [matrix, matrix of nodes, nodes={anchor=center},
                      row sep=0.1cm, column sep=0.1cm,]
                  {\GPSBaseBody{\smalldim}{\smalldim}{\middim}{Lh1b}}
            }   
        }   
    };  
\GPSShade{BLh}{\middim}{\middim}{\middim}{\middim}{0}{0}{pattern=crosshatch dots,pattern color=magenta};

\GPSPartM{\bigdim}{\middim}{\middim}{9}{outerloop};
\GPSShade{Bouterloop}{\middim}{\middim}{\middim}{\middim}{0}{0}{pattern=crosshatch dots,pattern color=magenta};

\GPSPartK{\smalldim}{\middim}{\middim}{3}{innerloopa};
\GPSShade{Binnerloopb}{\middim}{\middim}{\middim}{\middim}{0}{0}{pattern=crosshatch dots,pattern color=magenta};
\GPSShade{BLh1b}{\middim}{\smalldim}{\middim}{\smalldim}{0}{0}{pattern=crosshatch dots,pattern color=magenta};
\GPSShade{Cinnerloopb}{\smalldim}{\middim}{\smalldim}{\smalldim}{0}{0}{fill=cyan};
\GPSShade{CLh1b}{\smalldim}{\smalldim}{\smalldim}{\smalldim}{0}{0}{fill=cyan};
\GPSShade{Ainnerloopb}{\smalldim}{\middim}{\smalldim}{\middim}{0}{0}{pattern=crosshatch, pattern color=green};
\GPSShade{ALh1b}{\smalldim}{\middim}{\smalldim}{\middim}{0}{0}{pattern=crosshatch, pattern color=green};

\GPSPartN{\smalldim}{\middim}{\middim}{3}{innerloopb};
\GPSShade{Binnerloopa}{\middim}{\middim}{\middim}{\middim}{0}{0}{pattern=crosshatch dots,pattern color=magenta};
\GPSShade{BLh1a}{\smalldim}{\middim}{\smalldim}{\middim}{0}{0}{pattern=crosshatch dots,pattern color=magenta};
\GPSShade{Ainnerloopa}{\smalldim}{\middim}{\smalldim}{\smalldim}{0}{0}{fill=cyan};
\GPSShade{ALh1a}{\smalldim}{\smalldim}{\smalldim}{\smalldim}{0}{0}{fill=cyan};
\GPSShade{Cinnerloopa}{\smalldim}{\middim}{\smalldim}{\middim}{0}{0}{pattern=crosshatch, pattern color=green};
\GPSShade{CLh1a}{\smalldim}{\middim}{\smalldim}{\middim}{0}{0}{pattern=crosshatch, pattern color=green};

\matrix (PANELPANEL) [nodes={anchor=center},
            row sep=0.1cm,column sep=0.1cm,below=0.5cm of PANELBLOCK]
            {\GPSBaseBody{\middim}{\middim}{\bigdim}{Lh}}
    child{
        node [matrix, matrix of nodes, nodes={anchor=center},
              row sep=0.1cm, column sep=0.1cm,]
              {\GPSBaseBody{\middim}{\middim}{\bigdim}{outerloop}}
        child{
            node [matrix, matrix of nodes, nodes={anchor=center},
                  row sep=0.1cm, column sep=0.1cm,]
                  {\GPSBaseBody{\middim}{\middim}{\smalldim}{innerloopa}}
            child{
                node [matrix, matrix of nodes, nodes={anchor=center},
                      row sep=0.1cm, column sep=0.1cm,]
                  {\GPSBaseBody{\middim}{\smalldim}{\smalldim}{Lh1a}}
            }
        }
        child{
            node [matrix, matrix of nodes, nodes={anchor=center},
                  row sep=0.1cm, column sep=0.1cm,]
                  {\GPSBaseBody{\middim}{\middim}{\smalldim}{innerloopb}}
            child{
                node [matrix, matrix of nodes, nodes={anchor=center},
                      row sep=0.1cm, column sep=0.1cm,]
                  {\GPSBaseBody{\smalldim}{\middim}{\smalldim}{Lh1b}}
            }
        }
    };
\GPSShade{CLh}{\middim}{\middim}{\middim}{\middim}{0}{0}{pattern=crosshatch dots,pattern color=magenta};

\GPSPartK{\middim}{\middim}{\bigdim}{9}{outerloop};
\GPSShade{Couterloop}{\middim}{\middim}{\middim}{\middim}{0}{0}{pattern=crosshatch dots,pattern color=magenta};

\GPSPartN{\middim}{\middim}{\smalldim}{3}{innerloopa};
\GPSShade{Cinnerloopb}{\middim}{\middim}{\middim}{\middim}{0}{0}{pattern=crosshatch dots,pattern color=magenta};
\GPSShade{CLh1b}{\smalldim}{\middim}{\smalldim}{\middim}{0}{0}{pattern=crosshatch dots,pattern color=magenta};
\GPSShade{Ainnerloopb}{\middim}{\smalldim}{\smalldim}{\smalldim}{0}{0}{fill=cyan};
\GPSShade{ALh1b}{\smalldim}{\smalldim}{\smalldim}{\smalldim}{0}{0}{fill=cyan};
\GPSShade{Binnerloopb}{\smalldim}{\middim}{\smalldim}{\middim}{0}{0}{pattern=crosshatch, pattern color=green};
\GPSShade{BLh1b}{\smalldim}{\middim}{\smalldim}{\middim}{0}{0}{pattern=crosshatch, pattern color=green};

\GPSPartM{\middim}{\middim}{\smalldim}{3}{innerloopb};
\GPSShade{Cinnerloopa}{\middim}{\middim}{\middim}{\middim}{0}{0}{pattern=crosshatch dots,pattern color=magenta};
\GPSShade{CLh1a}{\middim}{\smalldim}{\middim}{\smalldim}{0}{0}{pattern=crosshatch dots,pattern color=magenta};
\GPSShade{Binnerloopa}{\smalldim}{\middim}{\smalldim}{\smalldim}{0}{0}{fill=cyan};
\GPSShade{BLh1a}{\smalldim}{\smalldim}{\smalldim}{\smalldim}{0}{0}{fill=cyan};
\GPSShade{Ainnerloopa}{\middim}{\smalldim}{\middim}{\smalldim}{0}{0}{pattern=crosshatch, pattern color=green};
\GPSShade{ALh1a}{\middim}{\smalldim}{\middim}{\smalldim}{0}{0}{pattern=crosshatch, pattern color=green};

\node (label1) [above=0.75cm of BLOCKPANEL, anchor=center] {$L_h$ subproblem.};
\node (label2) [right=6.0cm, anchor=center] at (label1.center) {$L_{h-1}$ outer loop.};
\node (label3) [right=5.0cm, anchor=center] at (label2.center) {$L_{h-1}$ inner loop.};
\node (label4) [right=4.0cm, anchor=center] at (label3.center) {$L_{h-1}$ subproblem.};
\end{tikzpicture}
} 

\begin{center}
\begin{tikzpicture}[>=latex,node distance=-0.25cm and .00cm]
\newcommand{\sWid}{0.4cm}
\node (resident) [pattern=crosshatch dots, pattern color=magenta,
    fit={(-\sWid/2, -\sWid/2) (\sWid/2, \sWid/2)}, inner sep=0pt, draw=black, thick] {};
\node (guest) [pattern=crosshatch, pattern color=green,%
    fit={(-\sWid/2, -\sWid/2) (\sWid/2, \sWid/2)}, inner sep=0pt, draw=black, thick, yshift=-0.3cm] at (resident.south){};
\node (stream) [fill=cyan,%
    fit={(-\sWid/2, -\sWid/2) (\sWid/2, \sWid/2)}, inner sep=0pt, draw=black, thick, yshift=-0.3cm] at (guest.south) {};
\node [right] at (resident.east) { {\scriptsize Resident block of $L_h$ cache (or part of it).} };  
\node [right] at (guest.east) { {\scriptsize Guest panel of $L_h$ cache.} };  
\node [right] at (stream.east) { {\scriptsize Resident block of $L_{h-1}$ cache.} };  
\end{tikzpicture}
\end{center}

\caption{Possible scenarios when partitioning for $L_h$ and $L_{h-1}$ 
when Resident~A (top), Resident~B (middle), or Resident~C (bottom) is encountered in $L_h$.}
\label{fig:onestep}
\end{figure*}
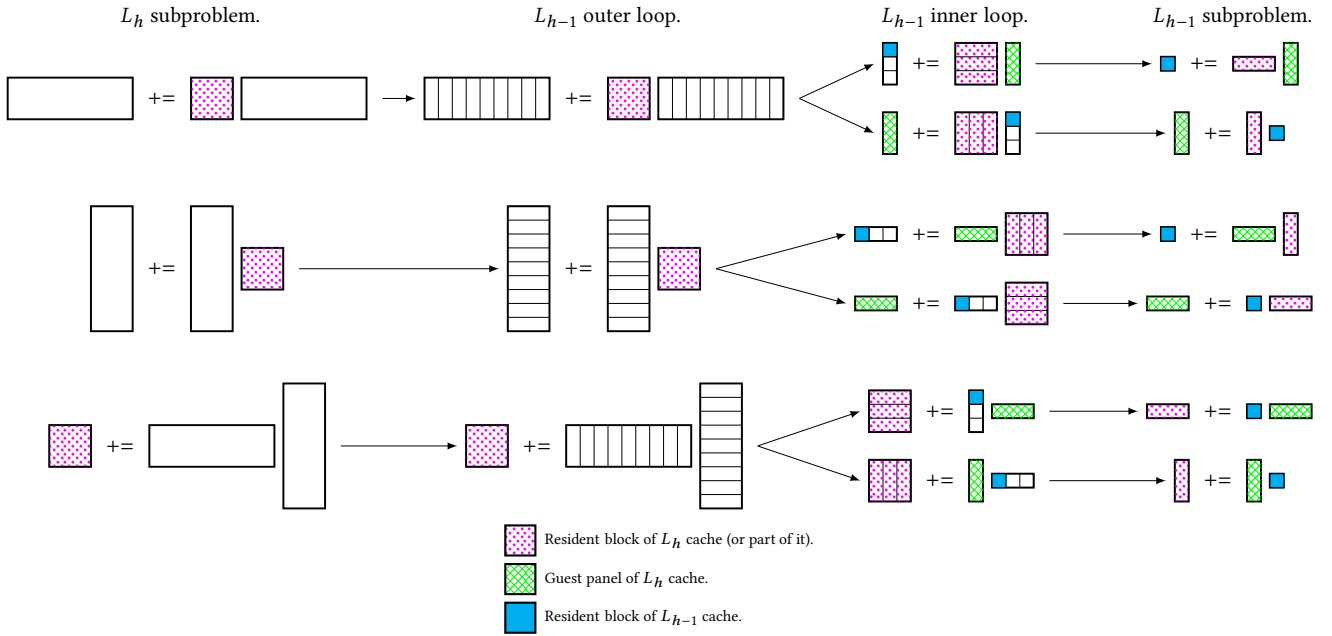

%% file: 04tradeoffs.tex
\label{sec:tradeoffs}

When simultaneously optimizing for multiple levels of cache,
there are tensions between I/O costs at the different levels of cache.

\subsection{Optimizing for $L_{h-1}$ impacts the $L_h$ I/O cost}
When simultaneously optimizing for both $L_{h-1}$ and $L_h$,
the size of the $L_h$ resident block is reduced relative to its size when optimizing only for $L_h$,
since larger portions of the $L_h$ streamed panels must fit in $L_h$.
When optimizing for both $L_h$ and $L_{h-1}$:
\begin{itemize}
    \item At minimum, the $L_h$ resident matrix and $L_h$ guest matrix must fit into $L_h$.
    \item If $L_h$ is inclusive, meaning that anything in $L_{h-1}$ must also be in $L_h$,
          then there must also be space in $L_h$ for the $L_{h-1}$ resident matrix. 
    \item If $L_h$ is inclusive and has a 
    LRU policy,
          then in order for an element to remain in $L_h$,
          fewer than $M_h$ elements may be accessed in between accesses of the element for it to remain in cache and hence every matrix partition  exposed by the $L_{h-1}$ outer loop must fit in $L_h$.
\end{itemize}
These conditions represent a tradeoff between optimizing for $L_h$ and $L_{h-1}$.
The larger $L_{h-1}$ is, the more data must fit into it,
and the smaller the $L_h$ resident block can be.
With the simplifying assumptions that the resident blocks of both $L_h$ and $L_{h-1}$ must be square,
the I/O cost for $L_h$ when optimizing for both $L_h$ and $L_{h-1}$ can be determined by the ratio $M_h / M_{h-1}$.

Sometimes, it is counter-productive or of limited value to optimize for $L_{h-1}$ when optimizing for $L_{h}$.
In this case:
\begin{enumerate}
\item A simple option is to treat $L_{h-1}$ as if it were smaller than it is, reducing the size of the $L_{h-1}$ resident block.
\item If $L_h$ is LRU, another option is to tweak the blocksizes for $L_{h-1}$ slightly. 
The portions of the $L_h$ streamed panels that must fit into $L_h$ alongside the $L_h$ resident block
depends on the tiling of the $L_{h-1}$ outer loop but not on the blocksize of the $L_{h-1}$ inner loop.
Therefore, one can tweak the shape of the $L_{h-1}$ resident block accordingly.
\item A third option is that 
one could ``skip'' optimizing for $L_{h-1}$ and instead simultaneously optimize for $L_h$ and $L_{h-2}$.
\end{enumerate}

Blocking for the $L_{h-1}$ cache adversely affects the number of transfers into and out of the $L_h$ cache
but blocking for further (smaller and faster) levels of cache does not, because the entire $L_{h-2}$ subproblem fits within the data that must be in the $L_h$ cache.

\subsection{Optimizing for $L_{h}$  impacts the $L_{h-1}$ I/O cost}

Simultaneously optimizing for $L_h$ and $L_{h-1}$ 
adversely effects transfers into and out of $L_h$.
We now argue that it also has an adverse effect on the transfers into and out of $L_h$.

When optimizing for only one cache of size $M$,
the streamed matrices are each associated with an aggregate I/O cost of $\approx {mnk}/{\sqrt{M}} $.
When optimizing for both $L_h$ and $L_{h-1}$, however, the I/O cost associated with the $L_{h-1}$ resident matrix becomes cubic
because each element of the $L_{h-1}$ resident matrix is moved into  $L_{h-1}$ once per $L_h$ subproblem.

When optimizing for both $L_h$ and $L_{h-1}$, 
the I/O cost associated with the $L_{h-1}$ resident matrix will be $\approx {mnk}/{\sqrt{M_h}}$,
whereas when only optimizing for the $L_{h-1}$ cache, the I/O cost associated with the $L_{h-1}$ resident matrix is 
equal to the number of compulsory reads and writes.
The I/O costs for the streamed matrices are not affected.

While optimizing for the $L_h$ cache has increased the $L_{h-1}$ I/O cost, 
optimizing for 
$L_{h+1}$, $L_{h+2}$, etc., does not affect it,
because optimizing for 
further levels of cache does not reduce the number of times each element is used every time it is brought into the $L_{h-1}$ cache.

\NoShow{
In Figure~\ref{fig:pareto_trade}, we consider a pair of caches, 
varying the $L_{h-1}$ cache sizes of an algorithm, and comparing its $L_h$ efficiency to its $L_{h-1}$ efficiency 
(defined in terms of flops per I/O).
If we assume that the resident matrices are square, and only consider algorithms within our family of algorithms,
then this plot gives us Pareto optimal solutions to the $L_h$ and $L_{h-1}$ tradeoff problem.
When trying to resolve multilevel cache tradeoffs, this can be done for every pair of levels in the memory hierarchy.
}

\subsection{Skipping caches}

We have seen that tradeoffs occur when simultaneously optimizing for the I/O cost of multiple levels of cache.
Sometimes these tradeoffs are too great, so instead of optimizing for both $L_h$ and $L_{h-1}$,
one may forego the $L_{h-1}$ cache, and instead simultaneously optimize for $L_h$ and $L_{h-2}$ I/O costs,
where the $L_{h-1}$ cache is intermediate between $L_h$ and $L_{h-2}$.
We call this {\textit{skipping}} the $L_{h-1}$ cache.

When the $L_{h-1}$ cache is skipped, an optimal subproblem is encountered at the $L_h$ level
and at the $L_{h-2}$ level, but not at the $L_{h-1}$ level,
However, this does not mean that the $L_{h-1}$ is not useful.
Recall that the $L_h$ guest matrix is reused during each iteration of the $L_{h-2}$ inner loop.
In the right circumstances, this $L_h$ guest matrix may be instead reused in the $L_{h-1}$ cache,
if that cache is skipped.
\begin{itemize}
\item In idealized circumstances, only the $L_h$ guest matrix should need to be in the $L_{h-1}$ cache.
\item If the $L_{h-1}$ cache is LRU, then a panel of the $L_h$ resident matrix must also fit into the $L_{h-1}$ cache.
\item If the $L_{h-1}$ cache is inclusive, then the $L_{h-2}$ resident block must also fit into the $L_{h-1}$ cache.
\end{itemize}
In this case, the $L_h$ guest matrix is reused from $L_{h-1}$,
but is not square, and the I/O cost associated with reading the other two operands is suboptimal.
Furthermore, in many cases this panel occupies only a fraction of  $L_{h-1}$,
reducing its size and further increasing the I/O cost.

Goto's algorithm is a member of the MOMMS family.
It skips optimizing for the $L_3$ and $L_1$ caches.
Since $A$ is the resident matrix of the $L_2$ cache, and $C$ is the resident matrix of the registers,
Goto's algorithm is named $A_2 C_0 $ according to the convention in Section~\ref{sec:classifying}.

%% file: 05experiments.tex
We now evaluate algorithms created by our methodology.
We do so by performing experiments on architectures with a varying number of levels of cache.

For current CPUs, Goto's algorithm attains excellent performance~\cite{BLIS2} that is difficult to exceed
despite the fact that it does not attain close to the I/O lower bound on computers with an $L_3$ cache.
In order to evaluate the I/O cost of different algorithms, we artificially vary the cost of accessing main memory.
In all experiments, we perform double precision MMM.

\subsection{Experimental setup}

We have implemented the described family of algorithms as the Multilevel Optimized Matrix-Matrix Multiplication Sandbox (MOMMS).
MOMMS implements algorithms for MMM by composing components
like matrix partitioning, packing, and parallelization at compile time.
MOMMS is written in Rust~\cite{rust},
a modern system programming language focusing on memory safety.
Most of this safety is enforced at compile-time through Rust's borrow checker.
In Rust, memory is freed when it goes out of scope, and there is no garbage collector.
From Rust, one can call C functions with very low overhead.
For low-level kernels, MOMMS calls the BLIS micro-kernel~\cite{BLIS1} coded in C and inline assembly language.

We custom built two computers.
One has an Intel i7-7700K CPU with two 8GB DIMMS of DDR4-3200 RAM and a motherboard with an Intel Z270 chipset.
The other has an Intel i7-5775C CPU with two 8GB DIMMS of DDR3-2400 RAM and a motherboard with an Intel Z97 chipset.
We refer to these computers by their processor names.
We chose the Z270 and Z97 chipset motherboards because these are enthusiast motherboards
for consumers interested in overclocking,
and they provide the ability to change the memory multiplier.
The i7-7700K computer has a 4-core Intel Kaby Lake CPU with  
64KB $L_1$, 256KB $L_2$ , and 6MB $L_3$ caches.
We chose this because it is a recent readily available Intel processor with an $L_3$ cache.
The i7-5775C is a 4-core Intel Broadwell CPU.
It also has 64KB $L_1$, 256KB $L_2$,and 6MB $L_3$ caches.
Most notably it has 128MB of eDRAM, functioning as an $L_4$ cache.

All experiments were performed with hyperthreading disabled.
A userspace CPU governor was used to set the CPUs to the nominal CPU frequency:
4.2 GHz for the i7-7700K and 3.3 GHz for the i7-5775C.

The bandwidth to main memory can be determined by the product of 
    the number of memory channels,
    the base clock rate,
    the number of bytes per transfer,
    and the memory multiplier.
With DDR RAM, this is doubled since it transfers on both the leading and trailing edges of the clock signal.
%
We increase the ratio of the rate of I/O to the rate of computation via the BIOS settings.
Reducing the memory multiplier and the number of memory channels decreases the rate of I/O without
changing the rate of computation.

\input fig_sidebyside_combined

\subsection{Optimizing for the $L_3$ cache}

We here describe an algorithm implemented in MOMMS
that optimizes for both $L_3$ and $L_2$ labeled $B_3 A_2 C_0$.
We compare this algorithm to our re-implementation of Goto's algorithm (also implemented in MOMMS),
and to vendor and state-of-the-art open source BLAS~\cite{BLAS3} implementations.
Figure~\ref{fig:sidebyside} compares Goto's algorithm
with other MOMMS algorithms optimized for both the I/O cost of $L_3$ and $L_2$.

We now describe the $B_3 A_2 C_0$ algorithm as implemented for the i7-7700K
and illustrated in Figure~\ref{fig:sidebyside} (second from the left).
First, we partition for $L_3$ cache.
The $L_3$ outer loop partitions the matrices in the $n$ dimension with blocksize 768.
Then the $L_3$ inner loop partitions in the $k$ dimension, also with blocksize 768.
This reveals a $768 \times 768$ block of $B$ that becomes the $ L_3 $ resident matrix. 
Next, we partition for the $L_2$ cache.
Since $B$ is the $L_3$ resident matrix, the $L_2$ outer loop must be in the $m$ dimension,
and it is with blocksize 120.
The $L_2$ inner loop then partitions the $k$ dimension with blocksize 192,
making a block of $A$ resident in $L_2$,
and a $120 \times 768$ panel of $C$ the guest matrix of $L_3$.
We skip $L_1$, since it is a quarter the size of $L_2$,
making it is not beneficial to optimize for both $L_2$ and $L_1$.
The next two loops make a $4 \times 12$ block of $C$ the resident matrix of registers,
and a $192 \times 12$ panel of $B$, the guest panel of $L_2$.
This guest panel of $L_2$ is designed to be reused in the (skipped) $L_1$ cache.
Finally, we call a $4 \times 12$ micro-kernel provided by BLIS~\cite{BLIS1}.

We compare this to Goto's algorithm with similar blocksizes as follows:
$n_c$ is 3000, $k_c$ is 192, $m_c$ is 120, $m_r$ is 4, and $n_r$ is 12.
Our implementation of Goto's algorithm uses the same micro-kernel from BLIS as does $B_3 A_2 C_0$.
For both algorithms, we parallelize the second loop around the micro-kernel with 4 threads.
This quadruples the bandwidth requirements of our algorithms without
increasing the amount of the $L_3$ cache that must be set-aside for elements of $A$~\cite{BLIS3}.

\input fig_rooflines2
\paragraph{Rooflines}

The roofline model is a simple model used to give an upper bound on performance based on the arithmetic intensity of an algorithm
for a specific computer~\cite{williams2009roofline}.
The computer is characterized by its rate of computation and the rate at which it can transfer data between main memory and cache.
The arithmetic intensity of an algorithm is the number of flops per byte transferred between memory and cache during the execution of that algorithm.
When the arithmetic intensity is low it is bandwidth bound, and when the arithmetic intensity is high it is compute bound.
The roofline model is thus a plot where the x-axis is the arithmetic intensity and the y-axis is maximum rate of computation for that arithmetic intensity.
The roofline that serves as an upper bound on performance is formed by two linear curves that intersect when the minimum time spent for computation
for an algorithm is equal to the minimum time spent for I/O.
Algorithms are plotted on the roofline model according to their arithmetic intensity and measured performance as a way to explain their performance
and to explain whether or not they could perform better.
One can either measure the arithmetic intensity of an algorithm or analyze it. We choose to analyze the arithmetic intensity of the algorithms plotted.

%
%
%
When the matrices are large, Goto's algorithm has an efficiency of $\left( \frac{1}{k_c} + \frac{1}{2 n_c} \right)^{-1}$ flops per element.
With the blocksizes we used, this is 23.26 flops per byte.
The algorithm $B_3 A_2 C_0$, with a $768 \times 768$ block of $B$ in the $L_3$ cache, 
has an efficiency of 64 flops per byte.

In Figure~\ref{fig:rooflines}, we show the roofline model
for the i7-7700K for the case of one channel of DDR4-800 RAM,
and for the case of two channels of DDR4-3200 RAM.
These cases represent the minimum and the maximum memory bandwidth that we configure the computer for.
We plot the modeled efficiency of Goto's algorithm and the algorithm $B_3 A_2 C_0$
against each algorithm's measured performance.
The roofline plot clearly shows that
in the high-bandwidth case, either algorithm is capable of achieving 
the peak performance of the CPU based on its arithmetic intensity,
but for the low-bandwidth case, only $B_3 A_2 C_0$ can.
The improved arithmetic intensity is caused by its more effective utilization of the $L_3$ cache.


\paragraph{Varying Bandwidth}
\ifthenelse{\boolean{showtikz}}{\input{fig_l3}}{}
Figure~\ref{fig:l3} reports the achieved performance of Goto's algorithm and $B_3 A_2 C_0$ for square matrices,
varying the amount of bandwidth to main memory.
Packing is often used to achieve spatial locality during an algorithm.
Otherwise blocks that are designed to reside in cache may not be able to do so due to cache conflict issues~\cite{packing}.
Packing incurs extra memory movements that do not fundamentally need to happen during MMM.
This paper is concerned with the fundamentals of temporal locality during MMM, and hence
we sidestep the spatial locality issue by storing matrices ``prepacked'' such that
every time a matrix is partitioned, the blocks are stored contiguously.
This lets us separate the issues of temporal and spatial locality in our experiments%
\footnote{Others have avoided packing for practical reasons.
BLASFEO~\cite{frison2018blasfeo} operates on so-called panel-major matrices for performance on small matrices.
The panel-major format is similar to the format used in Goto's algorithm for the packed panel of $B$.
Another library, libxsmm~\cite{heinecke2016libxsmm}, also targets small matrices,
and operates on column-major matrices, but does not perform packing.}.

At low bandwidth, $B_3 A_2 C_0$ outperforms Goto's algorithm by thirty to forty percent.
As the 
and the gap 
eventually disappears.

\paragraph{Comparing with existing implementations}
\ifthenelse{\boolean{showtikz}}{\input{fig_l3_packing}}{}
In Figure~\ref{fig:l3_packing}, we compare our implementations of Goto's algorithm and $B_3 A_2 C_0$ 
against the \dgemm routines in 
 ATLAS~\cite{ATLAS} (3.10.3), BLIS (0.2.1), and 
Intel's Math Kernel Library (MKL 2017 Release 2)~\cite{MKL}.
It would not be fair to compare against implementations of MMM
if we did not need to pack, so for this experiment, input matrices are stored in column major order,
and our implementations of Goto's algorithm and $B_3 A_2 C_0$ pack matrices the first time they become the resident or guest matrix at some level of cache.
This packing (and the fact that $C$ is not stored hierarchically for Goto's algorithm) account for the performance difference
seen for the Goto and $B_3 A_2 C_0$ curves between Figures~\ref{fig:l3} and~\ref{fig:l3_packing}.
We see that for high bandwidth scenarios, BLIS, the MOMMS implementation of Goto's algorithm,
and $B_3 A_2 C_0$ all attain roughly 75\% of peak,
and that MKL outperforms the other implementations.
For low bandwidth, implementations that use Goto's algorithm (or something similar) exhibit poorer performance
as they do not effectively utilize the $L_3$.
In this case, $C_3 A_2 C_0$ performs best, with $B_3 A_2 C_0$ close behind.
For large problem sizes, ATLAS performs almost as well as the algorithms implemented in MOMMS that optimize for the $L_3$ I/O cost but it does not perform nearly as well for the high bandwidth case.

\subsection{Optimizing for the $L_4$ cache}
\ifthenelse{\boolean{showtikz}}{\input{fig_l4_cache}}{}

In this section, we demonstrate that our methodology can be efficiently applied to the Intel i7-5775C,
which has four levels of cache, where the $L_4$ cache is 128MB of eDRAM.
We implemented an algorithm called $C_4 A_2 C_0$ for this architecture.
Figure~\ref{fig:l4_algorithm} shows the loop ordering and the blocksizes used for $C_4 A_2 C_0$.
In $C_4 A_2 C_0$, a $3600 \times 3600$ block of $C$ resides in the $L_4$ cache,
and a $120 \times 192$ block of $A$ resides in the $L_2$ cache.
We decided to skip blocking for the $L_3$ cache, as there is sufficient bandwidth from the $L_4$ cache without
optimizing for the number of $L_3$ cache misses.
Nevertheless, the guest matrix of the $L_4$ cache, a $192 \times 3600$ panel of $B$,
is appropriately sized to remain in the $L_3$.
$C_4 A_2 C_0$ uses the same inner kernel as $B_3 A_2 C_0$.

\ifthenelse{\boolean{showtikz}}{\input{fig_l4}}{}
\ifthenelse{\boolean{showtikz}}{\input{fig_l4_packing}}{}

In Figure~\ref{fig:l4}, we compare the performance of Goto's algorithm and $C_4 A_2 C_0$ for square matrices
across several bandwidths.
In this experiment, matrices are stored hierarchically, and so packing is not performed.
For high bandwidths, Goto's algorithm and $C_4 A_2 C_0$ exhibit similar performance,
but when bandwidth is low, $C_4 A_2 C_0$ outperforms Goto's algorithm for large problem sizes.

Figure~\ref{fig:l4_packing} compares the performance on square matrices of our implementations of Goto's algorithm and $C_4 A_2 C_0$.
Here, matrices are stored in column-major order and accordingly packing is performed
when partitions of $A$ and $B$ become resident or guest matrices of some level of cache.
In $C_4 A_2 C_0$, $C$ is unpacked when it is no longer resident in $L_4$.
In both Figures~\ref{fig:l4} and~\ref{fig:l4_packing}, the top of the graphs is the peak computational rate of the CPU.

Because $L_4$ is so large, we ran quite large problems since
otherwise the matrices would completely fit into cache.
Performance for Goto's algorithm and MKL do not fall off until the problem size becomes $m=n=k \approx 5000$.
We can see that Goto's algorithm does not optimally use $L_4$ 
and neither does Intel's MKL.

While BLIS's performance does not fall off as severely as for the other implementations when the problem size grows,
its overall performance is not as high.
The algorithmic differences between BLIS and the MOMMS implementation of Goto's algorithm
are parallelism and blocksizes.
BLIS uses a larger $k_c$ and a smaller $m_c$ than MOMMS
and parallelizes the 2nd and 3rd loops around the micro-kernel,
whereas MOMMS parallelizes the 2nd loop around the micro-kernel.
Modifying either the parallelism or the blocksizes so that they match that of the MOMMS implementation of Goto's algorithm
adversely affects performance for the low bandwidth case, causing a noticeable dropoff for larger matrices.
We postulate that somehow the way that data is shared by the threads within BLIS,
coupled with the larger value of $k_c$ within BLIS, (or the smaller $m_c$) fosters better reuse of data within the $L_4$ cache.

With DDR-800, all implementations of MMM on the i7-5775C outperform
those on the i7-7700K, despite the fact that the former processor is two generations older.
The large $L_4$ cache means that blocksizes for the $C_4 A_2 C_0$ algorithm can be very large,
so the algorithm does not need much bandwidth from main memory,
but even algorithms that do not take advantage of $L_4$ by using such large blocksizes
benefit from having the 128MB cache.
The large capacity cache can facilitate the hiding of latency to main memory,
through techniques such as hardware prefetching.

\subsection{Algorithms for different shapes of matrices}
Algorithm $A_3 B_2 C_0$ partitions the matrices such that a square block of $A$ is resident in  $L_3$
and a block of $B$ is resident in $L_2$.
It then calls an inner kernel updating a panel of $C$ whose elements are in $L_3$
by multiplying a block of $A$ whose elements are in $L_2$  times a panel of $B$
whose elements are in  $L_3$.
Algorithm $C_3 A_2 C_0$ partitions the matrices such that a square block of $C$ is resident in the $L_3$
and a block of $A$ is resident in $L_2$.
It then calls the same inner kernel as the algorithm $B_3 A_2 C_0$ does.
Blocksizes and loop orderings for algorithms $A_3 B_2 C_0$ and $C_3 A_2 C_0$ are shown in Figure~\ref{fig:sidebyside}.

Algorithms $A_3 B_2 C_0$, $B_3 A_2 C_0$, and $C_3 A_2 C_0$ represent three choices for blocking for $L_3$ .
In Section~\ref{sec:single_different_shapes}, we argued that each of these choices may be
optimal for a specific problem shape where two dimensions are equal to $\sqrt{M_3}$
and the other dimension is large, and selecting the wrong algorithm for a problem shape can result in an I/O cost that is $50\%$ greater.

On a computer with three levels of cache and low bandwidth, we claim the following:
$A_3 B_2 C_0$ casts its computation in terms of a block-panel multiply, 
with a block of $A$ in $L_3$, and so it should be the best choice of the three algorithms when 
$m = k \approx \sqrt{M_3}$, and $n$ is large.
Similarly, $B_3 A_2 C_0$ casts its computation in terms of a panel-block multiply, 
with a block of $B$ in $L_3$, and so it should be the best choice of the three algorithms when 
$n = k \approx \sqrt{M_3}$, and $m$ is large.
Finally, $C_3 A_2 C_0$ casts its computation in terms of a block dot product multiply, 
with a block of $C$ in $L_3$, and so it should be the best of the three algorithms when 
$m = n \approx \sqrt{M_3}$, and $k$ is large.

\input fig_shapes

Figure~\ref{fig:l3_shapes} reports results 
using $A_3 B_2 C_0$, $B_3 A_2 C_0$, $C_3 A_2 C_0$, and Goto's algorithm, with matrices stored hierarchically and no packing is performed.
We vary the shape of the matrices. In each case, two of the dimensions are set to $768$, 
and one of the dimensions is varied along the x-axis.
The experiments were performed on the Intel i7-7700K, with the DDR speed set to a single channel of DDR4-800.
When the dimension that is allowed to vary is large,
the predicted algorithm outperforms the others.
We also show performance when the matrices are square, and the size varies along the x-axis.
For our algorithms that optimize for the $L_3$ cache, 
there is very little performance difference between the square case and the case where an algorithm is the ``correct'' choice.
We conclude that when executing MMM, optimal I/O properties are attainable 
two dimensions are at least the square root of the last level cache size, 
as long as the third dimension is much larger.

The algorithms $A_3 B_2 C_0$, $B_3 A_2 C_0$, and $C_3 A_2 C_0$ outperform 
Goto's algorithm for larger problem sizes in this low bandwidth scenario.
This is because even when the algorithm is wrong for the problem shape,
the I/O cost is only $50\%$ higher.
In comparison, on this computer, Goto's algorithm has an I/O cost that is approximately two times greater than the optimal algorithm.

%% file: fig_sidebyside_combined.tex
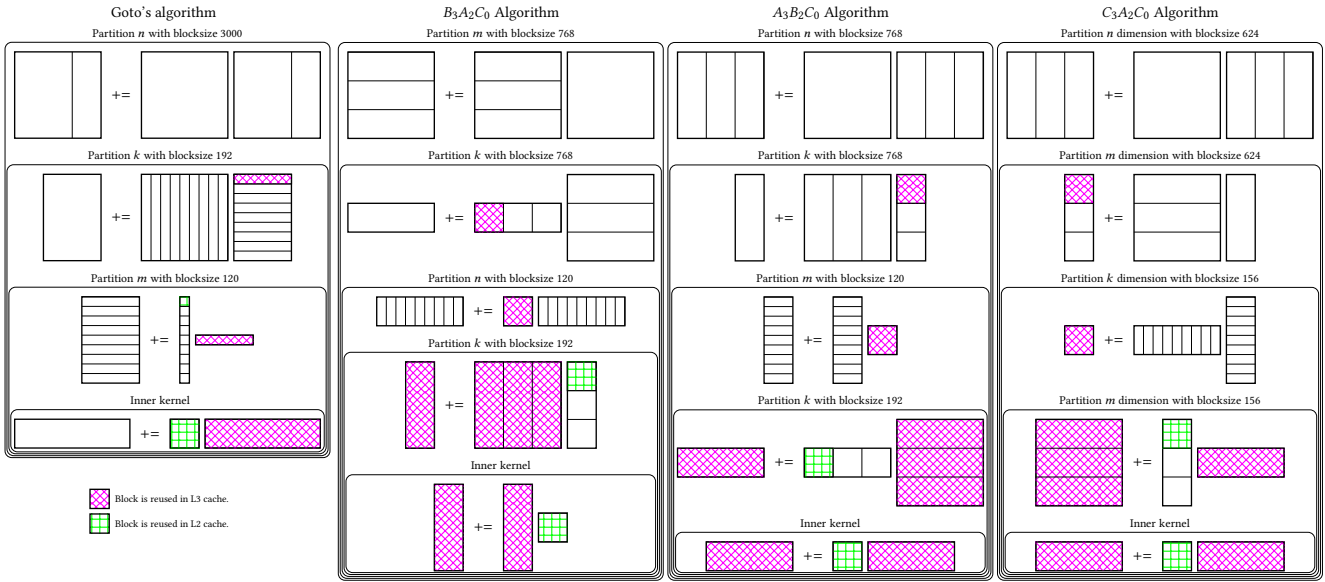
\begin{figure*}
\def\dbig{1.8cm}
\def\dmed{0.6cm}
\def\dsma{0.2cm}

\newcommand{\lthreeresident}{pattern=crosshatch}

\resizebox{\textwidth}{!}{
\begin{tabular}{@{} c @{\hspace{2pt}} c @{\hspace{2pt}} c @{\hspace{2pt}} c @{}}
Goto's algorithm & $B_3 A_2 C_0$ Algorithm & $A_3 B_2 C_0$ Algorithm & $C_3 A_2 C_0$ Algorithm\\
\begin{adjustbox}{valign=b}
\begin{tabular}{@{} c @{}}
\begin{tikzpicture}
\GPSBase{\dbig}{\dbig}{\dbig}{l3n}{}
\draw let \p1 = (Cl3n) in (\x1 - \dbig/2+2*\dbig/3,\y1-\dbig/2) -- (\x1-\dbig/2+2*\dbig/3,\y1+\dbig/2);
\draw let \p1 = (Bl3n) in (\x1 - \dbig/2+2*\dbig/3,\y1-\dbig/2) -- (\x1-\dbig/2+2*\dbig/3,\y1+\dbig/2);

\GPSBase{\dbig}{2/3 * \dbig}{\dbig}{l3k}{below=0.5cm of l3n}
\GPSPartK{\dbig}{2/3 * \dbig}{\dbig}{9}{l3k}
\GPSShade{Bl3k}{\dbig}{2/3 * \dbig}{\dbig / 9}{2/3 * \dbig}{0}{0}{pattern=crosshatch, pattern color=magenta}

\GPSBase{\dbig}{2/3 * \dbig}{\dsma}{l2m}{below=0.5cm of l3k}
\GPSShade{Bl2m}{\dsma}{2/3 * \dbig}{\dsma}{2/3 * \dbig}{0}{0}{pattern=crosshatch, pattern color=magenta}
\GPSPartM{\dbig}{2/3*\dbig}{\dsma}{9}{l2m}
\GPSShade{Al2m}{\dbig}{\dsma}{\dsma}{\dsma}{0}{0}{pattern=grid, pattern color=green}

\def\kernn{2.4cm}
\GPSBase{\dmed}{\kernn}{\dmed}{innerkernel}{below=0.5cm of l2m}
\GPSShade{Binnerkernel}{\dmed}{\kernn}{\dmed}{\kernn}{0}{0}{pattern=crosshatch, pattern color=magenta}
\GPSShade{Ainnerkernel}{\dmed}{\dmed}{\dmed}{\dmed}{0}{0}{pattern=grid, pattern color=green}

\path (Cinnerkernel) +(-\kernn/2, -\dmed/2) coordinate (bottomLeft);

\path (Bl3n) +(\dbig/2 + .2cm, \dbig/2 + .2cm) coordinate (topRightl3n);
\path (Bl3k) +(\dbig/3 + \dbig/3 + .2cm -.04cm, \dbig/2 +.2cm) coordinate (topRightl3k);
\path (Bl2m) +(\dbig/3 + \dbig - \dbig/6 - \dsma/2 + .2cm - .04cm*2, \dbig/2 + .2cm) coordinate (topRightl2m);
\path (Binnerkernel) +(\kernn/2 + .2cm - .04cm*3, \dmed/2 + .2cm) coordinate (topRightinnerkernel);

\path (Cl3n) +(-\dbig/2 - .2cm, \dbig/2 + .2cm) coordinate (topLeftl3n);
\path (Cl3k) +(-\dbig +\dmed/2 - .2cm +.04cm, \dbig/2 +.2cm) coordinate (topLeftl3k);
\path (Cl2m) +(-\dmed/2 - \dbig - .2cm + .04cm*2, \dbig/2 + .2cm) coordinate (topLeftl2m);
\path (Cinnerkernel) +(-\dbig + \dmed/2 - .2cm + .04cm*3, \dmed/2 + .2cm) coordinate (topLeftinnerkernel);

\draw [rounded corners] (bottomLeft) +(-.2cm, -.2cm) rectangle (topRightl3n);
\draw [rounded corners] (bottomLeft) +(-.2cm + 0.04cm, -.2cm + 0.04cm) rectangle (topRightl3k);
\draw [rounded corners] (bottomLeft) +(-.2cm + 0.04cm*2, -.2cm + 0.04cm*2) rectangle (topRightl2m);
\draw [rounded corners] (bottomLeft) +(-.2cm + 0.04cm*3, -.2cm + 0.04cm*3) rectangle (topRightinnerkernel);

\draw [draw=none] (topLeftl3n) -- (topRightl3n) node [midway, above] {\footnotesize{Partition $n$ with blocksize 3000}};
\draw [draw=none] (topLeftl3k) -- (topRightl3k) node [midway, above] {\footnotesize{Partition $k$ with blocksize 192}};
\draw [draw=none] (topLeftl2m) -- (topRightl2m) node [midway, above] {\footnotesize{Partition $m$ with blocksize 120}};
\draw [draw=none] (topLeftinnerkernel) -- (topRightinnerkernel) node [midway, above] {\footnotesize{Inner kernel}};
\end{tikzpicture} 
\vspace{0.5cm} \\
\begin{tikzpicture}[>=latex,node distance=-0.25cm and .00cm]
\newcommand{\sWid}{0.4cm}
\node (l3) [pattern=crosshatch, pattern color=magenta,
    fit={(-\sWid/2, -\sWid/2) (\sWid/2, \sWid/2)}, inner sep=0pt, draw=black, thick] {}; 
\node (l2) [pattern=grid, pattern color=green,
    fit={(-\sWid/2, -\sWid/2) (\sWid/2, \sWid/2)}, inner sep=0pt, draw=black, thick, yshift=-0.3cm] at (l3.south) {}; 
\node [right] at (l3.east) { {\scriptsize Block is reused in L3 cache.} };  
\node [right] at (l2.east) { {\scriptsize Block is reused in L2 cache.} };  
\end{tikzpicture} 
\vspace{0.95cm}
\end{tabular} 
\end{adjustbox}&
\begin{tikzpicture}
\GPSBase{\dbig}{\dbig}{\dbig}{l3n}{}
\GPSPartM{\dbig}{\dbig}{\dbig}{3}{l3n}
\GPSBase{\dmed}{\dbig}{\dbig}{l3k}{below=0.5cm of l3n}
\GPSPartK{\dmed}{\dbig}{\dbig}{3}{l3k}
\GPSShade{Al3k}{\dmed}{\dbig}{\dmed}{\dmed}{0}{0}{pattern=crosshatch,pattern color=magenta}
\GPSBase{\dmed}{\dbig}{\dmed}{l2m}{below=0.5cm of l3k}
\GPSShade{Al2m}{\dmed}{\dmed}{\dmed}{\dmed}{0}{0}{pattern=crosshatch,pattern color=magenta}
\GPSPartN{\dmed}{\dbig}{\dmed}{9}{l2m}
\GPSBase{\dbig}{\dmed}{\dbig}{l2k}{below=0.5cm of l2m}
\GPSShade{Al2k}{\dbig}{\dbig}{\dbig}{\dbig}{0}{0}{pattern=crosshatch, pattern color=magenta}
\GPSPartK{\dbig}{\dmed}{\dbig}{3}{l2k}
\GPSShade{Cl2k}{\dbig}{\dmed}{\dbig}{\dmed}{0}{0}{pattern=crosshatch, pattern color=magenta}
\GPSShade{Bl2k}{\dbig}{\dmed}{\dmed}{\dmed}{0}{0}{pattern=grid, pattern color=green}
\GPSBase{\dbig}{\dmed}{\dmed}{innerkernel}{below=0.5cm of l2k}
\GPSShade{Ainnerkernel}{\dbig}{\dmed}{\dbig}{\dmed}{0}{0}{pattern=crosshatch, pattern color=magenta}
\GPSShade{Cinnerkernel}{\dbig}{\dmed}{\dbig}{\dmed}{0}{0}{pattern=crosshatch, pattern color=magenta}
\GPSShade{Binnerkernel}{\dmed}{\dmed}{\dmed}{\dmed}{0}{0}{pattern=grid, pattern color=green}

\path (Cinnerkernel) +(-3*\dbig/2 + \dmed, -\dbig/2) coordinate (bottomLeft);

\path (Bl3n) +(\dbig/2 + .2cm, \dbig/2 + .2cm) coordinate (topRightl3n);
\path (Bl3k) +(\dbig/2 + .2cm -.04cm, \dbig/2 +.2cm) coordinate (topRightl3k);
\path (Bl2m) +(\dmed + \dbig/2 + .2cm - .04cm*2, \dmed/2 + .2cm) coordinate (topRightl2m);
\path (Bl2k) +(\dbig - \dmed/2 + .2cm - .04cm*3, \dbig/2 + .2cm)  coordinate (topRightl2k);
\path (Binnerkernel) +(3*\dbig/2 - \dmed + .2cm - .04cm*4, \dbig/2 + .2cm) coordinate (topRightinnerkernel);

\path (Cl3n) +(-\dbig/2 - .2cm, \dbig/2 + .2cm) coordinate (topLeftl3n);
\path (Cl3k) +(-\dbig/2 - .2cm +.04cm, \dbig/2 +.2cm) coordinate (topLeftl3k);
\path (Cl2m) +(-\dmed - \dbig/2 - .2cm + .04cm*2, \dmed/2 + .2cm) coordinate (topLeftl2m);
\path (Cl2k) +(-\dbig + \dmed/2 - .2cm + .04cm*3, \dbig/2 + .2cm)  coordinate (topLeftl2k);
\path (Cinnerkernel) +(-3*\dbig/2 + \dmed - .2cm + .04cm*4, \dbig/2 + .2cm) coordinate (topLeftinnerkernel);

\draw [rounded corners] (bottomLeft) +(-.2cm, -.2cm) rectangle (topRightl3n);
\draw [rounded corners] (bottomLeft) +(-.2cm + 0.04cm, -.2cm + 0.04cm) rectangle (topRightl3k);
\draw [rounded corners] (bottomLeft) +(-.2cm + 0.04cm*2, -.2cm + 0.04cm*2) rectangle (topRightl2m);
\draw [rounded corners] (bottomLeft) +(-.2cm + 0.04cm*3, -.2cm + 0.04cm*3) rectangle (topRightl2k);
\draw [rounded corners] (bottomLeft) +(-.2cm + 0.04cm*4, -.2cm + 0.04cm*4) rectangle (topRightinnerkernel);

\draw [draw=none] (topLeftl3n) -- (topRightl3n) node [midway, above] {\footnotesize{Partition $m$ with blocksize 768}};
\draw [draw=none] (topLeftl3k) -- (topRightl3k) node [midway, above] {\footnotesize{Partition $k$ with blocksize 768}};
\draw [draw=none] (topLeftl2m) -- (topRightl2m) node [midway, above] {\footnotesize{Partition $n$ with blocksize 120}};
\draw [draw=none] (topLeftl2k) -- (topRightl2k) node [midway, above] {\footnotesize{Partition $k$ with blocksize 192}};
\draw [draw=none] (topLeftinnerkernel) -- (topRightinnerkernel) node [midway, above] {\footnotesize{Inner kernel}};
\end{tikzpicture} &
\begin{tikzpicture}
\GPSBase{\dbig}{\dbig}{\dbig}{l3n}{}
\GPSPartN{\dbig}{\dbig}{\dbig}{3}{l3n}
\GPSBase{\dbig}{\dmed}{\dbig}{l3k}{below=0.5cm of l3n}
\GPSPartK{\dbig}{\dmed}{\dbig}{3}{l3k}
\GPSShade{Bl3k}{\dbig}{\dmed}{\dmed}{\dmed}{0}{0}{pattern=crosshatch,pattern color=magenta}
\GPSBase{\dbig}{\dmed}{\dmed}{l2m}{below=0.5cm of l3k}
\GPSShade{Bl2m}{\dmed}{\dmed}{\dmed}{\dmed}{0}{0}{pattern=crosshatch,pattern color=magenta}
\GPSPartM{\dbig}{\dmed}{\dmed}{9}{l2m}
\GPSBase{\dmed}{\dbig}{\dbig}{l2k}{below=0.5cm of l2m}
\GPSShade{Bl2k}{\dbig}{\dbig}{\dbig}{\dbig}{0}{0}{pattern=crosshatch, pattern color=magenta}
\GPSPartK{\dmed}{\dbig}{\dbig}{3}{l2k}
\GPSShade{Cl2k}{\dmed}{\dbig}{\dmed}{\dbig}{0}{0}{pattern=crosshatch, pattern color=magenta}
\GPSShade{Al2k}{\dmed}{\dbig}{\dmed}{\dmed}{0}{0}{pattern=grid, pattern color=green}
\GPSBase{\dmed}{\dbig}{\dmed}{innerkernel}{below=0.5cm of l2k}
\GPSShade{Binnerkernel}{\dmed}{\dbig}{\dmed}{\dbig}{0}{0}{pattern=crosshatch, pattern color=magenta}
\GPSShade{Cinnerkernel}{\dmed}{\dbig}{\dmed}{\dbig}{0}{0}{pattern=crosshatch, pattern color=magenta}
\GPSShade{Ainnerkernel}{\dmed}{\dmed}{\dmed}{\dmed}{0}{0}{pattern=grid, pattern color=green}

\path (Cinnerkernel) +(-\dbig + \dmed/2, -\dmed/2) coordinate (bottomLeft);

\path (Bl3n) +(\dbig/2 + .2cm, \dbig/2 + .2cm) coordinate (topRightl3n);
\path (Bl3k) +(\dbig - \dmed/2 + .2cm -.04cm, \dbig/2 +.2cm) coordinate (topRightl3k);
\path (Bl2m) +(\dmed/2 + \dbig + .2cm - .04cm*2, \dbig/2 + .2cm) coordinate (topRightl2m);
\path (Bl2k) +(\dbig/2 + .2cm - .04cm*3, \dbig/2 + .2cm)  coordinate (topRightl2k);
\path (Binnerkernel) +(\dbig - \dmed/2 + .2cm - .04cm*4, \dmed/2 + .2cm) coordinate (topRightinnerkernel);

\path (Cl3n) +(-\dbig/2 - .2cm, \dbig/2 + .2cm) coordinate (topLeftl3n);
\path (Cl3k) +(-\dbig +\dmed/2 - .2cm +.04cm, \dbig/2 +.2cm) coordinate (topLeftl3k);
\path (Cl2m) +(-\dmed/2 - \dbig - .2cm + .04cm*2, \dbig/2 + .2cm) coordinate (topLeftl2m);
\path (Cl2k) +(-\dbig/2 - .2cm + .04cm*3, \dbig/2 + .2cm)  coordinate (topLeftl2k);
\path (Cinnerkernel) +(-\dbig + \dmed/2 - .2cm + .04cm*4, \dmed/2 + .2cm) coordinate (topLeftinnerkernel);

\draw [rounded corners] (bottomLeft) +(-.2cm, -.2cm) rectangle (topRightl3n);
\draw [rounded corners] (bottomLeft) +(-.2cm + 0.04cm, -.2cm + 0.04cm) rectangle (topRightl3k);
\draw [rounded corners] (bottomLeft) +(-.2cm + 0.04cm*2, -.2cm + 0.04cm*2) rectangle (topRightl2m);
\draw [rounded corners] (bottomLeft) +(-.2cm + 0.04cm*3, -.2cm + 0.04cm*3) rectangle (topRightl2k);
\draw [rounded corners] (bottomLeft) +(-.2cm + 0.04cm*4, -.2cm + 0.04cm*4) rectangle (topRightinnerkernel);

\draw [draw=none] (topLeftl3n) -- (topRightl3n) node [midway, above] {\footnotesize{Partition $n$ with blocksize 768}};
\draw [draw=none] (topLeftl3k) -- (topRightl3k) node [midway, above] {\footnotesize{Partition $k$ with blocksize 768}};
\draw [draw=none] (topLeftl2m) -- (topRightl2m) node [midway, above] {\footnotesize{Partition $m$ with blocksize 120}};
\draw [draw=none] (topLeftl2k) -- (topRightl2k) node [midway, above] {\footnotesize{Partition $k$ with blocksize 192}};
\draw [draw=none] (topLeftinnerkernel) -- (topRightinnerkernel) node [midway, above] {\footnotesize{Inner kernel}};

\end{tikzpicture} &

\begin{tikzpicture}
\GPSBase{\dbig}{\dbig}{\dbig}{l3n}{}
\GPSPartN{\dbig}{\dbig}{\dbig}{3}{l3n}
\GPSBase{\dbig}{\dmed}{\dbig}{l3k}{below=0.5cm of l3n}
\GPSPartM{\dbig}{\dmed}{\dbig}{3}{l3k}
\GPSShade{Cl3k}{\dbig}{\dmed}{\dmed}{\dmed}{0}{0}{pattern=crosshatch,pattern color=magenta}
\GPSBase{\dmed}{\dmed}{\dbig}{l2m}{below=0.5cm of l3k}
\GPSShade{Cl2m}{\dmed}{\dmed}{\dmed}{\dmed}{0}{0}{pattern=crosshatch,pattern color=magenta}
\GPSPartK{\dmed}{\dmed}{\dbig}{9}{l2m}
\GPSBase{\dbig}{\dbig}{\dmed}{l2k}{below=0.5cm of l2m}
\GPSShade{Cl2k}{\dbig}{\dbig}{\dbig}{\dbig}{0}{0}{pattern=crosshatch, pattern color=magenta}
\GPSPartM{\dbig}{\dbig}{\dmed}{3}{l2k}
\GPSShade{Bl2k}{\dmed}{\dbig}{\dmed}{\dbig}{0}{0}{pattern=crosshatch, pattern color=magenta}
\GPSShade{Al2k}{\dbig}{\dmed}{\dmed}{\dmed}{0}{0}{pattern=grid, pattern color=green}
\GPSBase{\dmed}{\dbig}{\dmed}{innerkernel}{below=0.5cm of l2k}
\GPSShade{Binnerkernel}{\dmed}{\dbig}{\dmed}{\dbig}{0}{0}{pattern=crosshatch, pattern color=magenta}
\GPSShade{Cinnerkernel}{\dmed}{\dbig}{\dmed}{\dbig}{0}{0}{pattern=crosshatch, pattern color=magenta}
\GPSShade{Ainnerkernel}{\dmed}{\dmed}{\dmed}{\dmed}{0}{0}{pattern=grid, pattern color=green}

\path (Cinnerkernel) +(-\dbig + \dmed/2, -\dmed/2) coordinate (bottomLeft);

\path (Bl3n) +(\dbig/2 + .2cm, \dbig/2 + .2cm) coordinate (topRightl3n);
\path (Bl3k) +(\dbig - \dmed/2 + .2cm -.04cm, \dbig/2 +.2cm) coordinate (topRightl3k);
\path (Bl2m) +(\dbig - \dmed/2 + .2cm - .04cm*2, \dbig/2 + .2cm) coordinate (topRightl2m);
\path (Bl2k) +(\dbig - \dmed/2 + .2cm - .04cm*3, \dbig/2 + .2cm)  coordinate (topRightl2k);
\path (Binnerkernel) +(\dbig - \dmed/2 + .2cm - .04cm*4, \dmed/2 + .2cm) coordinate (topRightinnerkernel);

\path (Cl3n) +(-\dbig/2 - .2cm, \dbig/2 + .2cm) coordinate (topLeftl3n);
\path (Cl3k) +(-\dbig +\dmed/2 - .2cm +.04cm, \dbig/2 +.2cm) coordinate (topLeftl3k);
\path (Cl2m) +(-\dbig + \dmed/2 - .2cm + .04cm*2, \dbig/2 + .2cm) coordinate (topLeftl2m);
\path (Cl2k) +(-\dbig + \dmed/2 - .2cm + .04cm*3, \dbig/2 + .2cm)  coordinate (topLeftl2k);
\path (Cinnerkernel) +(-\dbig + \dmed/2 - .2cm + .04cm*4, \dmed/2 + .2cm) coordinate (topLeftinnerkernel);

\draw [rounded corners] (bottomLeft) +(-.2cm, -.2cm) rectangle (topRightl3n);
\draw [rounded corners] (bottomLeft) +(-.2cm + 0.04cm, -.2cm + 0.04cm) rectangle (topRightl3k);
\draw [rounded corners] (bottomLeft) +(-.2cm + 0.04cm*2, -.2cm + 0.04cm*2) rectangle (topRightl2m);
\draw [rounded corners] (bottomLeft) +(-.2cm + 0.04cm*3, -.2cm + 0.04cm*3) rectangle (topRightl2k);
\draw [rounded corners] (bottomLeft) +(-.2cm + 0.04cm*4, -.2cm + 0.04cm*4) rectangle (topRightinnerkernel);

\draw [draw=none] (topLeftl3n) -- (topRightl3n) node [midway, above] {\footnotesize{Partition $n$ dimension with blocksize 624}};
\draw [draw=none] (topLeftl3k) -- (topRightl3k) node [midway, above] {\footnotesize{Partition $m$ dimension with blocksize 624}};
\draw [draw=none] (topLeftl2m) -- (topRightl2m) node [midway, above] {\footnotesize{Partition $k$ dimension with blocksize 156}};
\draw [draw=none] (topLeftl2k) -- (topRightl2k) node [midway, above] {\footnotesize{Partition $m$ dimension with blocksize 156}};
\draw [draw=none] (topLeftinnerkernel) -- (topRightinnerkernel) node [midway, above] {\footnotesize{Inner kernel}};

\end{tikzpicture} \\
\end{tabular}
}

\caption{Four MOMMS algorithms for MMM.}
\label{fig:sidebyside}
\end{figure*}

%% file: fig_rooflines2.tex
\newcommand{\roofline}[3]{
    \addplot[ #3 ] coordinates {
        ( 1.0, #1 )
        ( #2 / #1, #2 )
        ( 512, #2 )
    };
}
\pgfdeclaredecoration{ignore}{final}
{
\state{final}{}
}

\pgfdeclaremetadecoration{middle}{initial}{
    \state{initial}[
        width={(\pgfmetadecoratedpathlength - \the\pgfdecorationsegmentlength)/2},
        next state=middle
    ]
    {\decoration{moveto}}

    \state{middle}[
        width={\the\pgfdecorationsegmentlength},
        next state=final
    ]
    {\decoration{curveto}}

    \state{final}
    {\decoration{ignore}}
}
\tikzset{middle segment/.style={decoration={middle},decorate, segment length=#1}}

\begin{figure}
\begin{center}
\begin{tikzpicture}
    \begin{axis}[ 
                 width=0.95\linewidth, height=2.1in,
                 solid,
                 xlabel={flops per byte},
                 ylabel={GFLOPS},
                 xmin=1,xmax=512,ymin=4,ymax=512,
                 xmode = log, ymode = log,
                 log ticks with fixed point,
                 xtick pos=left, ytick pos=left,
                 /pgfplots/xtick={1,2,4,8,16,32,64,128,256,512},
                 legend style={legend pos=south east, fill=white},
                 clip=true]
    \roofline{6.25}{262.5}{color=magenta,dashed,thick};
    \roofline{50}{262.5}{color=black,solid,thick};
    \addplot[only marks, black, mark = square*] coordinates { (23.26, 250) };
    \addplot[only marks, black, mark = triangle*] coordinates { (64, 245) };
    \addplot[only marks, magenta, mark = square*] coordinates { (23.26, 105) };
    \addplot[only marks, magenta, mark = triangle*] coordinates { (64, 145) };
    \draw[thick,->,middle segment=.5cm] (23.26, 250) to[bend right=-60] (23.26, 105);
    \draw[thick,->,middle segment=.3cm] (64, 245) to[bend right=-60] (64, 145);
    \legend{
        \begin{minipage}{3.0cm}{\scriptsize Roofline Single Channel DDR-800}\end{minipage},
        \begin{minipage}{3.0cm}{\scriptsize Roofline Dual Channel DDR-3200}\end{minipage},
        \begin{minipage}{3.0cm}{\scriptsize MOMMS Goto}\end{minipage},
        \begin{minipage}{3.0cm}{\scriptsize MOMMS $B_3 A_2 C_0$}\end{minipage}
    }
\end{axis}
\end{tikzpicture}
\end{center}
\caption{
    Roofline models for  i7-7700K at 4.2GHz with 4 cores under two bandwidth conditions.
    The y-axis is measured and the x-axis is theoretical.
    The higher points for each algorithm represent high-bandwidth performance
    and the lower points represent low-bandwidth performance.
}
\label{fig:rooflines}
\end{figure}
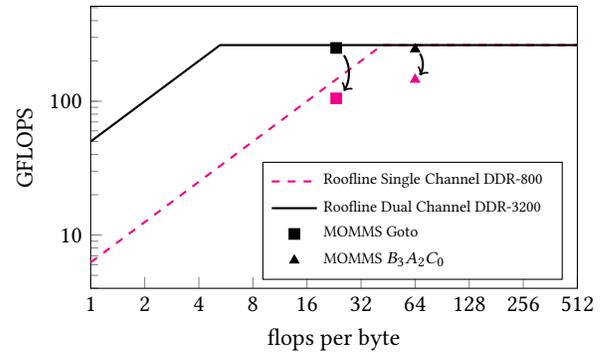

%% file: fig_l3.tex
\begin{figure}

\resizebox{\linewidth}{!}{
\begin{tabular}{@{}c @{\hspace{-1.75cm}} c@{}}
\input g_l3_800 & \input g_l3_1200 \\
\input g_l3_1600 & \input g_l3_3200x2 \\
\end{tabular}
}
\caption{Performance on  i7-7700K for square matrices,
varying problem size and available bandwidth.
Matrices are stored prepacked.}
\label{fig:l3}
\end{figure}

%% file: g_l3_800.tex
\begin{tikzpicture}
  \begin{axis}[width=2.5in, height=2in,
               solid,
               xlabel={$m=n=k$},
               ylabel={\small GFLOPS},
               title={
                   \begin{minipage}{3in}
                   \begin{center}
                    {\small 1 channel of DDR4-800} 
                   \end{center}
                   \end{minipage}
               },
               xmin=0,xmax=4000,ymin=0,ymax=262.5,
               legend style={legend pos=south east, fill=white},
               clip=false]
\addplot[color=black,solid,thick] coordinates {
    (   50  ,   2.8545  )
    (   100 ,   23.05396    )
    (   150 ,   64.09587    )
    (   200 ,   90.82291    )
    (   250 ,   117.94189   )
    (   300 ,   137.34971   )
    (   350 ,   144.42884   )
    (   400 ,   84.9422 )
    (   450 ,   96.08647    )
    (   500 ,   97.30587    )
    (   550 ,   105.16701   )
    (   600 ,   86.80498    )
    (   650 ,   86.74886    )
    (   700 ,   89.34553    )
    (   750 ,   91.04369    )
    (   800 ,   89.17903    )
    (   850 ,   85.79207    )
    (   900 ,   87.62235    )
    (   950 ,   90.06939    )
    (   1000    ,   83.05994    )
    (   1050    ,   86.22152    )
    (   1100    ,   90.46089    )
    (   1150    ,   94.51124    )
    (   1200    ,   87.1477 )
    (   1250    ,   89.38565    )
    (   1300    ,   93.37828    )
    (   1350    ,   87.38728    )
    (   1400    ,   89.86951    )
    (   1450    ,   93.04728    )
    (   1500    ,   96.27853    )
    (   1550    ,   89.84773    )
    (   1600    ,   92.59792    )
    (   1650    ,   95.12881    )
    (   1700    ,   97.9579 )
    (   1750    ,   92.29313    )
    (   1800    ,   94.68158    )
    (   1850    ,   95.90766    )
    (   1900    ,   98.86962    )
    (   1950    ,   93.73383    )
    (   2000    ,   96.97109    )
    (   2050    ,   99.65176    )
    (   2100    ,   102.36441   )
    (   2150    ,   96.25235    )
    (   2200    ,   98.56333    )
    (   2250    ,   100.97577   )
    (   2300    ,   103.32063   )
    (   2350    ,   97.82719    )
    (   2400    ,   99.58849    )
    (   2450    ,   100.42774   )
    (   2500    ,   87.03632    )
    (   2550    ,   98.79698    )
    (   2600    ,   100.77973   )
    (   2650    ,   102.8992    )
    (   2700    ,   98.53813    )
    (   2750    ,   99.7699 )
    (   2800    ,   101.53206   )
    (   2850    ,   103.2234    )
    (   2900    ,   99.10435    )
    (   2950    ,   100.91462   )
    (   3000    ,   102.80667   )
    (   3050    ,   102.08362   )
    (   3100    ,   97.73384    )
    (   3150    ,   97.31568    )
    (   3200    ,   101.11084   )
    (   3250    ,   102.11226   )
    (   3300    ,   99.41018    )
    (   3350    ,   100.53632   )
    (   3400    ,   101.70752   )
    (   3450    ,   102.19667   )
    (   3500    ,   100.07281   )
    (   3550    ,   101.20681   )
    (   3600    ,   103.16412   )
    (   3650    ,   99.65349    )
    (   3700    ,   101.09535   )
    (   3750    ,   102.33258   )
    (   3800    ,   103.0343    )
    (   3850    ,   100.8352    )
    (   3900    ,   101.84494   )
    (   3950    ,   102.61953   )
    (   4000    ,   104.35911   )
};
\addplot[color=magenta,dashed,thick] coordinates {
    (   50  ,   2.75349 )
    (   100 ,   23.97248    )
    (   150 ,   56.12652    )
    (   200 ,   107.93235   )
    (   250 ,   146.43175   )
    (   300 ,   175.12624   )
    (   350 ,   191.66937   )
    (   400 ,   138.05689   )
    (   450 ,   136.54866   )
    (   500 ,   129.6737    )
    (   550 ,   133.31122   )
    (   600 ,   118.10478   )
    (   650 ,   121.57724   )
    (   700 ,   125.64749   )
    (   750 ,   125.18767   )
    (   800 ,   88.3238 )
    (   850 ,   94.35388    )
    (   900 ,   102.3805    )
    (   950 ,   107.66286   )
    (   1000    ,   106.65083   )
    (   1050    ,   110.70155   )
    (   1100    ,   116.53801   )
    (   1150    ,   121.2603    )
    (   1200    ,   121.12671   )
    (   1250    ,   126.05385   )
    (   1300    ,   128.17625   )
    (   1350    ,   130.23464   )
    (   1400    ,   131.90797   )
    (   1450    ,   133.39721   )
    (   1500    ,   133.41779   )
    (   1550    ,   110.63106   )
    (   1600    ,   115.06873   )
    (   1650    ,   120.91473   )
    (   1700    ,   126.07246   )
    (   1750    ,   124.00105   )
    (   1800    ,   127.06264   )
    (   1850    ,   129.37267   )
    (   1900    ,   132.38224   )
    (   1950    ,   132.60029   )
    (   2000    ,   134.15867   )
    (   2050    ,   136.01166   )
    (   2100    ,   138.05845   )
    (   2150    ,   136.98609   )
    (   2200    ,   137.64608   )
    (   2250    ,   137.59866   )
    (   2300    ,   137.77883   )
    (   2350    ,   123.95301   )
    (   2400    ,   127.66173   )
    (   2450    ,   131.54821   )
    (   2500    ,   131.12738   )
    (   2550    ,   133.03775   )
    (   2600    ,   135.97811   )
    (   2650    ,   138.29591   )
    (   2700    ,   138.8544    )
    (   2750    ,   138.39528   )
    (   2800    ,   139.97848   )
    (   2850    ,   141.42211   )
    (   2900    ,   141.51021   )
    (   2950    ,   142.04479   )
    (   3000    ,   142.35269   )
    (   3050    ,   140.44992   )
    (   3100    ,   128.36085   )
    (   3150    ,   131.18855   )
    (   3200    ,   133.88549   )
    (   3250    ,   136.59218   )
    (   3300    ,   135.33002   )
    (   3350    ,   136.25178   )
    (   3400    ,   137.99936   )
    (   3450    ,   138.51485   )
    (   3500    ,   140.09716   )
    (   3550    ,   140.7777    )
    (   3600    ,   126.4625    )
    (   3650    ,   135.20222   )
    (   3700    ,   143.70678   )
    (   3750    ,   143.85026   )
    (   3800    ,   142.82617   )
    (   3850    ,   123.6747    )
    (   3900    ,   116.95714   )
    (   3950    ,   124.50861   )
    (   4000    ,   137.7425    )
};
\end{axis}
\end{tikzpicture}

%% file: g_l3_1200.tex
\begin{tikzpicture}
  \begin{axis}[width=2.5in, height=2in,
               solid,
               xlabel={$m=n=k$},
               title={
                   \begin{minipage}{3in}
                   \begin{center}
                    {\small 1 channel of DDR4-1200} 
                   \end{center}
                   \end{minipage}
               },
               xmin=0,xmax=4000,ymin=0,ymax=262.5,
               legend style={legend pos=south east, fill=white},
               clip=false]
\addplot[color=black,solid,thick] coordinates {
    (   50  ,   3.06207 )
    (   100 ,   23.94894    )
    (   150 ,   67.30817    )
    (   200 ,   85.51302    )
    (   250 ,   133.39423   )
    (   300 ,   150.53188   )
    (   350 ,   175.30804   )
    (   400 ,   123.73115   )
    (   450 ,   137.17468   )
    (   500 ,   139.55706   )
    (   550 ,   149.36361   )
    (   600 ,   125.47106   )
    (   650 ,   125.63018   )
    (   700 ,   129.6301    )
    (   750 ,   132.94993   )
    (   800 ,   124.72049   )
    (   850 ,   128.32378   )
    (   900 ,   130.9897    )
    (   950 ,   133.61345   )
    (   1000    ,   124.52622   )
    (   1050    ,   132.59249   )
    (   1100    ,   135.96516   )
    (   1150    ,   144.7898    )
    (   1200    ,   131.0198    )
    (   1250    ,   135.43447   )
    (   1300    ,   141.56042   )
    (   1350    ,   131.06907   )
    (   1400    ,   135.48124   )
    (   1450    ,   139.40604   )
    (   1500    ,   143.90557   )
    (   1550    ,   134.67267   )
    (   1600    ,   139.50079   )
    (   1650    ,   142.71385   )
    (   1700    ,   146.95308   )
    (   1750    ,   138.71563   )
    (   1800    ,   141.97496   )
    (   1850    ,   143.05587   )
    (   1900    ,   147.51042   )
    (   1950    ,   140.41819   )
    (   2000    ,   145.82134   )
    (   2050    ,   149.79725   )
    (   2100    ,   153.6523    )
    (   2150    ,   144.18667   )
    (   2200    ,   147.81858   )
    (   2250    ,   151.63527   )
    (   2300    ,   155.27965   )
    (   2350    ,   146.92506   )
    (   2400    ,   148.93407   )
    (   2450    ,   149.81481   )
    (   2500    ,   144.9622    )
    (   2550    ,   148.33522   )
    (   2600    ,   151.27705   )
    (   2650    ,   154.48123   )
    (   2700    ,   147.58523   )
    (   2750    ,   149.56577   )
    (   2800    ,   152.15696   )
    (   2850    ,   154.71356   )
    (   2900    ,   148.00767   )
    (   2950    ,   151.00405   )
    (   3000    ,   153.59979   )
    (   3050    ,   152.60258   )
    (   3100    ,   146.81424   )
    (   3150    ,   149.54615   )
    (   3200    ,   151.90558   )
    (   3250    ,   154.69637   )
    (   3300    ,   141.95049   )
    (   3350    ,   150.92177   )
    (   3400    ,   153.22893   )
    (   3450    ,   155.17206   )
    (   3500    ,   150.53418   )
    (   3550    ,   152.67247   )
    (   3600    ,   154.70674   )
    (   3650    ,   149.16207   )
    (   3700    ,   151.7554    )
    (   3750    ,   153.85084   )
    (   3800    ,   156.14876   )
    (   3850    ,   151.32918   )
    (   3900    ,   153.04631   )
    (   3950    ,   153.74138   )
    (   4000    ,   156.88302   )
};
\addplot[color=magenta,dashed,thick] coordinates {
    (   50  ,   2.93128 )
    (   100 ,   24.18994    )
    (   150 ,   70.00695    )
    (   200 ,   114.65096   )
    (   250 ,   148.55839   )
    (   300 ,   172.35491   )
    (   350 ,   190.53904   )
    (   400 ,   176.79998   )
    (   450 ,   176.29101   )
    (   500 ,   169.09533   )
    (   550 ,   174.59704   )
    (   600 ,   159.0459    )
    (   650 ,   161.02821   )
    (   700 ,   165.64919   )
    (   750 ,   166.41943   )
    (   800 ,   122.17101   )
    (   850 ,   132.67082   )
    (   900 ,   143.99235   )
    (   950 ,   146.75871   )
    (   1000    ,   144.31124   )
    (   1050    ,   150.80173   )
    (   1100    ,   158.43983   )
    (   1150    ,   163.96209   )
    (   1200    ,   161.48509   )
    (   1250    ,   168.22859   )
    (   1300    ,   171.06401   )
    (   1350    ,   174.67788   )
    (   1400    ,   175.87436   )
    (   1450    ,   176.39989   )
    (   1500    ,   177.55891   )
    (   1550    ,   148.2895    )
    (   1600    ,   153.6921    )
    (   1650    ,   162.67012   )
    (   1700    ,   168.45209   )
    (   1750    ,   166.60538   )
    (   1800    ,   170.28443   )
    (   1850    ,   170.8527    )
    (   1900    ,   173.87561   )
    (   1950    ,   173.96036   )
    (   2000    ,   175.23985   )
    (   2050    ,   176.9042    )
    (   2100    ,   179.05996   )
    (   2150    ,   176.67237   )
    (   2200    ,   177.81684   )
    (   2250    ,   177.94764   )
    (   2300    ,   178.13649   )
    (   2350    ,   163.44552   )
    (   2400    ,   167.72662   )
    (   2450    ,   166.65349   )
    (   2500    ,   167.26438   )
    (   2550    ,   170.02435   )
    (   2600    ,   173.06038   )
    (   2650    ,   177.46801   )
    (   2700    ,   180.80549   )
    (   2750    ,   180.08162   )
    (   2800    ,   180.9297    )
    (   2850    ,   182.03248   )
    (   2900    ,   182.43507   )
    (   2950    ,   182.59027   )
    (   3000    ,   183.05525   )
    (   3050    ,   180.86173   )
    (   3100    ,   167.72468   )
    (   3150    ,   171.74881   )
    (   3200    ,   172.65089   )
    (   3250    ,   177.95694   )
    (   3300    ,   176.52633   )
    (   3350    ,   177.04592   )
    (   3400    ,   178.82588   )
    (   3450    ,   180.39387   )
    (   3500    ,   179.69041   )
    (   3550    ,   180.86519   )
    (   3600    ,   181.94697   )
    (   3650    ,   184.39993   )
    (   3700    ,   184.45218   )
    (   3750    ,   184.66874   )
    (   3800    ,   184.06049   )
    (   3850    ,   170.64278   )
    (   3900    ,   173.60094   )
    (   3950    ,   175.57634   )
    (   4000    ,   179.09643   )
};
\end{axis}
\end{tikzpicture}

%% file: g_l3_1600.tex
\begin{tikzpicture}
  \begin{axis}[width=2.5in, height=2in,
               solid,
               xlabel={$m=n=k$},
               ylabel={\small GFLOPS},
               title={
                   \begin{minipage}{3in}
                   \begin{center}
                    {\small 1 channel of DDR4-1600} 
                   \end{center}
                   \end{minipage}
               },
               xmin=0,xmax=4000,ymin=0,ymax=262.5,
               legend style={legend pos=south east, fill=white},
               clip=false]
\addplot[color=black,solid,thick] coordinates {
    (   50  ,   3.04366 )
    (   100 ,   25.37234    )
    (   150 ,   71.65605    )
    (   200 ,   111.32216   )
    (   250 ,   143.22905   )
    (   300 ,   169.06118   )
    (   350 ,   183.69043   )
    (   400 ,   154.70428   )
    (   450 ,   172.9521    )
    (   500 ,   175.32516   )
    (   550 ,   185.42181   )
    (   600 ,   161.59651   )
    (   650 ,   160.90047   )
    (   700 ,   165.13897   )
    (   750 ,   170.09758   )
    (   800 ,   159.34655   )
    (   850 ,   163.52132   )
    (   900 ,   169.66249   )
    (   950 ,   178.18867   )
    (   1000    ,   162.36393   )
    (   1050    ,   172.85409   )
    (   1100    ,   177.88751   )
    (   1150    ,   185.00432   )
    (   1200    ,   172.89817   )
    (   1250    ,   175.55611   )
    (   1300    ,   183.21757   )
    (   1350    ,   171.45512   )
    (   1400    ,   176.30831   )
    (   1450    ,   182.75829   )
    (   1500    ,   189.3219    )
    (   1550    ,   177.40549   )
    (   1600    ,   183.0928    )
    (   1650    ,   187.49608   )
    (   1700    ,   193.32757   )
    (   1750    ,   182.02355   )
    (   1800    ,   186.76213   )
    (   1850    ,   188.11334   )
    (   1900    ,   194.10007   )
    (   1950    ,   185.10043   )
    (   2000    ,   191.36341   )
    (   2050    ,   196.95089   )
    (   2100    ,   202.01183   )
    (   2150    ,   190.17785   )
    (   2200    ,   194.42456   )
    (   2250    ,   199.146 )
    (   2300    ,   203.54663   )
    (   2350    ,   193.11864   )
    (   2400    ,   195.87969   )
    (   2450    ,   197.03854   )
    (   2500    ,   190.69648   )
    (   2550    ,   194.82441   )
    (   2600    ,   198.84931   )
    (   2650    ,   202.63734   )
    (   2700    ,   193.63251   )
    (   2750    ,   195.99613   )
    (   2800    ,   198.97194   )
    (   2850    ,   201.94585   )
    (   2900    ,   193.32923   )
    (   2950    ,   197.0135    )
    (   3000    ,   200.29764   )
    (   3050    ,   199.02057   )
    (   3100    ,   190.1641    )
    (   3150    ,   194.83625   )
    (   3200    ,   197.94497   )
    (   3250    ,   201.12283   )
    (   3300    ,   193.96518   )
    (   3350    ,   196.02403   )
    (   3400    ,   199.19543   )
    (   3450    ,   201.81394   )
    (   3500    ,   195.65933   )
    (   3550    ,   198.33311   )
    (   3600    ,   201.59287   )
    (   3650    ,   194.45186   )
    (   3700    ,   197.33358   )
    (   3750    ,   200.17977   )
    (   3800    ,   203.40777   )
    (   3850    ,   197.34506   )
    (   3900    ,   199.7541    )
    (   3950    ,   201.05483   )
    (   4000    ,   204.47686   )
};
\addplot[color=magenta,dashed,thick] coordinates {
    (   50  ,   3.12758 )
    (   100 ,   21.87729    )
    (   150 ,   74.56421    )
    (   200 ,   115.86646   )
    (   250 ,   153.80981   )
    (   300 ,   181.51505   )
    (   350 ,   200.35608   )
    (   400 ,   198.07864   )
    (   450 ,   202.96343   )
    (   500 ,   199.31277   )
    (   550 ,   208.58277   )
    (   600 ,   193.74404   )
    (   650 ,   190.71313   )
    (   700 ,   196.48003   )
    (   750 ,   193.75929   )
    (   800 ,   151.86885   )
    (   850 ,   163.53131   )
    (   900 ,   176.27841   )
    (   950 ,   177.51954   )
    (   1000    ,   174.55358   )
    (   1050    ,   180.52405   )
    (   1100    ,   184.68289   )
    (   1150    ,   189.92375   )
    (   1200    ,   189.88487   )
    (   1250    ,   194.20955   )
    (   1300    ,   197.46774   )
    (   1350    ,   198.88509   )
    (   1400    ,   199.39219   )
    (   1450    ,   200.41906   )
    (   1500    ,   200.58287   )
    (   1550    ,   176.20798   )
    (   1600    ,   182.10617   )
    (   1650    ,   185.20255   )
    (   1700    ,   192.12206   )
    (   1750    ,   193.99465   )
    (   1800    ,   198.68994   )
    (   1850    ,   199.66612   )
    (   1900    ,   202.44476   )
    (   1950    ,   202.68722   )
    (   2000    ,   203.7008    )
    (   2050    ,   204.80538   )
    (   2100    ,   206.95856   )
    (   2150    ,   203.74089   )
    (   2200    ,   204.76973   )
    (   2250    ,   205.04254   )
    (   2300    ,   205.87071   )
    (   2350    ,   191.94791   )
    (   2400    ,   196.60804   )
    (   2450    ,   195.14453   )
    (   2500    ,   195.62071   )
    (   2550    ,   197.87439   )
    (   2600    ,   200.96078   )
    (   2650    ,   201.89514   )
    (   2700    ,   202.42853   )
    (   2750    ,   207.74756   )
    (   2800    ,   208.73672   )
    (   2850    ,   210.02445   )
    (   2900    ,   210.03511   )
    (   2950    ,   210.39136   )
    (   3000    ,   211.14075   )
    (   3050    ,   208.51094   )
    (   3100    ,   195.55052   )
    (   3150    ,   200.17045   )
    (   3200    ,   202.9753    )
    (   3250    ,   206.06672   )
    (   3300    ,   205.46883   )
    (   3350    ,   205.18311   )
    (   3400    ,   206.50367   )
    (   3450    ,   207.8447    )
    (   3500    ,   207.28875   )
    (   3550    ,   207.9315    )
    (   3600    ,   208.95009   )
    (   3650    ,   205.59559   )
    (   3700    ,   205.61215   )
    (   3750    ,   205.34804   )
    (   3800    ,   205.16304   )
    (   3850    ,   197.49428   )
    (   3900    ,   201.66073   )
    (   3950    ,   200.66924   )
    (   4000    ,   203.94979   )
};
\end{axis}
\end{tikzpicture}

%% file: g_l3_3200x2.tex
\begin{tikzpicture}
  \begin{axis}[width=2.5in, height=2in,
               solid,
               xlabel={$m=n=k$},
               title={
                   \begin{minipage}{3in}
                   \begin{center}
                    {\small 2 channels of DDR4-3200} 
                   \end{center}
                   \end{minipage}
               },
               xmin=0,xmax=4000,ymin=0,ymax=262.5,
               legend style={legend pos=south east, fill=white},
               clip=false]
\addplot[color=black,solid,thick] coordinates {
    (   50  ,   3.09993 )
    (   100 ,   22.49997    )
    (   150 ,   74.54939    )
    (   200 ,   99.59973    )
    (   250 ,   150.80008   )
    (   300 ,   179.9892    )
    (   350 ,   196.60622   )
    (   400 ,   207.95357   )
    (   450 ,   218.1185    )
    (   500 ,   226.81765   )
    (   550 ,   233.23021   )
    (   600 ,   230.60404   )
    (   650 ,   233.05182   )
    (   700 ,   238.70019   )
    (   750 ,   240.49726   )
    (   800 ,   240.64629   )
    (   850 ,   243.4208    )
    (   900 ,   246.92286   )
    (   950 ,   245.35366   )
    (   1000    ,   243.20595   )
    (   1050    ,   245.59158   )
    (   1100    ,   246.97759   )
    (   1150    ,   249.47613   )
    (   1200    ,   247.24235   )
    (   1250    ,   231.31111   )
    (   1300    ,   237.88266   )
    (   1350    ,   239.58851   )
    (   1400    ,   244.92241   )
    (   1450    ,   245.83215   )
    (   1500    ,   247.13456   )
    (   1550    ,   244.56397   )
    (   1600    ,   248.03501   )
    (   1650    ,   248.70457   )
    (   1700    ,   250.0398    )
    (   1750    ,   247.63146   )
    (   1800    ,   250.19093   )
    (   1850    ,   251.31815   )
    (   1900    ,   252.36444   )
    (   1950    ,   249.77797   )
    (   2000    ,   252.34823   )
    (   2050    ,   252.89722   )
    (   2100    ,   253.85844   )
    (   2150    ,   249.93232   )
    (   2200    ,   252.05774   )
    (   2250    ,   252.42873   )
    (   2300    ,   253.15679   )
    (   2350    ,   251.42898   )
    (   2400    ,   252.74547   )
    (   2450    ,   247.94765   )
    (   2500    ,   245.11746   )
    (   2550    ,   248.0201    )
    (   2600    ,   248.99462   )
    (   2650    ,   249.09859   )
    (   2700    ,   246.92545   )
    (   2750    ,   249.626 )
    (   2800    ,   249.94425   )
    (   2850    ,   250.21909   )
    (   2900    ,   248.52735   )
    (   2950    ,   249.83919   )
    (   3000    ,   250.4136    )
    (   3050    ,   245.69214   )
    (   3100    ,   245.31201   )
    (   3150    ,   245.78807   )
    (   3200    ,   246.50595   )
    (   3250    ,   246.77689   )
    (   3300    ,   245.97742   )
    (   3350    ,   246.85569   )
    (   3400    ,   247.35615   )
    (   3450    ,   247.84425   )
    (   3500    ,   247.21879   )
    (   3550    ,   247.89498   )
    (   3600    ,   248.64836   )
    (   3650    ,   245.94673   )
    (   3700    ,   248.5608    )
    (   3750    ,   248.8582    )
    (   3800    ,   249.4798    )
    (   3850    ,   247.41119   )
    (   3900    ,   249.60161   )
    (   3950    ,   249.96825   )
    (   4000    ,   250.23413   )
};
\addplot[color=magenta,dashed,thick] coordinates {
    (   50  ,   3.24912 )
    (   100 ,   24.47381    )
    (   150 ,   73.95558    )
    (   200 ,   117.21783   )
    (   250 ,   153.80451   )
    (   300 ,   184.2777    )
    (   350 ,   201.68166   )
    (   400 ,   209.21699   )
    (   450 ,   219.52013   )
    (   500 ,   224.26754   )
    (   550 ,   232.93404   )
    (   600 ,   233.8514    )
    (   650 ,   236.05432   )
    (   700 ,   239.49303   )
    (   750 ,   240.83452   )
    (   800 ,   230.62835   )
    (   850 ,   235.15994   )
    (   900 ,   238.68271   )
    (   950 ,   241.37929   )
    (   1000    ,   238.99668   )
    (   1050    ,   240.67798   )
    (   1100    ,   242.08436   )
    (   1150    ,   246.15385   )
    (   1200    ,   233.18722   )
    (   1250    ,   236.53826   )
    (   1300    ,   237.78501   )
    (   1350    ,   238.60806   )
    (   1400    ,   239.5963    )
    (   1450    ,   239.58326   )
    (   1500    ,   240.60395   )
    (   1550    ,   235.33869   )
    (   1600    ,   239.70033   )
    (   1650    ,   240.60949   )
    (   1700    ,   242.41521   )
    (   1750    ,   241.7644    )
    (   1800    ,   243.33357   )
    (   1850    ,   244.23691   )
    (   1900    ,   246.22606   )
    (   1950    ,   245.13739   )
    (   2000    ,   246.12747   )
    (   2050    ,   246.31702   )
    (   2100    ,   248.10046   )
    (   2150    ,   246.29712   )
    (   2200    ,   247.02332   )
    (   2250    ,   247.10851   )
    (   2300    ,   247.97523   )
    (   2350    ,   241.91342   )
    (   2400    ,   246.85501   )
    (   2450    ,   242.73749   )
    (   2500    ,   242.46513   )
    (   2550    ,   243.22769   )
    (   2600    ,   244.375 )
    (   2650    ,   244.64491   )
    (   2700    ,   244.80046   )
    (   2750    ,   245.23858   )
    (   2800    ,   245.80926   )
    (   2850    ,   246.30038   )
    (   2900    ,   245.99453   )
    (   2950    ,   246.16009   )
    (   3000    ,   246.53408   )
    (   3050    ,   246.54116   )
    (   3100    ,   244.78012   )
    (   3150    ,   245.84358   )
    (   3200    ,   246.631 )
    (   3250    ,   247.46133   )
    (   3300    ,   247.16782   )
    (   3350    ,   247.57922   )
    (   3400    ,   248.07479   )
    (   3450    ,   248.94966   )
    (   3500    ,   248.29401   )
    (   3550    ,   248.75507   )
    (   3600    ,   249.44757   )
    (   3650    ,   245.61941   )
    (   3700    ,   245.91349   )
    (   3750    ,   245.78667   )
    (   3800    ,   246.07987   )
    (   3850    ,   243.61143   )
    (   3900    ,   245.76108   )
    (   3950    ,   245.77779   )
    (   4000    ,   246.81511   )
};
\legend{
    \begin{minipage}{1.0in}{\scriptsize  MOMMS Goto}\end{minipage},
    \begin{minipage}{1.0in}{\scriptsize MOMMS $B_3 A_2 C_0$}\end{minipage}
}
\end{axis}
\end{tikzpicture}

%% file: fig_l3_packing.tex
\begin{figure}
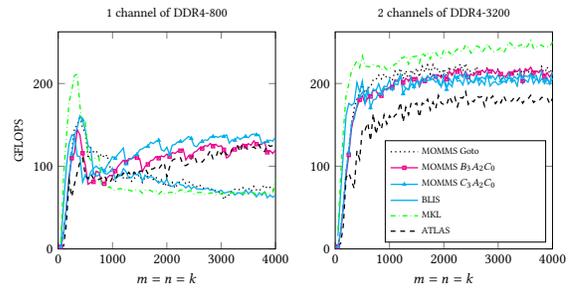


\resizebox{\linewidth}{!}{
\begin{tabular}{@{}c @{\hspace{-1.75cm}} c@{}}
\input g_l3_packing_800 & \input g_l3_packing_3200x2 \\
\end{tabular}}
\caption{
Comparison with state-of-the-art open source and vendor libraries,
for both high and low bandwidth scenarios.
Matrices stored in column-major order.}
\label{fig:l3_packing}
\end{figure}

%% file: g_l3_packing_800.tex
\begin{tikzpicture}
  \begin{axis}[width=2.5in, height=2.5in,
               solid,
               xlabel={$m=n=k$},
               ylabel={\small GFLOPS},
               title={
                   \begin{minipage}{3in}
                   \begin{center}
                    {\small 1 channel of DDR4-800}
                   \end{center}
                   \end{minipage}
               },
               xmin=0,xmax=4000,ymin=0,ymax=262.5,
               legend style={legend pos=south east, fill=white},
               clip=false]
\addplot[color=black,dotted,thick] coordinates {
    (   50  ,   2.83682 )
    (   100 ,   20.13531    )
    (   150 ,   49.18821    )
    (   200 ,   83.87854    )
    (   250 ,   118.81949   )
    (   300 ,   146.25149   )
    (   350 ,   149.15923   )
    (   400 ,   161.11471   )
    (   450 ,   145.23752   )
    (   500 ,   141.65012   )
    (   550 ,   109.24184   )
    (   600 ,   122.14355   )
    (   650 ,   117.34928   )
    (   700 ,   118.82014   )
    (   750 ,   110.67545   )
    (   800 ,   114.98699   )
    (   850 ,   110.41671   )
    (   900 ,   113.78317   )
    (   950 ,   100.35809   )
    (   1000    ,   85.60086    )
    (   1050    ,   80.32817    )
    (   1100    ,   89.82297    )
    (   1150    ,   88.95691    )
    (   1200    ,   85.83049    )
    (   1250    ,   84.51274    )
    (   1300    ,   88.41581    )
    (   1350    ,   81.84037    )
    (   1400    ,   85.53972    )
    (   1450    ,   87.12534    )
    (   1500    ,   89.44569    )
    (   1550    ,   80.4537 )
    (   1600    ,   86.85733    )
    (   1650    ,   87.78379    )
    (   1700    ,   89.91487    )
    (   1750    ,   83.98151    )
    (   1800    ,   86.88975    )
    (   1850    ,   80.63893    )
    (   1900    ,   83.61943    )
    (   1950    ,   75.8469 )
    (   2000    ,   79.62408    )
    (   2050    ,   74.06139    )
    (   2100    ,   80.35416    )
    (   2150    ,   73.38735    )
    (   2200    ,   78.70814    )
    (   2250    ,   81.95239    )
    (   2300    ,   83.91953    )
    (   2350    ,   79.14343    )
    (   2400    ,   81.07894    )
    (   2450    ,   79.22854    )
    (   2500    ,   76.14337    )
    (   2550    ,   75.37591    )
    (   2600    ,   77.74674    )
    (   2650    ,   73.62772    )
    (   2700    ,   73.5072 )
    (   2750    ,   73.34238    )
    (   2800    ,   74.1486 )
    (   2850    ,   70.03286    )
    (   2900    ,   70.62576    )
    (   2950    ,   68.45839    )
    (   3000    ,   69.32473    )
    (   3050    ,   59.65569    )
    (   3100    ,   64.90067    )
    (   3150    ,   65.25927    )
    (   3200    ,   71.16858    )
    (   3250    ,   72.16361    )
    (   3300    ,   72.10177    )
    (   3350    ,   71.52895    )
    (   3400    ,   74.13336    )
    (   3450    ,   70.26765    )
    (   3500    ,   73.55419    )
    (   3550    ,   69.54425    )
    (   3600    ,   75.69977    )
    (   3650    ,   71.3189 )
    (   3700    ,   70.24562    )
    (   3750    ,   73.93207    )
    (   3800    ,   76.91303    )
    (   3850    ,   73.20249    )
    (   3900    ,   74.52726    )
    (   3950    ,   70.88363    )
    (   4000    ,   73.43104    )
};
\addplot[color=magenta,solid,thick,mark=square,mark repeat={4},mark options={scale=.5}] coordinates {
    (   50  ,   2.72337 )
    (   100 ,   18.97911    )
    (   150 ,   50.09239    )
    (   200 ,   86.96787    )
    (   250 ,   109.71495   )
    (   300 ,   115.87013   )
    (   350 ,   144.11353   )
    (   400 ,   137.90726   )
    (   450 ,   115.69049   )
    (   500 ,   101.11583   )
    (   550 ,   78.24292    )
    (   600 ,   80.52429    )
    (   650 ,   81.14903    )
    (   700 ,   94.02985    )
    (   750 ,   92.51798    )
    (   800 ,   81.57541    )
    (   850 ,   78.81933    )
    (   900 ,   85.79452    )
    (   950 ,   89.07427    )
    (   1000    ,   87.97257    )
    (   1050    ,   94.09658    )
    (   1100    ,   99.6691 )
    (   1150    ,   103.59441   )
    (   1200    ,   107.27079   )
    (   1250    ,   104.95087   )
    (   1300    ,   109.21251   )
    (   1350    ,   108.98613   )
    (   1400    ,   110.75878   )
    (   1450    ,   110.4866    )
    (   1500    ,   109.1856    )
    (   1550    ,   93.45452    )
    (   1600    ,   99.79259    )
    (   1650    ,   102.24194   )
    (   1700    ,   106.62616   )
    (   1750    ,   107.74672   )
    (   1800    ,   111.66363   )
    (   1850    ,   109.67091   )
    (   1900    ,   112.49443   )
    (   1950    ,   111.30054   )
    (   2000    ,   114.76203   )
    (   2050    ,   111.66278   )
    (   2100    ,   116.71346   )
    (   2150    ,   115.44448   )
    (   2200    ,   118.14377   )
    (   2250    ,   118.80701   )
    (   2300    ,   120.73102   )
    (   2350    ,   107.53749   )
    (   2400    ,   112.71597   )
    (   2450    ,   111.70659   )
    (   2500    ,   116.73996   )
    (   2550    ,   118.29964   )
    (   2600    ,   122.02279   )
    (   2650    ,   121.40657   )
    (   2700    ,   124.90663   )
    (   2750    ,   124.17477   )
    (   2800    ,   127.54346   )
    (   2850    ,   124.21495   )
    (   2900    ,   126.58538   )
    (   2950    ,   123.75005   )
    (   3000    ,   124.8036    )
    (   3050    ,   119.86021   )
    (   3100    ,   110.00086   )
    (   3150    ,   112.22491   )
    (   3200    ,   119.32692   )
    (   3250    ,   120.33343   )
    (   3300    ,   124.30728   )
    (   3350    ,   122.72435   )
    (   3400    ,   126.9487    )
    (   3450    ,   124.17969   )
    (   3500    ,   127.27385   )
    (   3550    ,   125.32914   )
    (   3600    ,   130.701 )
    (   3650    ,   126.69886   )
    (   3700    ,   126.21818   )
    (   3750    ,   126.72923   )
    (   3800    ,   129.26516   )
    (   3850    ,   116.80948   )
    (   3900    ,   118.8733    )
    (   3950    ,   115.41089   )
    (   4000    ,   119.48744   )
};
\addplot[color=cyan,solid,thick,mark=triangle, mark repeat={4},mark options={scale=.5}] coordinates {
    (   50  ,   2.81126 )
    (   100 ,   18.95932    )
    (   150 ,   48.85923    )
    (   200 ,   84.33525    )
    (   250 ,   114.89604   )
    (   300 ,   137.48924   )
    (   350 ,   146.38352   )
    (   400 ,   160.87152   )
    (   450 ,   155.90248   )
    (   500 ,   143.39393   )
    (   550 ,   124.81629   )
    (   600 ,   121.66413   )
    (   650 ,   91.09056    )
    (   700 ,   87.94721    )
    (   750 ,   89.27853    )
    (   800 ,   96.6449 )
    (   850 ,   94.32151    )
    (   900 ,   100.74818   )
    (   950 ,   105.5658    )
    (   1000    ,   111.61634   )
    (   1050    ,   109.48877   )
    (   1100    ,   114.37474   )
    (   1150    ,   117.97742   )
    (   1200    ,   121.90699   )
    (   1250    ,   102.71797   )
    (   1300    ,   107.51754   )
    (   1350    ,   108.81248   )
    (   1400    ,   113.26724   )
    (   1450    ,   114.23816   )
    (   1500    ,   119.83996   )
    (   1550    ,   121.17103   )
    (   1600    ,   123.24675   )
    (   1650    ,   123.89226   )
    (   1700    ,   128.96735   )
    (   1750    ,   127.27331   )
    (   1800    ,   130.21759   )
    (   1850    ,   131.37796   )
    (   1900    ,   119.91797   )
    (   1950    ,   118.22849   )
    (   2000    ,   123.02504   )
    (   2050    ,   114.30017   )
    (   2100    ,   125.48821   )
    (   2150    ,   126.24432   )
    (   2200    ,   129.48698   )
    (   2250    ,   129.97615   )
    (   2300    ,   130.23891   )
    (   2350    ,   133.1952    )
    (   2400    ,   135.9252    )
    (   2450    ,   133.50273   )
    (   2500    ,   124.83729   )
    (   2550    ,   123.62059   )
    (   2600    ,   128.13712   )
    (   2650    ,   130.04355   )
    (   2700    ,   131.72211   )
    (   2750    ,   131.28016   )
    (   2800    ,   135.45842   )
    (   2850    ,   135.83183   )
    (   2900    ,   135.97493   )
    (   2950    ,   134.92405   )
    (   3000    ,   136.58883   )
    (   3050    ,   135.88058   )
    (   3100    ,   137.93898   )
    (   3150    ,   127.87591   )
    (   3200    ,   131.91504   )
    (   3250    ,   131.35532   )
    (   3300    ,   135.27636   )
    (   3350    ,   132.24293   )
    (   3400    ,   134.61407   )
    (   3450    ,   135.97737   )
    (   3500    ,   134.44903   )
    (   3550    ,   138.10834   )
    (   3600    ,   138.94944   )
    (   3650    ,   137.92626   )
    (   3700    ,   137.97515   )
    (   3750    ,   126.78893   )
    (   3800    ,   129.5792    )
    (   3850    ,   129.83044   )
    (   3900    ,   131.49421   )
    (   3950    ,   130.29623   )
    (   4000    ,   135.173     )
};
\addplot[color=cyan,solid,thick] coordinates {
    (   50  ,   6.75402 )
    (   100 ,   55.14503    )
    (   150 ,   99.07384    )
    (   200 ,   126.96397   )
    (   250 ,   137.88448   )
    (   300 ,   136.16283   )
    (   350 ,   112.38517   )
    (   400 ,   111.86152   )
    (   450 ,   80.69119    )
    (   500 ,   89.85608    )
    (   550 ,   83.15979    )
    (   600 ,   93.21456    )
    (   650 ,   97.34783    )
    (   700 ,   100.16451   )
    (   750 ,   96.95366    )
    (   800 ,   98.69892    )
    (   850 ,   91.25739    )
    (   900 ,   90.47723    )
    (   950 ,   93.30181    )
    (   1000    ,   97.66678    )
    (   1050    ,   84.39899    )
    (   1100    ,   87.38938    )
    (   1150    ,   88.79856    )
    (   1200    ,   97.16322    )
    (   1250    ,   93.27048    )
    (   1300    ,   85.61875    )
    (   1350    ,   86.25666    )
    (   1400    ,   90.52101    )
    (   1450    ,   89.46734    )
    (   1500    ,   92.81949    )
    (   1550    ,   81.92448    )
    (   1600    ,   89.35026    )
    (   1650    ,   85.55204    )
    (   1700    ,   87.59688    )
    (   1750    ,   86.28762    )
    (   1800    ,   83.0627 )
    (   1850    ,   79.61261    )
    (   1900    ,   82.3797 )
    (   1950    ,   79.27591    )
    (   2000    ,   84.33147    )
    (   2050    ,   76.62363    )
    (   2100    ,   75.64828    )
    (   2150    ,   77.57614    )
    (   2200    ,   81.06885    )
    (   2250    ,   77.68538    )
    (   2300    ,   77.06742    )
    (   2350    ,   73.73859    )
    (   2400    ,   76.7116 )
    (   2450    ,   72.62047    )
    (   2500    ,   75.4382 )
    (   2550    ,   72.52474    )
    (   2600    ,   72.49812    )
    (   2650    ,   71.93667    )
    (   2700    ,   72.52781    )
    (   2750    ,   73.66296    )
    (   2800    ,   70.85765    )
    (   2850    ,   71.58714    )
    (   2900    ,   68.46877    )
    (   2950    ,   69.39994    )
    (   3000    ,   71.08054    )
    (   3050    ,   67.44237    )
    (   3100    ,   68.29186    )
    (   3150    ,   69.44806    )
    (   3200    ,   68.8341 )
    (   3250    ,   65.10931    )
    (   3300    ,   67.99418    )
    (   3350    ,   67.66706    )
    (   3400    ,   65.26303    )
    (   3450    ,   71.56904    )
    (   3500    ,   66.69493    )
    (   3550    ,   69.72074    )
    (   3600    ,   66.76778    )
    (   3650    ,   62.9798 )
    (   3700    ,   67.94322    )
    (   3750    ,   64.97369    )
    (   3800    ,   69.00226    )
    (   3850    ,   61.83454    )
    (   3900    ,   65.09799    )
    (   3950    ,   62.48656    )
    (   4000    ,   65.01367    )
};
\addplot[color=green,dashdotted,thick] coordinates {
    (   50  ,   7.9811  )
    (   100 ,   88.5975 )
    (   150 ,   134.2669    )
    (   200 ,   177.38949   )
    (   250 ,   191.49106   )
    (   300 ,   209.54761   )
    (   350 ,   211.43083   )
    (   400 ,   172.83961   )
    (   450 ,   149.91573   )
    (   500 ,   152.74069   )
    (   550 ,   112.78617   )
    (   600 ,   107.16295   )
    (   650 ,   107.60308   )
    (   700 ,   110.4846    )
    (   750 ,   110.82045   )
    (   800 ,   78.05201    )
    (   850 ,   78.12012    )
    (   900 ,   74.12821    )
    (   950 ,   75.16478    )
    (   1000    ,   70.94078    )
    (   1050    ,   71.01156    )
    (   1100    ,   72.4355 )
    (   1150    ,   73.11571    )
    (   1200    ,   70.77899    )
    (   1250    ,   72.52593    )
    (   1300    ,   71.1777 )
    (   1350    ,   70.08392    )
    (   1400    ,   71.64088    )
    (   1450    ,   71.10924    )
    (   1500    ,   70.70008    )
    (   1550    ,   67.30399    )
    (   1600    ,   70.83734    )
    (   1650    ,   70.8535 )
    (   1700    ,   70.88697    )
    (   1750    ,   68.5682 )
    (   1800    ,   70.03763    )
    (   1850    ,   66.28213    )
    (   1900    ,   69.4871 )
    (   1950    ,   66.45789    )
    (   2000    ,   67.41955    )
    (   2050    ,   68.39863    )
    (   2100    ,   68.41497    )
    (   2150    ,   66.33427    )
    (   2200    ,   69.2262 )
    (   2250    ,   71.44408    )
    (   2300    ,   71.1745 )
    (   2350    ,   67.04272    )
    (   2400    ,   69.97845    )
    (   2450    ,   70.48666    )
    (   2500    ,   66.38939    )
    (   2550    ,   67.97345    )
    (   2600    ,   68.33573    )
    (   2650    ,   68.30701    )
    (   2700    ,   66.08441    )
    (   2750    ,   66.71696    )
    (   2800    ,   67.6397 )
    (   2850    ,   67.74736    )
    (   2900    ,   65.73489    )
    (   2950    ,   66.73166    )
    (   3000    ,   66.59019    )
    (   3050    ,   67.82709    )
    (   3100    ,   65.25482    )
    (   3150    ,   66.30686    )
    (   3200    ,   68.24104    )
    (   3250    ,   68.01815    )
    (   3300    ,   66.36779    )
    (   3350    ,   67.23742    )
    (   3400    ,   68.24857    )
    (   3450    ,   67.47568    )
    (   3500    ,   66.52977    )
    (   3550    ,   66.47899    )
    (   3600    ,   67.84733    )
    (   3650    ,   65.28293    )
    (   3700    ,   63.88355    )
    (   3750    ,   67.50226    )
    (   3800    ,   67.68627    )
    (   3850    ,   66.4968 )
    (   3900    ,   66.76543    )
    (   3950    ,   67.2221 )
    (   4000    ,   84.31261    )
};
\addplot[color=black,dashed,thick] coordinates {
    (   50  ,   2.543   )
    (   100 ,   3.755   )
    (   150 ,   30.189  )
    (   200 ,   82.703  )
    (   250 ,   63.067  )
    (   300 ,   78.406  )
    (   350 ,   92.093  )
    (   400 ,   104.914 )
    (   450 ,   109.701 )
    (   500 ,   88.598  )
    (   550 ,   84.881  )
    (   600 ,   90.878  )
    (   650 ,   84.826  )
    (   700 ,   83.027  )
    (   750 ,   83.991  )
    (   800 ,   85.835  )
    (   850 ,   89.155  )
    (   900 ,   81.647  )
    (   950 ,   79.87   )
    (   1000    ,   86.334  )
    (   1050    ,   84.127  )
    (   1100    ,   90.03   )
    (   1150    ,   86.524  )
    (   1200    ,   91.837  )
    (   1250    ,   88.382  )
    (   1300    ,   92.663  )
    (   1350    ,   87.793  )
    (   1400    ,   92.912  )
    (   1450    ,   89.4    )
    (   1500    ,   94.045  )
    (   1550    ,   94.338  )
    (   1600    ,   97.364  )
    (   1650    ,   91.264  )
    (   1700    ,   96.224  )
    (   1750    ,   96.891  )
    (   1800    ,   99.052  )
    (   1850    ,   96.674  )
    (   1900    ,   101.355 )
    (   1950    ,   104.525 )
    (   2000    ,   87.263  )
    (   2050    ,   98.888  )
    (   2100    ,   107.25  )
    (   2150    ,   108.904 )
    (   2200    ,   109.347 )
    (   2250    ,   110.374 )
    (   2300    ,   101.224 )
    (   2350    ,   100.747 )
    (   2400    ,   103.048 )
    (   2450    ,   109.345 )
    (   2500    ,   114.073 )
    (   2550    ,   104.848 )
    (   2600    ,   106.746 )
    (   2650    ,   105.017 )
    (   2700    ,   108.997 )
    (   2750    ,   112.928 )
    (   2800    ,   118.893 )
    (   2850    ,   108.902 )
    (   2900    ,   110.432 )
    (   2950    ,   119.956 )
    (   3000    ,   114.545 )
    (   3050    ,   119.948 )
    (   3100    ,   114.127 )
    (   3150    ,   112.836 )
    (   3200    ,   120.424 )
    (   3250    ,   123.532 )
    (   3300    ,   124.788 )
    (   3350    ,   117.269 )
    (   3400    ,   127.563 )
    (   3450    ,   122.274 )
    (   3500    ,   119.976 )
    (   3550    ,   121.878 )
    (   3600    ,   123.406 )
    (   3650    ,   120.394 )
    (   3700    ,   125.646 )
    (   3750    ,   118.267 )
    (   3800    ,   117.691 )
    (   3850    ,   126.292 )
    (   3900    ,   124.106 )
    (   3950    ,   124.824 )
    (   4000    ,   129.226 )
};
\end{axis}
\end{tikzpicture}

%% file: g_l3_packing_3200x2.tex
\begin{tikzpicture}
  \begin{axis}[width=2.5in, height=2.5in,
               solid,
               xlabel={$m=n=k$},
               title={
                   \begin{minipage}{3in}
                   \begin{center}
                    {\small 2 channels of DDR4-3200}
                   \end{center}
                   \end{minipage}
               },
               xmin=0,xmax=4000,ymin=0,ymax=262.5,
               legend style={legend pos=south east, fill=white},
               clip=false]
\addplot[color=black,dotted,thick] coordinates {
    (   50  ,   2.99211 )
    (   100 ,   21.14321    )
    (   150 ,   53.91848    )
    (   200 ,   92.37982    )
    (   250 ,   120.31216   )
    (   300 ,   151.59129   )
    (   350 ,   170.00565   )
    (   400 ,   181.16968   )
    (   450 ,   188.90337   )
    (   500 ,   198.49052   )
    (   550 ,   188.44406   )
    (   600 ,   194.9053    )
    (   650 ,   195.60666   )
    (   700 ,   204.57673   )
    (   750 ,   204.76243   )
    (   800 ,   210.15259   )
    (   850 ,   205.61954   )
    (   900 ,   211.63367   )
    (   950 ,   207.03538   )
    (   1000    ,   207.53069   )
    (   1050    ,   203.71137   )
    (   1100    ,   214.96542   )
    (   1150    ,   210.81916   )
    (   1200    ,   218.2813    )
    (   1250    ,   205.91138   )
    (   1300    ,   215.66074   )
    (   1350    ,   203.79684   )
    (   1400    ,   216.44678   )
    (   1450    ,   210.28135   )
    (   1500    ,   218.65954   )
    (   1550    ,   207.9906    )
    (   1600    ,   220.82557   )
    (   1650    ,   213.96773   )
    (   1700    ,   222.45832   )
    (   1750    ,   211.7189    )
    (   1800    ,   222.72596   )
    (   1850    ,   215.87037   )
    (   1900    ,   217.9903    )
    (   1950    ,   209.5544    )
    (   2000    ,   217.68307   )
    (   2050    ,   209.03823   )
    (   2100    ,   220.28627   )
    (   2150    ,   209.2812    )
    (   2200    ,   220.7161    )
    (   2250    ,   216.43892   )
    (   2300    ,   223.47767   )
    (   2350    ,   215.14199   )
    (   2400    ,   224.18438   )
    (   2450    ,   214.05845   )
    (   2500    ,   215.33083   )
    (   2550    ,   212.31915   )
    (   2600    ,   220.33732   )
    (   2650    ,   213.66959   )
    (   2700    ,   216.52847   )
    (   2750    ,   212.52084   )
    (   2800    ,   219.67068   )
    (   2850    ,   214.00206   )
    (   2900    ,   215.85959   )
    (   2950    ,   212.53314   )
    (   3000    ,   218.31135   )
    (   3050    ,   205.22391   )
    (   3100    ,   207.38618   )
    (   3150    ,   207.46556   )
    (   3200    ,   214.79094   )
    (   3250    ,   209.88809   )
    (   3300    ,   215.19807   )
    (   3350    ,   210.32979   )
    (   3400    ,   217.65828   )
    (   3450    ,   212.14957   )
    (   3500    ,   217.2618    )
    (   3550    ,   212.43263   )
    (   3600    ,   219.6254    )
    (   3650    ,   209.56241   )
    (   3700    ,   216.3047    )
    (   3750    ,   212.43703   )
    (   3800    ,   220.43279   )
    (   3850    ,   210.61042   )
    (   3900    ,   220.08038   )
    (   3950    ,   212.13437   )
    (   4000    ,   219.02409   )
};
\addplot[color=magenta,solid,thick,mark=square,mark repeat={4},mark options={scale=.5}] coordinates {
    (   50  ,   2.74795 )
    (   100 ,   19.85131    )
    (   150 ,   51.10849    )
    (   200 ,   88.02139    )
    (   250 ,   113.60166   )
    (   300 ,   145.32067   )
    (   350 ,   161.70763   )
    (   400 ,   171.21249   )
    (   450 ,   180.26385   )
    (   500 ,   185.08729   )
    (   550 ,   175.92509   )
    (   600 ,   182.59166   )
    (   650 ,   184.32389   )
    (   700 ,   191.53622   )
    (   750 ,   191.28901   )
    (   800 ,   185.6163    )
    (   850 ,   185.21035   )
    (   900 ,   193.06251   )
    (   950 ,   193.57108   )
    (   1000    ,   195.52083   )
    (   1050    ,   195.53778   )
    (   1100    ,   204.29588   )
    (   1150    ,   203.40614   )
    (   1200    ,   208.73324   )
    (   1250    ,   201.18159   )
    (   1300    ,   206.16775   )
    (   1350    ,   203.39708   )
    (   1400    ,   207.68817   )
    (   1450    ,   205.16279   )
    (   1500    ,   207.46678   )
    (   1550    ,   196.20745   )
    (   1600    ,   203.87522   )
    (   1650    ,   203.00332   )
    (   1700    ,   208.48278   )
    (   1750    ,   205.65529   )
    (   1800    ,   210.83066   )
    (   1850    ,   208.48657   )
    (   1900    ,   210.42933   )
    (   1950    ,   206.91559   )
    (   2000    ,   211.18252   )
    (   2050    ,   205.433 )
    (   2100    ,   214.74945   )
    (   2150    ,   209.8111    )
    (   2200    ,   215.79345   )
    (   2250    ,   214.22268   )
    (   2300    ,   217.75425   )
    (   2350    ,   208.14331   )
    (   2400    ,   214.56021   )
    (   2450    ,   207.91469   )
    (   2500    ,   211.31326   )
    (   2550    ,   209.64952   )
    (   2600    ,   214.5423    )
    (   2650    ,   212.76025   )
    (   2700    ,   215.83756   )
    (   2750    ,   213.52738   )
    (   2800    ,   217.53171   )
    (   2850    ,   214.68986   )
    (   2900    ,   216.16493   )
    (   2950    ,   212.15233   )
    (   3000    ,   216.18056   )
    (   3050    ,   211.42348   )
    (   3100    ,   208.43063   )
    (   3150    ,   209.18432   )
    (   3200    ,   214.29963   )
    (   3250    ,   211.92953   )
    (   3300    ,   216.37396   )
    (   3350    ,   214.1475    )
    (   3400    ,   217.72478   )
    (   3450    ,   203.16236   )
    (   3500    ,   217.90748   )
    (   3550    ,   215.74329   )
    (   3600    ,   219.65832   )
    (   3650    ,   213.11782   )
    (   3700    ,   215.20672   )
    (   3750    ,   213.24576   )
    (   3800    ,   217.25119   )
    (   3850    ,   208.50149   )
    (   3900    ,   213.05154   )
    (   3950    ,   207.96594   )
    (   4000    ,   213.53716   )
};
\addplot[color=cyan,solid,thick,mark=triangle,mark repeat={4},mark options={scale=.5}] coordinates {
    (   50  ,   2.91681 )
    (   100 ,   19.9112 )
    (   150 ,   52.32842    )
    (   200 ,   94.58948    )
    (   250 ,   117.24319   )
    (   300 ,   151.66835   )
    (   350 ,   157.05301   )
    (   400 ,   173.68296   )
    (   450 ,   183.89587   )
    (   500 ,   189.48854   )
    (   550 ,   184.09721   )
    (   600 ,   194.23107   )
    (   650 ,   170.60161   )
    (   700 ,   180.3017    )
    (   750 ,   181.25878   )
    (   800 ,   188.76241   )
    (   850 ,   187.4151    )
    (   900 ,   194.53971   )
    (   950 ,   191.79881   )
    (   1000    ,   194.73163   )
    (   1050    ,   193.46951   )
    (   1100    ,   202.34191   )
    (   1150    ,   202.55112   )
    (   1200    ,   210.83714   )
    (   1250    ,   187.50947   )
    (   1300    ,   194.61103   )
    (   1350    ,   192.37796   )
    (   1400    ,   199.58296   )
    (   1450    ,   196.30534   )
    (   1500    ,   202.65689   )
    (   1550    ,   201.23061   )
    (   1600    ,   206.30149   )
    (   1650    ,   203.71512   )
    (   1700    ,   209.42798   )
    (   1750    ,   205.33382   )
    (   1800    ,   210.23385   )
    (   1850    ,   208.62503   )
    (   1900    ,   199.94124   )
    (   1950    ,   196.84772   )
    (   2000    ,   203.80057   )
    (   2050    ,   192.8315    )
    (   2100    ,   206.07212   )
    (   2150    ,   200.82087   )
    (   2200    ,   208.04204   )
    (   2250    ,   207.19794   )
    (   2300    ,   210.80247   )
    (   2350    ,   207.90032   )
    (   2400    ,   213.37133   )
    (   2450    ,   207.71544   )
    (   2500    ,   202.79461   )
    (   2550    ,   200.66888   )
    (   2600    ,   205.92668   )
    (   2650    ,   204.85045   )
    (   2700    ,   207.687 )
    (   2750    ,   205.63395   )
    (   2800    ,   210.22205   )
    (   2850    ,   207.43883   )
    (   2900    ,   209.18719   )
    (   2950    ,   204.24911   )
    (   3000    ,   210.87216   )
    (   3050    ,   202.84621   )
    (   3100    ,   208.13253   )
    (   3150    ,   201.83186   )
    (   3200    ,   207.15119   )
    (   3250    ,   204.11893   )
    (   3300    ,   208.39901   )
    (   3350    ,   205.38308   )
    (   3400    ,   209.80454   )
    (   3450    ,   206.77009   )
    (   3500    ,   209.25251   )
    (   3550    ,   208.25179   )
    (   3600    ,   212.35633   )
    (   3650    ,   207.49869   )
    (   3700    ,   209.70983   )
    (   3750    ,   200.56743   )
    (   3800    ,   205.54482   )
    (   3850    ,   203.13731   )
    (   3900    ,   207.08381   )
    (   3950    ,   198.92299   )
    (   4000    ,   206.21653   )
};
\addplot[color=cyan,solid,thick] coordinates {
    (   50  ,   15.42258    )
    (   100 ,   67.90019    )
    (   150 ,   116.24303   )
    (   200 ,   161.3505    )
    (   250 ,   175.98693   )
    (   300 ,   173.6033    )
    (   350 ,   176.50643   )
    (   400 ,   201.13294   )
    (   450 ,   189.97357   )
    (   500 ,   202.89145   )
    (   550 ,   175.72108   )
    (   600 ,   189.11188   )
    (   650 ,   184.91081   )
    (   700 ,   193.30772   )
    (   750 ,   191.19361   )
    (   800 ,   199.62488   )
    (   850 ,   189.65749   )
    (   900 ,   196.87336   )
    (   950 ,   195.9719    )
    (   1000    ,   204.79109   )
    (   1050    ,   192.13253   )
    (   1100    ,   199.77954   )
    (   1150    ,   198.39885   )
    (   1200    ,   208.84771   )
    (   1250    ,   198.55447   )
    (   1300    ,   193.76516   )
    (   1350    ,   192.54902   )
    (   1400    ,   207.47147   )
    (   1450    ,   194.66763   )
    (   1500    ,   203.77016   )
    (   1550    ,   197.10593   )
    (   1600    ,   207.61825   )
    (   1650    ,   201.47144   )
    (   1700    ,   203.83587   )
    (   1750    ,   204.0754    )
    (   1800    ,   205.64071   )
    (   1850    ,   201.52188   )
    (   1900    ,   204.643 )
    (   1950    ,   195.9825    )
    (   2000    ,   210.81907   )
    (   2050    ,   195.61377   )
    (   2100    ,   200.31131   )
    (   2150    ,   196.21492   )
    (   2200    ,   210.3101    )
    (   2250    ,   202.63921   )
    (   2300    ,   204.01146   )
    (   2350    ,   198.36433   )
    (   2400    ,   210.89143   )
    (   2450    ,   199.11653   )
    (   2500    ,   203.48705   )
    (   2550    ,   201.54619   )
    (   2600    ,   210.42013   )
    (   2650    ,   198.2339    )
    (   2700    ,   203.50892   )
    (   2750    ,   198.39975   )
    (   2800    ,   212.82174   )
    (   2850    ,   196.96139   )
    (   2900    ,   201.24301   )
    (   2950    ,   202.02571   )
    (   3000    ,   198.42776   )
    (   3050    ,   199.99367   )
    (   3100    ,   202.88662   )
    (   3150    ,   200.95495   )
    (   3200    ,   212.31101   )
    (   3250    ,   201.3994    )
    (   3300    ,   204.28538   )
    (   3350    ,   197.05789   )
    (   3400    ,   213.94821   )
    (   3450    ,   199.84949   )
    (   3500    ,   204.19128   )
    (   3550    ,   201.61819   )
    (   3600    ,   211.32929   )
    (   3650    ,   199.70855   )
    (   3700    ,   205.81122   )
    (   3750    ,   202.07001   )
    (   3800    ,   214.30389   )
    (   3850    ,   199.85346   )
    (   3900    ,   205.05273   )
    (   3950    ,   201.04315   )
    (   4000    ,   213.5431    )
};
\addplot[color=green,dashdotted,thick] coordinates {
    (   50  ,   22.96739    )
    (   100 ,   87.53118    )
    (   150 ,   142.7514    )
    (   200 ,   185.7657    )
    (   250 ,   198.92802   )
    (   300 ,   215.1223    )
    (   350 ,   221.44576   )
    (   400 ,   227.49287   )
    (   450 ,   219.35394   )
    (   500 ,   233.19873   )
    (   550 ,   221.03496   )
    (   600 ,   220.51111   )
    (   650 ,   220.15246   )
    (   700 ,   222.61914   )
    (   750 ,   220.51583   )
    (   800 ,   221.50224   )
    (   850 ,   216.43874   )
    (   900 ,   220.45839   )
    (   950 ,   216.79757   )
    (   1000    ,   217.10015   )
    (   1050    ,   220.55045   )
    (   1100    ,   223.52018   )
    (   1150    ,   224.36078   )
    (   1200    ,   230.86566   )
    (   1250    ,   230.83371   )
    (   1300    ,   233.81383   )
    (   1350    ,   223.50331   )
    (   1400    ,   232.33398   )
    (   1450    ,   232.42673   )
    (   1500    ,   233.20881   )
    (   1550    ,   224.81415   )
    (   1600    ,   237.25537   )
    (   1650    ,   233.285 )
    (   1700    ,   237.06815   )
    (   1750    ,   229.67073   )
    (   1800    ,   238.045 )
    (   1850    ,   237.30062   )
    (   1900    ,   232.33989   )
    (   1950    ,   231.05258   )
    (   2000    ,   240.82807   )
    (   2050    ,   235.76723   )
    (   2100    ,   238.27115   )
    (   2150    ,   233.03661   )
    (   2200    ,   238.08794   )
    (   2250    ,   241.72724   )
    (   2300    ,   238.8846    )
    (   2350    ,   239.24918   )
    (   2400    ,   246.78616   )
    (   2450    ,   243.17903   )
    (   2500    ,   236.71357   )
    (   2550    ,   240.4869    )
    (   2600    ,   245.34412   )
    (   2650    ,   244.36774   )
    (   2700    ,   236.65707   )
    (   2750    ,   241.88104   )
    (   2800    ,   248.14606   )
    (   2850    ,   242.96934   )
    (   2900    ,   238.24535   )
    (   2950    ,   238.77819   )
    (   3000    ,   241.00117   )
    (   3050    ,   243.52595   )
    (   3100    ,   236.91905   )
    (   3150    ,   243.00279   )
    (   3200    ,   250.1056    )
    (   3250    ,   247.0977    )
    (   3300    ,   242.61803   )
    (   3350    ,   244.9172    )
    (   3400    ,   246.1192    )
    (   3450    ,   247.28752   )
    (   3500    ,   242.23682   )
    (   3550    ,   246.07754   )
    (   3600    ,   252.0323    )
    (   3650    ,   242.57763   )
    (   3700    ,   243.79841   )
    (   3750    ,   246.33055   )
    (   3800    ,   246.28537   )
    (   3850    ,   244.1068    )
    (   3900    ,   244.2921    )
    (   3950    ,   244.48915   )
    (   4000    ,   250.20468   )
};
\addplot[color=black,dashed,thick] coordinates {
    (   50  ,   3.509   )
    (   100 ,   6.645   )
    (   150 ,   28.674  )
    (   200 ,   86.346  )
    (   250 ,   91.814  )
    (   300 ,   82.672  )
    (   350 ,   72.95   )
    (   400 ,   118.81  )
    (   450 ,   108.248 )
    (   500 ,   145.703 )
    (   550 ,   143.944 )
    (   600 ,   152.577 )
    (   650 ,   130.552 )
    (   700 ,   163.702 )
    (   750 ,   138.462 )
    (   800 ,   153.005 )
    (   850 ,   144.56  )
    (   900 ,   161.86  )
    (   950 ,   151.06  )
    (   1000    ,   160.982 )
    (   1050    ,   156.558 )
    (   1100    ,   161.529 )
    (   1150    ,   163.984 )
    (   1200    ,   177.409 )
    (   1250    ,   153.392 )
    (   1300    ,   171.186 )
    (   1350    ,   162.367 )
    (   1400    ,   176.981 )
    (   1450    ,   174.158 )
    (   1500    ,   176.242 )
    (   1550    ,   158.988 )
    (   1600    ,   176.268 )
    (   1650    ,   174.562 )
    (   1700    ,   180.925 )
    (   1750    ,   168.042 )
    (   1800    ,   177.466 )
    (   1850    ,   169.536 )
    (   1900    ,   185.499 )
    (   1950    ,   176.453 )
    (   2000    ,   174.482 )
    (   2050    ,   173.605 )
    (   2100    ,   184.461 )
    (   2150    ,   177.649 )
    (   2200    ,   183.604 )
    (   2250    ,   172.121 )
    (   2300    ,   182.75  )
    (   2350    ,   182.443 )
    (   2400    ,   190.271 )
    (   2450    ,   174.811 )
    (   2500    ,   179.578 )
    (   2550    ,   177.837 )
    (   2600    ,   184.87  )
    (   2650    ,   183.271 )
    (   2700    ,   184.01  )
    (   2750    ,   172.698 )
    (   2800    ,   184.099 )
    (   2850    ,   181.03  )
    (   2900    ,   184.953 )
    (   2950    ,   175.75  )
    (   3000    ,   180.508 )
    (   3050    ,   176.681 )
    (   3100    ,   185.201 )
    (   3150    ,   178.263 )
    (   3200    ,   178.021 )
    (   3250    ,   178.935 )
    (   3300    ,   183.32  )
    (   3350    ,   179.207 )
    (   3400    ,   183.366 )
    (   3450    ,   176.206 )
    (   3500    ,   180.935 )
    (   3550    ,   181.547 )
    (   3600    ,   185.807 )
    (   3650    ,   177.052 )
    (   3700    ,   179.647 )
    (   3750    ,   178.953 )
    (   3800    ,   183.226 )
    (   3850    ,   182.835 )
    (   3900    ,   182.923 )
    (   3950    ,   175.304 )
    (   4000    ,   181.942 )
};
\legend{
    \begin{minipage}{1.0in}{\scriptsize  MOMMS Goto}\end{minipage},
    \begin{minipage}{1.0in}{\scriptsize MOMMS $B_3 A_2 C_0$}\end{minipage},
    \begin{minipage}{1.0in}{\scriptsize MOMMS $C_3 A_2 C_0$}\end{minipage},
    \begin{minipage}{1.0in}{\scriptsize  BLIS}\end{minipage},
    \begin{minipage}{1.0in}{\scriptsize MKL}\end{minipage},
    \begin{minipage}{1.0in}{\scriptsize ATLAS}\end{minipage}
}
\end{axis}
\end{tikzpicture}

%% file: fig_l4_cache.tex
\begin{figure}
\def\dbig{1.8cm}
\def\rsFour{1.2cm}
\def\dmed{0.6cm}
\def\dsma{0.2cm}

\newcommand{\lthreeresident}{pattern=crosshatch}

\begin{tikzpicture}[
    >=latex,
    grow=right,
    level 1/.style={sibling distance=1cm,level distance=5.0cm},
    level 2/.style={sibling distance=2.5cm, level distance=5.0cm},
    level 3/.style={sibling distance=2.5cm, level distance=4.5cm},
    level 4/.style={sibling distance=1.0cm, level distance=4.0cm},
    edge from parent/.style={->,draw},
    edge from parent path={(\tikzparentnode.east) -- (\tikzchildnode.west)}  
]
\GPSBase{\dbig}{\dbig}{\dbig}{l4m}{}
\draw let \p1 = (Cl4m) in (\x1 - \dbig/2,\y1+\dbig/2-\rsFour) -- (\x1+\dbig/2,\y1+\dbig/2-\rsFour);
\draw let \p1 = (Al4m) in (\x1 - \dbig/2,\y1+\dbig/2-\rsFour) -- (\x1+\dbig/2,\y1+\dbig/2-\rsFour);

\GPSBase{2*\dbig/3}{\dbig}{\dbig}{l4n}{below=.5cm of l4m}
\GPSShade{Cl4n}{\rsFour}{\dbig}{\rsFour}{\rsFour}{0}{0}{pattern=dots, pattern color=cyan}
\draw let \p1 = (Cl4n) in (\x1 - \dbig/2+\rsFour,\y1-\rsFour/2) -- (\x1-\dbig/2+\rsFour,\y1+\rsFour/2);
\draw let \p1 = (Bl4n) in (\x1 - \dbig/2+\rsFour,\y1-\dbig/2) -- (\x1-\dbig/2+\rsFour,\y1+\dbig/2);

\GPSBase{\rsFour}{\rsFour}{\dbig}{l3k}{below=0.5cm of l4n}
\GPSPartK{\rsFour}{\rsFour}{\dbig}{9}{l3k}
\GPSShade{Bl3k}{\dbig}{\rsFour}{\dbig / 9}{\rsFour}{0}{0}{pattern=crosshatch, pattern color=magenta}
\GPSShade{Cl3k}{\rsFour}{\rsFour}{\rsFour}{\rsFour}{0}{0}{pattern=dots, pattern color=cyan}

\GPSBase{\rsFour}{\rsFour}{\dsma}{l2m}{below=0.5cm of l3k}
\GPSShade{Bl2m}{\dsma}{\rsFour}{\dsma}{\rsFour}{0}{0}{pattern=crosshatch, pattern color=magenta}
\GPSShade{Cl2m}{\rsFour}{\rsFour}{\rsFour}{\rsFour}{0}{0}{pattern=dots, pattern color=cyan}
\GPSShade{Al2m}{\rsFour}{\dsma}{\dsma}{\dsma}{0}{0}{pattern=grid, pattern color=green}
\GPSPartM{\rsFour}{\rsFour}{\dsma}{6}{l2m}


\def\kernn{2.4cm}
\GPSBase{\dmed}{\kernn}{\dmed}{innerkernel}{below=0.5cm of l2m}
\GPSShade{Binnerkernel}{\dmed}{\kernn}{\dmed}{\kernn}{0}{0}{pattern=crosshatch, pattern color=magenta}
\GPSShade{Ainnerkernel}{\dmed}{\dmed}{\dmed}{\dmed}{0}{0}{pattern=grid, pattern color=green}
\GPSShade{Cinnerkernel}{\dmed}{\kernn}{\dmed}{\kernn}{0}{0}{pattern=dots, pattern color=cyan}

\path (Cinnerkernel) +(-\kernn/2, -\dmed/2) coordinate (bottomLeft);

\path (Bl4m) +(\dbig/2 + .24cm, \dbig/2 + .24cm) coordinate (topRightl4m);
\path (Bl4n) +(\dbig/2 + .24cm -.04cm, \dbig/2 + .24cm) coordinate (topRightl4n);
\path (Bl3k) +(\dbig/3 + \dbig/3 + .24cm -.04cm*2, \dbig/2 +.24cm) coordinate (topRightl3k);
\path (Bl2m) +(\dbig/3 + \dbig - \dbig/6 - \dsma/2 + .24cm - .04cm*3, \rsFour/2 + .24cm) coordinate (topRightl2m);
\path (Binnerkernel) +(\kernn/2 + .24cm - .04cm*4, \dmed/2 + .24cm) coordinate (topRightinnerkernel);

\path (Cl4m) +(-\dbig/2 - .24cm, \dbig/2 + .24cm) coordinate (topLeftl4m);
\path (Cl4n) +(-\dbig/2 - .24cm, \dbig/2 + .24cm + .04cm) coordinate (topLeftl4n);
\path (Cl3k) +(-\dbig +\dmed/2 - .24cm +.04cm*2, \dbig/2 +.24cm) coordinate (topLeftl3k);
\path (Cl2m) +(-\dmed/2 - \dbig - .24cm + .04cm*3, \rsFour/2 + .24cm) coordinate (topLeftl2m);
\path (Cinnerkernel) +(-\dbig + \dmed/2 - .24cm + .04cm*4, \dmed/2 + .24cm) coordinate (topLeftinnerkernel);

\draw [rounded corners] (bottomLeft) +(-.24cm, -.24cm) rectangle (topRightl4m);
\draw [rounded corners] (bottomLeft) +(-.24cm + 0.04cm*1, -.24cm + 0.04cm*1) rectangle (topRightl4n);
\draw [rounded corners] (bottomLeft) +(-.24cm + 0.04cm*2, -.24cm + 0.04cm*2) rectangle (topRightl3k);
\draw [rounded corners] (bottomLeft) +(-.24cm + 0.04cm*3, -.24cm + 0.04cm*3) rectangle (topRightl2m);
\draw [rounded corners] (bottomLeft) +(-.24cm + 0.04cm*4, -.24cm + 0.04cm*4) rectangle (topRightinnerkernel);

\draw [draw=none] (topLeftl4m) -- (topRightl4m) node [midway, above] {\footnotesize{Partition $m$ dimension with blocksize 3600}};
\draw [draw=none] (topLeftl4n) -- (topRightl4n) node [midway, above] {\footnotesize{Partition $n$ dimension with blocksize 3600}};
\draw [draw=none] (topLeftl3k) -- (topRightl3k) node [midway, above] {\footnotesize{Partition $k$ dimension with blocksize 192}};
\draw [draw=none] (topLeftl2m) -- (topRightl2m) node [midway, above] {\footnotesize{Partition $m$ dimension with blocksize 120}};
\draw [draw=none] (topLeftinnerkernel) -- (topRightinnerkernel) node [midway, above] {\footnotesize{Inner kernel}};

\end{tikzpicture} 

\begin{center}
\begin{tikzpicture}[>=latex,node distance=-0.25cm and .00cm]
\newcommand{\sWid}{0.4cm}
\node (l4) [pattern=dots, pattern color=cyan,
    fit={(-\sWid/2, -\sWid/2) (\sWid/2, \sWid/2)}, inner sep=0pt, draw=black, thick] {}; 
\node (l3) [pattern=crosshatch, pattern color=magenta,
    fit={(-\sWid/2, -\sWid/2) (\sWid/2, \sWid/2)}, inner sep=0pt, draw=black, thick, yshift=-0.3cm] at (l4.south) {}; 
\node (l2) [pattern=grid, pattern color=green,
    fit={(-\sWid/2, -\sWid/2) (\sWid/2, \sWid/2)}, inner sep=0pt, draw=black, thick, yshift=-0.3cm] at (l3.south) {}; 
\node [right] at (l4.east) { {\scriptsize Block is reused in L4 cache.} };  
\node [right] at (l3.east) { {\scriptsize Block is reused in L3 cache.} };  
\node [right] at (l2.east) { {\scriptsize Block is reused in L2 cache.} };  
\end{tikzpicture}
\end{center}

\caption{Algorithm $C_4 A_2 C_0$ that optimizes for both $L_4$ and $L_2$.
It places a square block of $C$ in $L_4$,
and a square block of $A$ in $L_2$.
$ L_3 $ is ``skipped'', with the and the $L_4$ guest panel, $B$, reused from $L_3$.}
\label{fig:l4_algorithm}
\end{figure}
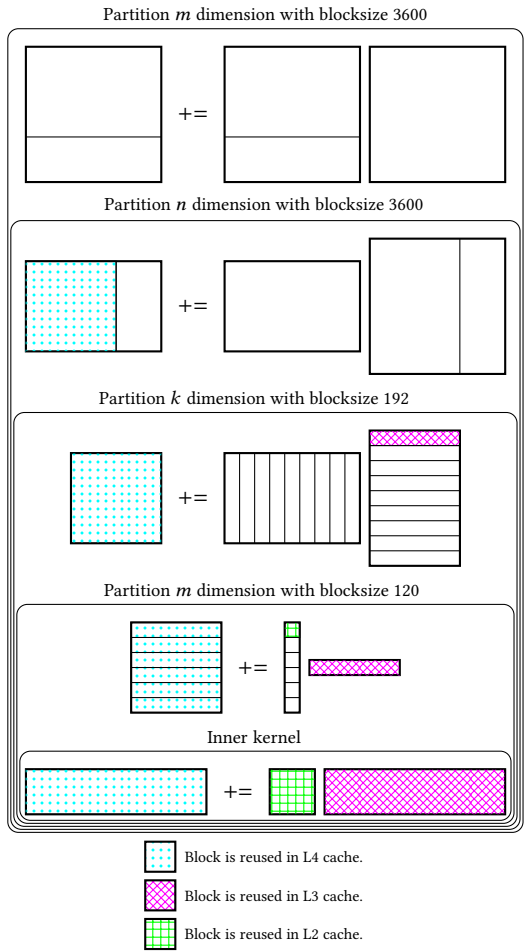

%% file: fig_l4.tex
\begin{figure}

\resizebox{\linewidth}{!}{
\begin{tabular}{@{}c @{\hspace{-1.75cm}} c@{}}
\input g_l4_800 & \input g_l4_1066 \\
\input g_l4_1333 & \input g_l4_2400x2 \\
\end{tabular}}
\caption{Performance for square matrices,
varying problem size and available bandwidth.
Matrices are stored prepacked.}
\label{fig:l4}
\end{figure}

%% file: g_l4_800.tex
\begin{tikzpicture}
  \begin{axis}[width=2.5in, height=2in,
               solid,
               xlabel={$m=n=k$},
               ylabel={\small GFLOPS},
               title={
                   \begin{minipage}{3in}
                   \begin{center}
                    {\small 1 channel of DDR3-800} 
                   \end{center}
                   \end{minipage}
               },
               xmin=0,xmax=15000,ymin=0,ymax=211.2,
               legend style={legend pos=south east, fill=white},
               x tick label style={rotate=45},
               clip=false]
\addplot[color=black,solid,thick] coordinates {
    (   500 ,   112.89434   )
    (   1000    ,   150.10123   )
    (   1500    ,   178.32459   )
    (   2000    ,   177.54769   )
    (   2500    ,   180.87958   )
    (   3000    ,   180.45529   )
    (   3500    ,   176.78984   )
    (   4000    ,   172.58959   )
    (   4500    ,   178.81486   )
    (   5000    ,   170.54605   )
    (   5500    ,   166.12609   )
    (   6000    ,   154.55433   )
    (   6500    ,   151.00299   )
    (   7000    ,   147.87379   )
    (   7500    ,   147.02221   )
    (   8000    ,   143.38303   )
    (   8500    ,   136.95924   )
    (   9000    ,   136.20892   )
    (   9500    ,   135.19061   )
    (   10000   ,   135.42836   )
    (   10500   ,   136.51121   )
    (   11000   ,   132.87191   )
    (   11500   ,   133.52235   )
    (   12000   ,   132.85942   )
    (   12500   ,   133.33099   )
    (   13000   ,   130.67885   )
    (   13500   ,   132.89597   )
    (   14000   ,   129.62714   )
    (   14500   ,   131.23899   )
    (   15000   ,   132.6622    )
};
\addplot[color=magenta,dashed,thick] coordinates {
    (   500 ,   138.84623   )
    (   1000    ,   154.58357   )
    (   1500    ,   177.4404    )
    (   2000    ,   177.12519   )
    (   2500    ,   181.93208   )
    (   3000    ,   181.58971   )
    (   3500    ,   174.65094   )
    (   4000    ,   172.03968   )
    (   4500    ,   175.79488   )
    (   5000    ,   176.27125   )
    (   5500    ,   178.6601    )
    (   6000    ,   179.36608   )
    (   6500    ,   178.08214   )
    (   7000    ,   171.7472    )
    (   7500    ,   171.37165   )
    (   8000    ,   173.35414   )
    (   8500    ,   176.15671   )
    (   9000    ,   177.11467   )
    (   9500    ,   173.51528   )
    (   10000   ,   172.25027   )
    (   10500   ,   170.8773    )
    (   11000   ,   165.37657   )
    (   11500   ,   166.24483   )
    (   12000   ,   172.89293   )
    (   12500   ,   173.88384   )
    (   13000   ,   175.27189   )
    (   13500   ,   175.91332   )
    (   14000   ,   174.89105   )
    (   14500   ,   170.86401   )
    (   15000   ,   173.46723   )
};
\end{axis}
\end{tikzpicture}

%% file: g_l4_1066.tex
\begin{tikzpicture}
  \begin{axis}[width=2.5in, height=2in,
               solid,
               xlabel={$m=n=k$},
               title={
                   \begin{minipage}{3in}
                   \begin{center}
                    {\small 1 channel of DDR3-1066} 
                   \end{center}
                   \end{minipage}
               },
               xmin=0,xmax=15000,ymin=0,ymax=211.2,
               legend style={legend pos=south east, fill=white},
               xticklabel style={rotate=45},
               clip=false]
\addplot[color=black,solid,thick] coordinates {
    (   500 ,   140.08365   )
    (   1000    ,   164.13299   )
    (   1500    ,   178.22905   )
    (   2000    ,   185.03241   )
    (   2500    ,   180.68727   )
    (   3000    ,   182.01479   )
    (   3500    ,   179.35118   )
    (   4000    ,   181.44585   )
    (   4500    ,   182.70883   )
    (   5000    ,   179.50827   )
    (   5500    ,   176.39217   )
    (   6000    ,   169.19148   )
    (   6500    ,   168.91615   )
    (   7000    ,   167.35987   )
    (   7500    ,   165.39092   )
    (   8000    ,   165.42307   )
    (   8500    ,   160.67948   )
    (   9000    ,   161.24004   )
    (   9500    ,   160.13473   )
    (   10000   ,   159.79151   )
    (   10500   ,   162.00307   )
    (   11000   ,   158.25512   )
    (   11500   ,   159.62303   )
    (   12000   ,   159.18233   )
    (   12500   ,   158.55508   )
    (   13000   ,   160.44786   )
    (   13500   ,   158.77297   )
    (   14000   ,   159.34641   )
    (   14500   ,   159.75088   )
    (   15000   ,   158.75302   )
};
\addplot[color=magenta,dashed,thick] coordinates {
    (   500 ,   152.59227   )
    (   1000    ,   169.42878   )
    (   1500    ,   179.46986   )
    (   2000    ,   183.13748   )
    (   2500    ,   178.82007   )
    (   3000    ,   181.7189    )
    (   3500    ,   173.60644   )
    (   4000    ,   175.21497   )
    (   4500    ,   179.37208   )
    (   5000    ,   178.97198   )
    (   5500    ,   181.12325   )
    (   6000    ,   180.58125   )
    (   6500    ,   178.85809   )
    (   7000    ,   175.49662   )
    (   7500    ,   175.51988   )
    (   8000    ,   176.75099   )
    (   8500    ,   179.90981   )
    (   9000    ,   179.34636   )
    (   9500    ,   178.37417   )
    (   10000   ,   176.96462   )
    (   10500   ,   177.0885    )
    (   11000   ,   175.16663   )
    (   11500   ,   176.17149   )
    (   12000   ,   177.15592   )
    (   12500   ,   177.47611   )
    (   13000   ,   177.51879   )
    (   13500   ,   177.01272   )
    (   14000   ,   176.47962   )
    (   14500   ,   173.77739   )
    (   15000   ,   176.51481   )
};
\end{axis}
\end{tikzpicture}

%% file: g_l4_1333.tex
\begin{tikzpicture}
  \begin{axis}[width=2.5in, height=2in,
               solid,
               xlabel={$m=n=k$},
               ylabel={\small GFLOPS},
               title={
                   \begin{minipage}{3in}
                   \begin{center}
                    {\small 1 channel of DDR3-1333} 
                   \end{center}
                   \end{minipage}
               },
               xmin=0,xmax=15000,ymin=0,ymax=211.2,
               legend style={legend pos=south east, fill=white},
               xticklabel style={rotate=45},
               clip=false]
\addplot[color=black,solid,thick] coordinates {
    (   500 ,   138.92671   )
    (   1000    ,   176.80166   )
    (   1500    ,   180.35836   )
    (   2000    ,   185.39558   )
    (   2500    ,   178.90422   )
    (   3000    ,   179.86248   )
    (   3500    ,   179.6879    )
    (   4000    ,   182.32608   )
    (   4500    ,   181.56041   )
    (   5000    ,   182.99269   )
    (   5500    ,   178.77295   )
    (   6000    ,   174.96019   )
    (   6500    ,   175.11241   )
    (   7000    ,   174.79985   )
    (   7500    ,   173.60561   )
    (   8000    ,   174.7474    )
    (   8500    ,   171.24913   )
    (   9000    ,   171.61289   )
    (   9500    ,   171.10788   )
    (   10000   ,   170.46702   )
    (   10500   ,   166.91364   )
    (   11000   ,   169.83504   )
    (   11500   ,   171.30934   )
    (   12000   ,   170.81223   )
    (   12500   ,   169.80081   )
    (   13000   ,   171.59642   )
    (   13500   ,   170.49352   )
    (   14000   ,   170.70047   )
    (   14500   ,   170.4328    )
    (   15000   ,   170.00549   )
};
\addplot[color=magenta,dashed,thick] coordinates {
    (   500 ,   154.10439   )
    (   1000    ,   169.61575   )
    (   1500    ,   181.88705   )
    (   2000    ,   184.77449   )
    (   2500    ,   179.91874   )
    (   3000    ,   183.56426   )
    (   3500    ,   182.38786   )
    (   4000    ,   179.61351   )
    (   4500    ,   183.63131   )
    (   5000    ,   182.97061   )
    (   5500    ,   185.06565   )
    (   6000    ,   184.20956   )
    (   6500    ,   182.95009   )
    (   7000    ,   182.93222   )
    (   7500    ,   182.21271   )
    (   8000    ,   183.45694   )
    (   8500    ,   183.91281   )
    (   9000    ,   184.65001   )
    (   9500    ,   184.45565   )
    (   10000   ,   183.89296   )
    (   10500   ,   182.99865   )
    (   11000   ,   181.70097   )
    (   11500   ,   183.4473    )
    (   12000   ,   183.40045   )
    (   12500   ,   183.90257   )
    (   13000   ,   184.21875   )
    (   13500   ,   181.13871   )
    (   14000   ,   183.19289   )
    (   14500   ,   180.09536   )
    (   15000   ,   183.08357   )
};
\end{axis}
\end{tikzpicture}

%% file: g_l4_2400x2.tex
\begin{tikzpicture}
  \begin{axis}[width=2.5in, height=2in,
               solid,
               xlabel={$m=n=k$},
               title={
                   \begin{minipage}{3in}
                   \begin{center}
                    {\small 2 channels of DDR3-2400} 
                   \end{center}
                   \end{minipage}
               },
               xmin=0,xmax=15000,ymin=0,ymax=211.2,
               legend style={legend pos=south east, fill=white},
               xticklabel style={rotate=45},
               clip=false]
\addplot[color=black,solid,thick] coordinates {
    (   500 ,   169.62907   )
    (   1000    ,   188.02011   )
    (   1500    ,   187.32132   )
    (   2000    ,   191.05167   )
    (   2500    ,   182.42714   )
    (   3000    ,   183.17918   )
    (   3500    ,   182.24247   )
    (   4000    ,   184.03834   )
    (   4500    ,   185.02293   )
    (   5000    ,   185.73413   )
    (   5500    ,   183.01937   )
    (   6000    ,   183.34706   )
    (   6500    ,   183.63898   )
    (   7000    ,   184.34293   )
    (   7500    ,   184.00905   )
    (   8000    ,   182.82397   )
    (   8500    ,   183.14039   )
    (   9000    ,   181.81753   )
    (   9500    ,   181.99481   )
    (   10000   ,   183.13588   )
    (   10500   ,   182.42992   )
    (   11000   ,   182.81289   )
    (   11500   ,   182.66891   )
    (   12000   ,   181.57878   )
    (   12500   ,   182.05014   )
    (   13000   ,   181.70873   )
    (   13500   ,   182.35258   )
    (   14000   ,   182.44116   )
    (   14500   ,   182.02353   )
    (   15000   ,   182.44727   )
};
\addplot[color=magenta,dashed,thick] coordinates {
    (   500 ,   170.29312   )
    (   1000    ,   187.83461   )
    (   1500    ,   187.39644   )
    (   2000    ,   190.94388   )
    (   2500    ,   182.25212   )
    (   3000    ,   182.87881   )
    (   3500    ,   182.61035   )
    (   4000    ,   182.89871   )
    (   4500    ,   185.10491   )
    (   5000    ,   185.8175    )
    (   5500    ,   186.84329   )
    (   6000    ,   186.96907   )
    (   6500    ,   185.12393   )
    (   7000    ,   185.24698   )
    (   7500    ,   184.39697   )
    (   8000    ,   185.95155   )
    (   8500    ,   186.08157   )
    (   9000    ,   186.67729   )
    (   9500    ,   185.77603   )
    (   10000   ,   184.99812   )
    (   10500   ,   185.39625   )
    (   11000   ,   184.84165   )
    (   11500   ,   186.11934   )
    (   12000   ,   186.2475    )
    (   12500   ,   186.21261   )
    (   13000   ,   186.30476   )
    (   13500   ,   185.52683   )
    (   14000   ,   184.91708   )
    (   14500   ,   185.01012   )
    (   15000   ,   185.38562   )
};
\legend{
    \begin{minipage}{1.0in}{\scriptsize  MOMMS Goto}\end{minipage},
    \begin{minipage}{1.0in}{\scriptsize MOMMS $C_4 A_2 C_0$}\end{minipage}
}
\end{axis}
\end{tikzpicture}

%% file: fig_l4_packing.tex
\begin{figure}
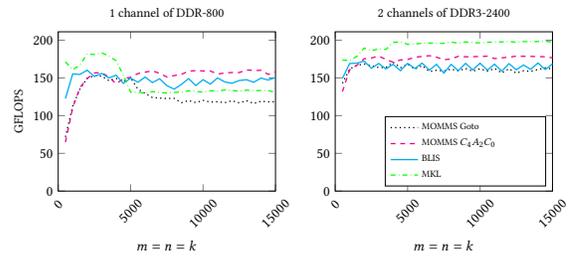


\resizebox{\linewidth}{!}{
\begin{tabular}{@{}c @{\hspace{-1.75cm}} c@{}}
\input g_l4_packing_800 & \input g_l4_packing_2400x2 \\
\end{tabular}}
\caption{Performance on i7-5775C
with an 128MB L4 cache
for low bandwidth scenario (1 channel of DDR3-800)
and a high bandwidth scenario (2 channels of DDR3-2400).}
\label{fig:l4_packing}
\end{figure}

%% file: g_l4_packing_800.tex
\begin{tikzpicture}
  \begin{axis}[width=2.5in, height=2in,
               solid,
               xlabel={$m=n=k$},
               ylabel={\small GFLOPS},
               title={
                   \begin{minipage}{3in}
                   \begin{center}
                    {\small 1 channel of DDR-800}
                   \end{center}
                   \end{minipage}
               },
               xmin=0,xmax=15000,ymin=0,ymax=211.2,
               legend style={legend pos=south east, fill=white},
               xticklabel style={rotate=45},
               clip=false]
\addplot[color=black,dotted,thick] coordinates {
    (   500 ,   71.81805    )
    (   1000    ,   112.78329   )
    (   1500    ,   135.1651    )
    (   2000    ,   149.49145   )
    (   2500    ,   153.00366   )
    (   3000    ,   152.39865   )
    (   3500    ,   146.33811   )
    (   4000    ,   149.27377   )
    (   4500    ,   145.42041   )
    (   5000    ,   150.05111   )
    (   5500    ,   135.29746   )
    (   6000    ,   129.52964   )
    (   6500    ,   124.07672   )
    (   7000    ,   123.28432   )
    (   7500    ,   122.72825   )
    (   8000    ,   122.99456   )
    (   8500    ,   117.19833   )
    (   9000    ,   119.97673   )
    (   9500    ,   117.32158   )
    (   10000   ,   120.24313   )
    (   10500   ,   117.96418   )
    (   11000   ,   118.48529   )
    (   11500   ,   117.14427   )
    (   12000   ,   119.7574    )
    (   12500   ,   116.95777   )
    (   13000   ,   119.82172   )
    (   13500   ,   116.13199   )
    (   14000   ,   119.07718   )
    (   14500   ,   118.31895   )
    (   15000   ,   118.3747    )
};
\addplot[color=magenta,dashed,thick] coordinates {
    (   500 ,   64.90187    )
    (   1000    ,   110.32623   )
    (   1500    ,   137.20176   )
    (   2000    ,   149.2138    )
    (   2500    ,   156.68239   )
    (   3000    ,   156.88157   )
    (   3500    ,   151.27731   )
    (   4000    ,   142.10541   )
    (   4500    ,   149.07579   )
    (   5000    ,   150.85052   )
    (   5500    ,   155.7372    )
    (   6000    ,   157.66688   )
    (   6500    ,   159.38737   )
    (   7000    ,   156.12479   )
    (   7500    ,   151.29614   )
    (   8000    ,   153.10837   )
    (   8500    ,   154.70357   )
    (   9000    ,   158.97495   )
    (   9500    ,   159.71882   )
    (   10000   ,   159.24112   )
    (   10500   ,   159.20781   )
    (   11000   ,   155.14816   )
    (   11500   ,   156.57005   )
    (   12000   ,   157.16841   )
    (   12500   ,   157.15341   )
    (   13000   ,   160.613 )
    (   13500   ,   159.9212    )
    (   14000   ,   160.05043   )
    (   14500   ,   154.7209    )
    (   15000   ,   157.89237   )
};
\addplot[color=cyan,solid,thick] coordinates {
    (   500 ,   122.57161   )
    (   1000    ,   155.29829   )
    (   1500    ,   154.62533   )
    (   2000    ,   160.26794   )
    (   2500    ,   151.8506    )
    (   3000    ,   155.97179   )
    (   3500    ,   150.01427   )
    (   4000    ,   153.61382   )
    (   4500    ,   142.77429   )
    (   5000    ,   149.76281   )
    (   5500    ,   144.31961   )
    (   6000    ,   151.22337   )
    (   6500    ,   143.85818   )
    (   7000    ,   148.85947   )
    (   7500    ,   139.92526   )
    (   8000    ,   135.10727   )
    (   8500    ,   141.38318   )
    (   9000    ,   149.03756   )
    (   9500    ,   139.95566   )
    (   10000   ,   147.66962   )
    (   10500   ,   141.81708   )
    (   11000   ,   150.00268   )
    (   11500   ,   144.70273   )
    (   12000   ,   142.90322   )
    (   12500   ,   146.53725   )
    (   13000   ,   147.62898   )
    (   13500   ,   144.25448   )
    (   14000   ,   149.83323   )
    (   14500   ,   147.23199   )
    (   15000   ,   150.49797   )
};
\addplot[color=green,dashdotted,thick] coordinates {
    (   500 ,   171.29658   )
    (   1000    ,   160.96244   )
    (   1500    ,   167.04949   )
    (   2000    ,   181.84362   )
    (   2500    ,   181.01701   )
    (   3000    ,   183.30087   )
    (   3500    ,   178.93046   )
    (   4000    ,   172.88487   )
    (   4500    ,   153.40564   )
    (   5000    ,   131.04208   )
    (   5500    ,   130.94007   )
    (   6000    ,   129.73756   )
    (   6500    ,   131.85111   )
    (   7000    ,   130.6996    )
    (   7500    ,   129.87925   )
    (   8000    ,   130.91907   )
    (   8500    ,   131.08332   )
    (   9000    ,   132.85634   )
    (   9500    ,   132.17784   )
    (   10000   ,   131.21251   )
    (   10500   ,   133.0284    )
    (   11000   ,   132.35034   )
    (   11500   ,   134.05058   )
    (   12000   ,   133.34866   )
    (   12500   ,   132.5308    )
    (   13000   ,   133.76421   )
    (   13500   ,   133.13131   )
    (   14000   ,   133.31723   )
    (   14500   ,   133.06246   )
    (   15000   ,   131.11505   )
};
\end{axis}
\end{tikzpicture}

%% file: g_l4_packing_2400x2.tex
\begin{tikzpicture}
  \begin{axis}[width=2.5in, height=2in,
               solid,
               xlabel={$m=n=k$},
               title={
                   \begin{minipage}{3in}
                   \begin{center}
                    {\small 2 channels of DDR3-2400}
                   \end{center}
                   \end{minipage}
               },
               xmin=0,xmax=15000,ymin=0,ymax=211.2,
               legend style={legend pos=south east, fill=white},
               xticklabel style={rotate=45},
               clip=false]
\addplot[color=black,dotted,thick] coordinates {
    (   500 ,   142.93931   )
    (   1000    ,   162.03851   )
    (   1500    ,   166.61094   )
    (   2000    ,   168.50609   )
    (   2500    ,   166.14941   )
    (   3000    ,   163.78219   )
    (   3500    ,   163.83784   )
    (   4000    ,   170.86765   )
    (   4500    ,   163.59121   )
    (   5000    ,   168.24718   )
    (   5500    ,   161.71871   )
    (   6000    ,   166.36377   )
    (   6500    ,   158.45544   )
    (   7000    ,   160.20385   )
    (   7500    ,   159.17456   )
    (   8000    ,   161.83301   )
    (   8500    ,   159.2826    )
    (   9000    ,   162.53274   )
    (   9500    ,   158.82347   )
    (   10000   ,   162.17541   )
    (   10500   ,   158.93406   )
    (   11000   ,   162.59079   )
    (   11500   ,   156.63  )
    (   12000   ,   162.21462   )
    (   12500   ,   156.30186   )
    (   13000   ,   159.57508   )
    (   13500   ,   158.61039   )
    (   14000   ,   161.25024   )
    (   14500   ,   163.49745   )
    (   15000   ,   161.8135    )
};
\addplot[color=magenta,dashed,thick] coordinates {
    (   500 ,   131.97452   )
    (   1000    ,   160.89337   )
    (   1500    ,   166.899 )
    (   2000    ,   175.37588   )
    (   2500    ,   176.45958   )
    (   3000    ,   178.91966   )
    (   3500    ,   174.62981   )
    (   4000    ,   170.71166   )
    (   4500    ,   173.82595   )
    (   5000    ,   174.47969   )
    (   5500    ,   177.36369   )
    (   6000    ,   178.52636   )
    (   6500    ,   179.48358   )
    (   7000    ,   177.48885   )
    (   7500    ,   174.09615   )
    (   8000    ,   175.14407   )
    (   8500    ,   178.66  )
    (   9000    ,   178.03399   )
    (   9500    ,   178.20439   )
    (   10000   ,   178.38462   )
    (   10500   ,   178.83156   )
    (   11000   ,   175.64265   )
    (   11500   ,   176.64799   )
    (   12000   ,   177.56832   )
    (   12500   ,   178.84434   )
    (   13000   ,   178.69336   )
    (   13500   ,   178.52838   )
    (   14000   ,   178.34596   )
    (   14500   ,   178.02172   )
    (   15000   ,   176.7011    )
};
\addplot[color=cyan,solid,thick] coordinates {
    (   500 ,   149.48345   )
    (   1000    ,   169.14987   )
    (   1500    ,   169.67367   )
    (   2000    ,   172.18326   )
    (   2500    ,   162.83941   )
    (   3000    ,   169.18602   )
    (   3500    ,   161.97953   )
    (   4000    ,   168.04158   )
    (   4500    ,   159.63814   )
    (   5000    ,   169.32733   )
    (   5500    ,   162.5676    )
    (   6000    ,   169.65257   )
    (   6500    ,   160.36906   )
    (   7000    ,   168.05904   )
    (   7500    ,   156.36949   )
    (   8000    ,   167.9565    )
    (   8500    ,   159.69064   )
    (   9000    ,   169.29752   )
    (   9500    ,   160.60903   )
    (   10000   ,   169.50725   )
    (   10500   ,   160.32766   )
    (   11000   ,   168.24266   )
    (   11500   ,   158.99166   )
    (   12000   ,   168.93426   )
    (   12500   ,   160.80527   )
    (   13000   ,   167.1282    )
    (   13500   ,   161.46052   )
    (   14000   ,   169.82685   )
    (   14500   ,   161.10274   )
    (   15000   ,   168.59689   )
};
\addplot[color=green,dashdotted,thick] coordinates {
    (   500 ,   173.67033   )
    (   1000    ,   172.96636   )
    (   1500    ,   178.74604   )
    (   2000    ,   189.56019   )
    (   2500    ,   185.96706   )
    (   3000    ,   188.61274   )
    (   3500    ,   187.757 )
    (   4000    ,   197.0776    )
    (   4500    ,   197.8614    )
    (   5000    ,   194.52143   )
    (   5500    ,   194.87946   )
    (   6000    ,   196.05846   )
    (   6500    ,   196.31679   )
    (   7000    ,   195.63555   )
    (   7500    ,   195.92325   )
    (   8000    ,   197.51647   )
    (   8500    ,   196.10481   )
    (   9000    ,   197.7407    )
    (   9500    ,   196.79415   )
    (   10000   ,   196.48348   )
    (   10500   ,   197.29677   )
    (   11000   ,   197.53911   )
    (   11500   ,   197.76918   )
    (   12000   ,   198.38587   )
    (   12500   ,   197.27213   )
    (   13000   ,   198.2818    )
    (   13500   ,   198.18908   )
    (   14000   ,   198.29156   )
    (   14500   ,   198.58053   )
    (   15000   ,   197.82247   )
};
\legend{
    \begin{minipage}{1.0in}{\scriptsize  MOMMS Goto}\end{minipage},
    \begin{minipage}{1.0in}{\scriptsize MOMMS $C_4 A_2 C_0$}\end{minipage},
    \begin{minipage}{1.0in}{\scriptsize  BLIS}\end{minipage},
    \begin{minipage}{1.0in}{\scriptsize MKL}\end{minipage}
}
\end{axis}
\end{tikzpicture}

%% file: fig_shapes.tex
\begin{figure}
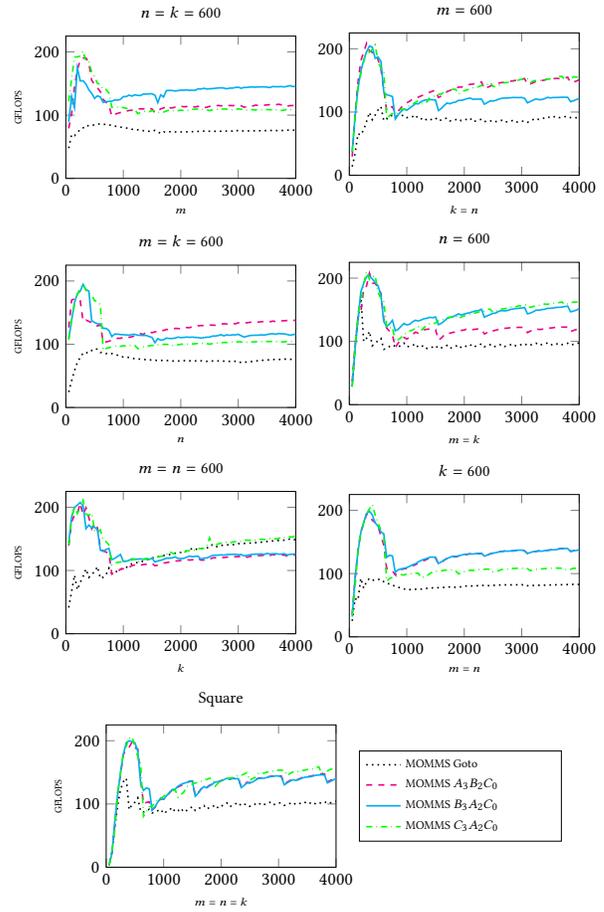

\begin{center}
\resizebox{.45\textwidth}{!}{
\begin{tabular}{@{}c@{}}
 \begin{tabular}{@{}c @{\hspace{-0.125cm}} c@{}}
    \input g_shapes_800_m & \input g_shapes_800_kn \\ 
    \input g_shapes_800_n & \input g_shapes_800_mk \\
    \input g_shapes_800_k & \input g_shapes_800_mn \\
    \end{tabular} \\
\input g_shapes_800_square
\end{tabular}
}
\end{center}
\caption{Comparison for different shapes
on i7-7700K with low-bandwidth scenario (1 channel of DDR4-800).}
\label{fig:l3_shapes}
\end{figure}

%% file: g_shapes_800_m.tex
\begin{tikzpicture}
  \begin{axis}[width=2.2in, height=1.60in,
               solid,
               xlabel={\tiny $m$},
               ylabel={\tiny  GFLOPS},
               title={\small  $n=k=600$},
               xmin=0,xmax=4000,ymin=0,ymax=225,
               legend style={legend pos=south east, fill=white},
               clip=false]
\addplot[color=black,dotted,thick] coordinates {
    (   50  ,   47.83519    )
    (   100 ,   69.37873    )
    (   150 ,   67.65129    )
    (   200 ,   72.22686    )
    (   250 ,   77.51858    )
    (   300 ,   79.66147    )
    (   350 ,   83.98995    )
    (   400 ,   82.8254 )
    (   450 ,   83.42163    )
    (   500 ,   84.87663    )
    (   550 ,   85.54076    )
    (   600 ,   86.33411    )
    (   650 ,   85.1639 )
    (   700 ,   85.62261    )
    (   750 ,   84.62389    )
    (   800 ,   83.67287    )
    (   850 ,   82.73368    )
    (   900 ,   81.80688    )
    (   950 ,   81.14614    )
    (   1000    ,   80.18893    )
    (   1050    ,   79.67052    )
    (   1100    ,   79.01822    )
    (   1150    ,   77.59433    )
    (   1200    ,   77.94883    )
    (   1250    ,   76.81586    )
    (   1300    ,   76.09401    )
    (   1350    ,   75.44793    )
    (   1400    ,   75.11702    )
    (   1450    ,   74.39502    )
    (   1500    ,   74.37424    )
    (   1550    ,   76.01478    )
    (   1600    ,   73.88821    )
    (   1650    ,   70.8006 )
    (   1700    ,   73.4397 )
    (   1750    ,   72.92447    )
    (   1800    ,   73.77011    )
    (   1850    ,   73.47677    )
    (   1900    ,   73.42677    )
    (   1950    ,   73.82868    )
    (   2000    ,   73.47155    )
    (   2050    ,   73.54452    )
    (   2100    ,   73.57545    )
    (   2150    ,   73.88   )
    (   2200    ,   73.86837    )
    (   2250    ,   73.94127    )
    (   2300    ,   74.29092    )
    (   2350    ,   73.99487    )
    (   2400    ,   74.78716    )
    (   2450    ,   74.15202    )
    (   2500    ,   74.56724    )
    (   2550    ,   74.58802    )
    (   2600    ,   74.42684    )
    (   2650    ,   74.46253    )
    (   2700    ,   74.5309 )
    (   2750    ,   74.66162    )
    (   2800    ,   74.8672 )
    (   2850    ,   74.64721    )
    (   2900    ,   74.89765    )
    (   2950    ,   76.45347    )
    (   3000    ,   75.03007    )
    (   3050    ,   74.94047    )
    (   3100    ,   74.99778    )
    (   3150    ,   74.97763    )
    (   3200    ,   75.05867    )
    (   3250    ,   75.30885    )
    (   3300    ,   75.16989    )
    (   3350    ,   75.27464    )
    (   3400    ,   75.38753    )
    (   3450    ,   75.27394    )
    (   3500    ,   75.75967    )
    (   3550    ,   75.73454    )
    (   3600    ,   75.73685    )
    (   3650    ,   75.46437    )
    (   3700    ,   75.76918    )
    (   3750    ,   75.67021    )
    (   3800    ,   75.93123    )
    (   3850    ,   76.31822    )
    (   3900    ,   75.98889    )
    (   3950    ,   76.24107    )
    (   4000    ,   75.97395    )
};
\addplot[color=magenta,dashed,thick] coordinates {
(   50  ,   79.14175    )
(   100 ,   101.86973   )
(   150 ,   144.3175    )
(   200 ,   171.77908   )
(   250 ,   187.18023   )
(   300 ,   190.20866   )
(   350 ,   190.4736    )
(   400 ,   182.451 )
(   450 ,   154.87897   )
(   500 ,   148.675 )
(   550 ,   143.05097   )
(   600 ,   137.25109   )
(   650 ,   124.68094   )
(   700 ,   120.50301   )
(   750 ,   114.96272   )
(   800 ,   98.30122    )
(   850 ,   101.54702   )
(   900 ,   102.75305   )
(   950 ,   104.14097   )
(   1000    ,   104.8101    )
(   1050    ,   106.18344   )
(   1100    ,   107.36999   )
(   1150    ,   107.67734   )
(   1200    ,   109.39633   )
(   1250    ,   110.00265   )
(   1300    ,   109.90032   )
(   1350    ,   109.79635   )
(   1400    ,   111.98048   )
(   1450    ,   110.58733   )
(   1500    ,   109.68202   )
(   1550    ,   103.64876   )
(   1600    ,   106.1717    )
(   1650    ,   109.98124   )
(   1700    ,   111.20968   )
(   1750    ,   114.79186   )
(   1800    ,   111.42138   )
(   1850    ,   111.45192   )
(   1900    ,   112.59  )
(   1950    ,   112.8122    )
(   2000    ,   114.03514   )
(   2050    ,   113.29226   )
(   2100    ,   113.53812   )
(   2150    ,   111.78607   )
(   2200    ,   113.36294   )
(   2250    ,   112.92756   )
(   2300    ,   113.9549    )
(   2350    ,   111.25866   )
(   2400    ,   111.82292   )
(   2450    ,   112.41441   )
(   2500    ,   113.87665   )
(   2550    ,   113.69328   )
(   2600    ,   113.97611   )
(   2650    ,   115.46167   )
(   2700    ,   114.58559   )
(   2750    ,   115.85115   )
(   2800    ,   115.01981   )
(   2850    ,   115.19094   )
(   2900    ,   115.04136   )
(   2950    ,   115.28914   )
(   3000    ,   115.30473   )
(   3050    ,   114.77549   )
(   3100    ,   112.54662   )
(   3150    ,   113.13889   )
(   3200    ,   114.55223   )
(   3250    ,   114.47061   )
(   3300    ,   115.79117   )
(   3350    ,   115.33038   )
(   3400    ,   115.64653   )
(   3450    ,   115.66525   )
(   3500    ,   116.18074   )
(   3550    ,   116.5368    )
(   3600    ,   115.62283   )
(   3650    ,   116.28936   )
(   3700    ,   117.27325   )
(   3750    ,   116.26255   )
(   3800    ,   116.64501   )
(   3850    ,   114.15126   )
(   3900    ,   115.05036   )
(   3950    ,   115.16348   )
(   4000    ,   116.09051   )
};
\addplot[color=cyan,solid,thick] coordinates {
    (   50  ,   89.31784    )
    (   100 ,   129.68323   )
    (   150 ,   111.93229   )
    (   200 ,   174.62652   )
    (   250 ,   154.13031   )
    (   300 ,   154.38584   )
    (   350 ,   149.11048   )
    (   400 ,   140.03328   )
    (   450 ,   134.35766   )
    (   500 ,   126.46724   )
    (   550 ,   130.51328   )
    (   600 ,   128.71662   )
    (   650 ,   120.03564   )
    (   700 ,   121.04799   )
    (   750 ,   121.90874   )
    (   800 ,   122.04665   )
    (   850 ,   124.03211   )
    (   900 ,   123.61682   )
    (   950 ,   124.89127   )
    (   1000    ,   129.28617   )
    (   1050    ,   129.32778   )
    (   1100    ,   131.61649   )
    (   1150    ,   130.08516   )
    (   1200    ,   132.09945   )
    (   1250    ,   131.17287   )
    (   1300    ,   131.04749   )
    (   1350    ,   133.37832   )
    (   1400    ,   133.18227   )
    (   1450    ,   133.67731   )
    (   1500    ,   133.84117   )
    (   1550    ,   131.53301   )
    (   1600    ,   119.83983   )
    (   1650    ,   132.04305   )
    (   1700    ,   128.64028   )
    (   1750    ,   138.8617    )
    (   1800    ,   139.10959   )
    (   1850    ,   138.91774   )
    (   1900    ,   138.14499   )
    (   1950    ,   138.64532   )
    (   2000    ,   139.6005    )
    (   2050    ,   139.26173   )
    (   2100    ,   139.94169   )
    (   2150    ,   139.23074   )
    (   2200    ,   139.41752   )
    (   2250    ,   139.28538   )
    (   2300    ,   141.58893   )
    (   2350    ,   141.21562   )
    (   2400    ,   142.46384   )
    (   2450    ,   141.31675   )
    (   2500    ,   141.71628   )
    (   2550    ,   141.89709   )
    (   2600    ,   142.44026   )
    (   2650    ,   142.97108   )
    (   2700    ,   143.15743   )
    (   2750    ,   142.77048   )
    (   2800    ,   141.65462   )
    (   2850    ,   142.68326   )
    (   2900    ,   142.75592   )
    (   2950    ,   143.74536   )
    (   3000    ,   144.16799   )
    (   3050    ,   142.48053   )
    (   3100    ,   143.79057   )
    (   3150    ,   143.78442   )
    (   3200    ,   144.41318   )
    (   3250    ,   144.56839   )
    (   3300    ,   144.14904   )
    (   3350    ,   144.65161   )
    (   3400    ,   144.63866   )
    (   3450    ,   144.72566   )
    (   3500    ,   144.28213   )
    (   3550    ,   143.66925   )
    (   3600    ,   145.23055   )
    (   3650    ,   144.89379   )
    (   3700    ,   144.8863    )
    (   3750    ,   145.1913    )
    (   3800    ,   144.56289   )
    (   3850    ,   145.64315   )
    (   3900    ,   146.6189    )
    (   3950    ,   145.18418   )
    (   4000    ,   145.54056   )
};
\addplot[color=green,dashdotted,thick] coordinates {
    (   50  ,   121.86246   )
    (   100 ,   157.2667    )
    (   150 ,   192.10211   )
    (   200 ,   191.93142   )
    (   250 ,   193.57274   )
    (   300 ,   201.2438    )
    (   350 ,   184.99853   )
    (   400 ,   184.10014   )
    (   450 ,   167.7557    )
    (   500 ,   160.87825   )
    (   550 ,   150.1378    )
    (   600 ,   148.76709   )
    (   650 ,   140.52197   )
    (   700 ,   140.52349   )
    (   750 ,   134.59245   )
    (   800 ,   112.24202   )
    (   850 ,   112.02636   )
    (   900 ,   112.35291   )
    (   950 ,   111.78332   )
    (   1000    ,   111.96244   )
    (   1050    ,   110.71904   )
    (   1100    ,   110.06553   )
    (   1150    ,   108.94751   )
    (   1200    ,   108.72665   )
    (   1250    ,   102.34943   )
    (   1300    ,   104.16171   )
    (   1350    ,   105.71681   )
    (   1400    ,   106.53935   )
    (   1450    ,   105.42861   )
    (   1500    ,   107.17929   )
    (   1550    ,   106.3893    )
    (   1600    ,   104.30328   )
    (   1650    ,   104.17001   )
    (   1700    ,   104.82902   )
    (   1750    ,   106.37264   )
    (   1800    ,   108.13709   )
    (   1850    ,   107.01896   )
    (   1900    ,   107.29481   )
    (   1950    ,   107.01923   )
    (   2000    ,   108.07329   )
    (   2050    ,   107.14712   )
    (   2100    ,   107.70191   )
    (   2150    ,   107.25037   )
    (   2200    ,   109.10446   )
    (   2250    ,   109.38634   )
    (   2300    ,   109.74705   )
    (   2350    ,   108.81897   )
    (   2400    ,   108.64511   )
    (   2450    ,   108.73166   )
    (   2500    ,   107.15172   )
    (   2550    ,   108.40581   )
    (   2600    ,   109.1643    )
    (   2650    ,   109.39277   )
    (   2700    ,   109.64899   )
    (   2750    ,   109.56013   )
    (   2800    ,   109.63021   )
    (   2850    ,   108.37621   )
    (   2900    ,   108.48287   )
    (   2950    ,   108.75467   )
    (   3000    ,   109.86917   )
    (   3050    ,   109.313 )
    (   3100    ,   109.32561   )
    (   3150    ,   108.19036   )
    (   3200    ,   108.80671   )
    (   3250    ,   108.5949    )
    (   3300    ,   108.09748   )
    (   3350    ,   108.46012   )
    (   3400    ,   108.40455   )
    (   3450    ,   108.62009   )
    (   3500    ,   108.87518   )
    (   3550    ,   108.40062   )
    (   3600    ,   108.63001   )
    (   3650    ,   108.22442   )
    (   3700    ,   108.53332   )
    (   3750    ,   107.40905   )
    (   3800    ,   107.98495   )
    (   3850    ,   109.26423   )
    (   3900    ,   108.85079   )
    (   3950    ,   108.85351   )
    (   4000    ,   108.21852   )
};
\end{axis}
\end{tikzpicture}

%% file: g_shapes_800_kn.tex
\begin{tikzpicture}
  \begin{axis}[width=2.2in, height=1.60in,
               solid,
               xlabel={\tiny $k=n$},
               title={ \small  $m=600$ },
               xmin=0,xmax=4000,ymin=0,ymax=225,
               legend style={legend pos=south east, fill=white},
               clip=false]
\addplot[color=black,dotted,thick] coordinates {
    (   50  ,   13.73237    )
    (   100 ,   34.30855    )
    (   150 ,   65.71837    )
    (   200 ,   63.68463    )
    (   250 ,   71.95639    )
    (   300 ,   85.14644    )
    (   350 ,   99.44063    )
    (   400 ,   87.91688    )
    (   450 ,   95.28265    )
    (   500 ,   103.32254   )
    (   550 ,   108.45245   )
    (   600 ,   86.35123    )
    (   650 ,   88.92242    )
    (   700 ,   93.51756    )
    (   750 ,   94.66664    )
    (   800 ,   92.87591    )
    (   850 ,   93.83638    )
    (   900 ,   94.12072    )
    (   950 ,   95.02335    )
    (   1000    ,   91.1282 )
    (   1050    ,   92.03766    )
    (   1100    ,   92.66722    )
    (   1150    ,   95.54504    )
    (   1200    ,   89.21635    )
    (   1250    ,   89.20071    )
    (   1300    ,   91.20936    )
    (   1350    ,   86.63492    )
    (   1400    ,   87.55303    )
    (   1450    ,   88.88151    )
    (   1500    ,   90.81487    )
    (   1550    ,   85.83336    )
    (   1600    ,   87.59766    )
    (   1650    ,   89.02083    )
    (   1700    ,   90.52522    )
    (   1750    ,   85.97827    )
    (   1800    ,   87.16142    )
    (   1850    ,   87.86914    )
    (   1900    ,   89.50819    )
    (   1950    ,   86.11445    )
    (   2000    ,   87.51802    )
    (   2050    ,   88.64658    )
    (   2100    ,   90.57564    )
    (   2150    ,   84.69429    )
    (   2200    ,   85.88359    )
    (   2250    ,   87.24117    )
    (   2300    ,   88.65255    )
    (   2350    ,   84.80691    )
    (   2400    ,   85.83469    )
    (   2450    ,   86.32995    )
    (   2500    ,   83.85488    )
    (   2550    ,   84.891  )
    (   2600    ,   85.9196 )
    (   2650    ,   86.94291    )
    (   2700    ,   83.89808    )
    (   2750    ,   84.76505    )
    (   2800    ,   85.63314    )
    (   2850    ,   86.78428    )
    (   2900    ,   83.82264    )
    (   2950    ,   84.73972    )
    (   3000    ,   85.4732 )
    (   3050    ,   85.65285    )
    (   3100    ,   83.68963    )
    (   3150    ,   86.31731    )
    (   3200    ,   88.20281    )
    (   3250    ,   90.05642    )
    (   3300    ,   88.18454    )
    (   3350    ,   89.29347    )
    (   3400    ,   90.91568    )
    (   3450    ,   92.61267    )
    (   3500    ,   90.7415 )
    (   3550    ,   92.18805    )
    (   3600    ,   93.39162    )
    (   3650    ,   91.18828    )
    (   3700    ,   92.02313    )
    (   3750    ,   92.66489    )
    (   3800    ,   93.27938    )
    (   3850    ,   91.11477    )
    (   3900    ,   91.31718    )
    (   3950    ,   90.99363    )
    (   4000    ,   91.74457    )
};
\addplot[color=magenta,dashed,thick] coordinates {
    (   50  ,   29.58784    )
    (   100 ,   95.99539    )
    (   150 ,   146.20754   )
    (   200 ,   181.09381   )
    (   250 ,   194.47988   )
    (   300 ,   208.45598   )
    (   350 ,   196.88704   )
    (   400 ,   194.2164    )
    (   450 ,   193.41144   )
    (   500 ,   182.77337   )
    (   550 ,   167.90786   )
    (   600 ,   134.83201   )
    (   650 ,   102.15381   )
    (   700 ,   107.06603   )
    (   750 ,   108.33791   )
    (   800 ,   98.37183    )
    (   850 ,   101.28364   )
    (   900 ,   106.67517   )
    (   950 ,   112.6266    )
    (   1000    ,   114.60079   )
    (   1050    ,   117.03988   )
    (   1100    ,   120.27027   )
    (   1150    ,   122.62494   )
    (   1200    ,   123.51001   )
    (   1250    ,   125.569 )
    (   1300    ,   127.36546   )
    (   1350    ,   129.77171   )
    (   1400    ,   131.2523    )
    (   1450    ,   133.14342   )
    (   1500    ,   134.28772   )
    (   1550    ,   122.92765   )
    (   1600    ,   125.36359   )
    (   1650    ,   128.59037   )
    (   1700    ,   132.07818   )
    (   1750    ,   134.13897   )
    (   1800    ,   135.98463   )
    (   1850    ,   137.15518   )
    (   1900    ,   139.42026   )
    (   1950    ,   139.39144   )
    (   2000    ,   141.0264    )
    (   2050    ,   142.28616   )
    (   2100    ,   143.80288   )
    (   2150    ,   144.143 )
    (   2200    ,   145.61884   )
    (   2250    ,   146.11294   )
    (   2300    ,   147.74118   )
    (   2350    ,   136.27051   )
    (   2400    ,   138.84665   )
    (   2450    ,   140.896 )
    (   2500    ,   142.71686   )
    (   2550    ,   143.73892   )
    (   2600    ,   145.74845   )
    (   2650    ,   146.67338   )
    (   2700    ,   146.6541    )
    (   2750    ,   147.75528   )
    (   2800    ,   148.50696   )
    (   2850    ,   149.44369   )
    (   2900    ,   150.13879   )
    (   2950    ,   151.00128   )
    (   3000    ,   152.03178   )
    (   3050    ,   152.91399   )
    (   3100    ,   143.2909    )
    (   3150    ,   144.47025   )
    (   3200    ,   146.36872   )
    (   3250    ,   148.48188   )
    (   3300    ,   149.10075   )
    (   3350    ,   149.70512   )
    (   3400    ,   150.61468   )
    (   3450    ,   152.06791   )
    (   3500    ,   151.69121   )
    (   3550    ,   152.59101   )
    (   3600    ,   153.18133   )
    (   3650    ,   152.8951    )
    (   3700    ,   153.84061   )
    (   3750    ,   153.96568   )
    (   3800    ,   155.04197   )
    (   3850    ,   146.53834   )
    (   3900    ,   147.42865   )
    (   3950    ,   148.95345   )
    (   4000    ,   151.00045   )
};
\addplot[color=cyan,solid,thick] coordinates {
    (   50  ,   36.85232    )
    (   100 ,   92.86848    )
    (   150 ,   136.32921   )
    (   200 ,   164.13174   )
    (   250 ,   182.07065   )
    (   300 ,   195.19425   )
    (   350 ,   204.38723   )
    (   400 ,   202.49192   )
    (   450 ,   184.87326   )
    (   500 ,   187.04723   )
    (   550 ,   180.07721   )
    (   600 ,   131.68106   )
    (   650 ,   127.36085   )
    (   700 ,   126.13635   )
    (   750 ,   124.48785   )
    (   800 ,   88.88354    )
    (   850 ,   95.74035    )
    (   900 ,   100.97455   )
    (   950 ,   104.65065   )
    (   1000    ,   102.71698   )
    (   1050    ,   106.35218   )
    (   1100    ,   109.47187   )
    (   1150    ,   113.37613   )
    (   1200    ,   112.9072    )
    (   1250    ,   116.56782   )
    (   1300    ,   118.57598   )
    (   1350    ,   119.25235   )
    (   1400    ,   119.77855   )
    (   1450    ,   119.43333   )
    (   1500    ,   119.02571   )
    (   1550    ,   102.37671   )
    (   1600    ,   105.58691   )
    (   1650    ,   110.41858   )
    (   1700    ,   114.2185    )
    (   1750    ,   113.05688   )
    (   1800    ,   114.93292   )
    (   1850    ,   115.65804   )
    (   1900    ,   118.46496   )
    (   1950    ,   118.51945   )
    (   2000    ,   119.60154   )
    (   2050    ,   121.20127   )
    (   2100    ,   122.17481   )
    (   2150    ,   120.06766   )
    (   2200    ,   120.33883   )
    (   2250    ,   120.43044   )
    (   2300    ,   120.50589   )
    (   2350    ,   112.33513   )
    (   2400    ,   114.90883   )
    (   2450    ,   117.69318   )
    (   2500    ,   117.30684   )
    (   2550    ,   118.50546   )
    (   2600    ,   120.35036   )
    (   2650    ,   121.71178   )
    (   2700    ,   122.06547   )
    (   2750    ,   122.1195    )
    (   2800    ,   122.53049   )
    (   2850    ,   123.08053   )
    (   2900    ,   122.93796   )
    (   2950    ,   122.81279   )
    (   3000    ,   123.12189   )
    (   3050    ,   121.73121   )
    (   3100    ,   114.44378   )
    (   3150    ,   116.6871    )
    (   3200    ,   119.00376   )
    (   3250    ,   120.70447   )
    (   3300    ,   119.56853   )
    (   3350    ,   119.71876   )
    (   3400    ,   121.18198   )
    (   3450    ,   123.26429   )
    (   3500    ,   122.37359   )
    (   3550    ,   122.96681   )
    (   3600    ,   123.04297   )
    (   3650    ,   123.12008   )
    (   3700    ,   123.40997   )
    (   3750    ,   123.3679    )
    (   3800    ,   123.29344   )
    (   3850    ,   116.36954   )
    (   3900    ,   117.77415   )
    (   3950    ,   119.69566   )
    (   4000    ,   121.33029   )
};
\addplot[color=green,dashdotted,thick] coordinates {
    (   50  ,   37.75294    )
    (   100 ,   93.0146 )
    (   150 ,   139.26571   )
    (   200 ,   160.0096    )
    (   250 ,   179.26801   )
    (   300 ,   197.38826   )
    (   350 ,   194.55356   )
    (   400 ,   198.92065   )
    (   450 ,   205.88719   )
    (   500 ,   179.54299   )
    (   550 ,   169.30038   )
    (   600 ,   164.03938   )
    (   650 ,   88.66157    )
    (   700 ,   94.38299    )
    (   750 ,   97.55312    )
    (   800 ,   96.0483 )
    (   850 ,   99.41226    )
    (   900 ,   104.6651    )
    (   950 ,   104.05992   )
    (   1000    ,   107.65789   )
    (   1050    ,   110.82308   )
    (   1100    ,   112.41297   )
    (   1150    ,   115.33128   )
    (   1200    ,   117.95727   )
    (   1250    ,   114.43296   )
    (   1300    ,   114.24291   )
    (   1350    ,   115.92713   )
    (   1400    ,   122.41771   )
    (   1450    ,   124.14156   )
    (   1500    ,   126.58099   )
    (   1550    ,   127.99943   )
    (   1600    ,   128.38754   )
    (   1650    ,   131.73088   )
    (   1700    ,   133.64177   )
    (   1750    ,   135.14332   )
    (   1800    ,   136.02865   )
    (   1850    ,   135.90123   )
    (   1900    ,   128.29872   )
    (   1950    ,   131.44127   )
    (   2000    ,   133.73015   )
    (   2050    ,   135.00217   )
    (   2100    ,   137.05268   )
    (   2150    ,   138.36454   )
    (   2200    ,   138.64319   )
    (   2250    ,   140.12588   )
    (   2300    ,   142.08959   )
    (   2350    ,   142.73449   )
    (   2400    ,   143.76067   )
    (   2450    ,   146.71315   )
    (   2500    ,   140.43855   )
    (   2550    ,   140.23041   )
    (   2600    ,   141.29646   )
    (   2650    ,   145.97905   )
    (   2700    ,   146.30743   )
    (   2750    ,   146.7167    )
    (   2800    ,   148.12942   )
    (   2850    ,   148.07371   )
    (   2900    ,   149.48391   )
    (   2950    ,   150.64162   )
    (   3000    ,   151.86711   )
    (   3050    ,   150.95944   )
    (   3100    ,   152.19235   )
    (   3150    ,   146.60755   )
    (   3200    ,   148.93546   )
    (   3250    ,   150.06714   )
    (   3300    ,   149.92788   )
    (   3350    ,   149.77066   )
    (   3400    ,   150.9441    )
    (   3450    ,   151.68329   )
    (   3500    ,   152.52382   )
    (   3550    ,   153.84716   )
    (   3600    ,   153.84094   )
    (   3650    ,   155.995 )
    (   3700    ,   156.49833   )
    (   3750    ,   151.09424   )
    (   3800    ,   151.67227   )
    (   3850    ,   152.12348   )
    (   3900    ,   155.73257   )
    (   3950    ,   155.23097   )
    (   4000    ,   155.67606   )
};
\end{axis}
\end{tikzpicture}

%% file: g_shapes_800_n.tex
\begin{tikzpicture}
  \begin{axis}[width=2.2in, height=1.60in,
               solid,
               xlabel={\tiny $n$},
               ylabel={\tiny  GFLOPS},
               title={ \small  $m=k=600$},
               xmin=0,xmax=4000,ymin=0,ymax=225,
               legend style={legend pos=south east, fill=white},
               clip=false]
\addplot[color=black,dotted,thick] coordinates {
    (   50  ,   24.43881    )
    (   100 ,   42.10718    )
    (   150 ,   58.04807    )
    (   200 ,   71.16298    )
    (   250 ,   79.28523    )
    (   300 ,   85.77692    )
    (   350 ,   85.09882    )
    (   400 ,   89.78636    )
    (   450 ,   90.01748    )
    (   500 ,   93.06322    )
    (   550 ,   91.98173    )
    (   600 ,   87.08419    )
    (   650 ,   86.94762    )
    (   700 ,   85.2244 )
    (   750 ,   86.17333    )
    (   800 ,   85.20304    )
    (   850 ,   83.4253 )
    (   900 ,   82.69385    )
    (   950 ,   80.54635    )
    (   1000    ,   79.85777    )
    (   1050    ,   79.97177    )
    (   1100    ,   78.96674    )
    (   1150    ,   78.17602    )
    (   1200    ,   76.81961    )
    (   1250    ,   76.80884    )
    (   1300    ,   76.33111    )
    (   1350    ,   75.88078    )
    (   1400    ,   75.69397    )
    (   1450    ,   74.98703    )
    (   1500    ,   75.50679    )
    (   1550    ,   74.68054    )
    (   1600    ,   74.37638    )
    (   1650    ,   74.58683    )
    (   1700    ,   74.16884    )
    (   1750    ,   74.23039    )
    (   1800    ,   74.24326    )
    (   1850    ,   73.6045 )
    (   1900    ,   73.33523    )
    (   1950    ,   73.4432 )
    (   2000    ,   73.81494    )
    (   2050    ,   74.02602    )
    (   2100    ,   74.18431    )
    (   2150    ,   73.71448    )
    (   2200    ,   74.11901    )
    (   2250    ,   73.6553 )
    (   2300    ,   73.77309    )
    (   2350    ,   73.76647    )
    (   2400    ,   73.57495    )
    (   2450    ,   72.95219    )
    (   2500    ,   73.5856 )
    (   2550    ,   73.51686    )
    (   2600    ,   73.718  )
    (   2650    ,   73.6018 )
    (   2700    ,   73.17148    )
    (   2750    ,   73.32646    )
    (   2800    ,   73.3481 )
    (   2850    ,   72.86504    )
    (   2900    ,   72.99163    )
    (   2950    ,   73.04484    )
    (   3000    ,   73.15656    )
    (   3050    ,   70.79695    )
    (   3100    ,   71.46079    )
    (   3150    ,   71.94358    )
    (   3200    ,   72.44017    )
    (   3250    ,   72.78127    )
    (   3300    ,   73.4497 )
    (   3350    ,   73.95445    )
    (   3400    ,   74.27699    )
    (   3450    ,   74.96648    )
    (   3500    ,   75.05704    )
    (   3550    ,   75.43354    )
    (   3600    ,   76.04448    )
    (   3650    ,   75.79843    )
    (   3700    ,   76.01049    )
    (   3750    ,   75.89995    )
    (   3800    ,   76.54887    )
    (   3850    ,   76.29039    )
    (   3900    ,   76.33521    )
    (   3950    ,   75.95235    )
    (   4000    ,   75.68451    )
};
\addplot[color=magenta,dashed,thick] coordinates {
    (   50  ,   126.53779   )
    (   100 ,   169.62329   )
    (   150 ,   171.91266   )
    (   200 ,   170.47775   )
    (   250 ,   168.8934    )
    (   300 ,   136.22509   )
    (   350 ,   138.16751   )
    (   400 ,   136.22451   )
    (   450 ,   134.14172   )
    (   500 ,   131.46809   )
    (   550 ,   128.79695   )
    (   600 ,   131.04979   )
    (   650 ,   102.98786   )
    (   700 ,   103.65881   )
    (   750 ,   104.7933    )
    (   800 ,   106.57016   )
    (   850 ,   107.26484   )
    (   900 ,   109.45865   )
    (   950 ,   109.37696   )
    (   1000    ,   110.90124   )
    (   1050    ,   110.08044   )
    (   1100    ,   110.94462   )
    (   1150    ,   111.93819   )
    (   1200    ,   113.34712   )
    (   1250    ,   114.87755   )
    (   1300    ,   116.88516   )
    (   1350    ,   115.28309   )
    (   1400    ,   119.74627   )
    (   1450    ,   119.29613   )
    (   1500    ,   118.29768   )
    (   1550    ,   119.60765   )
    (   1600    ,   120.28771   )
    (   1650    ,   122.34532   )
    (   1700    ,   121.74255   )
    (   1750    ,   123.35852   )
    (   1800    ,   123.32148   )
    (   1850    ,   123.11439   )
    (   1900    ,   125.11177   )
    (   1950    ,   124.90144   )
    (   2000    ,   124.29806   )
    (   2050    ,   125.07261   )
    (   2100    ,   126.46992   )
    (   2150    ,   126.08419   )
    (   2200    ,   127.74836   )
    (   2250    ,   127.72865   )
    (   2300    ,   129.07009   )
    (   2350    ,   129.54394   )
    (   2400    ,   129.54722   )
    (   2450    ,   130.11886   )
    (   2500    ,   130.33333   )
    (   2550    ,   130.21505   )
    (   2600    ,   129.97885   )
    (   2650    ,   130.61071   )
    (   2700    ,   131.56482   )
    (   2750    ,   132.54432   )
    (   2800    ,   131.951 )
    (   2850    ,   133.0303    )
    (   2900    ,   132.81219   )
    (   2950    ,   132.29148   )
    (   3000    ,   133.92031   )
    (   3050    ,   133.94314   )
    (   3100    ,   133.74826   )
    (   3150    ,   133.95797   )
    (   3200    ,   134.57508   )
    (   3250    ,   135.2764    )
    (   3300    ,   135.3901    )
    (   3350    ,   134.97797   )
    (   3400    ,   135.6393    )
    (   3450    ,   135.91107   )
    (   3500    ,   135.64549   )
    (   3550    ,   135.4995    )
    (   3600    ,   136.55873   )
    (   3650    ,   135.89918   )
    (   3700    ,   137.38846   )
    (   3750    ,   137.00941   )
    (   3800    ,   137.11093   )
    (   3850    ,   137.55155   )
    (   3900    ,   137.81908   )
    (   3950    ,   137.48564   )
    (   4000    ,   137.99825   )
};
\addplot[color=cyan,solid,thick] coordinates {
    (   50  ,   108.83498   )
    (   100 ,   137.34665   )
    (   150 ,   166.38012   )
    (   200 ,   179.73893   )
    (   250 ,   183.90466   )
    (   300 ,   194.57146   )
    (   350 ,   186.2664    )
    (   400 ,   184.0892    )
    (   450 ,   136.97721   )
    (   500 ,   134.4081    )
    (   550 ,   132.37391   )
    (   600 ,   131.9982    )
    (   650 ,   125.0984    )
    (   700 ,   125.05968   )
    (   750 ,   121.81109   )
    (   800 ,   112.08036   )
    (   850 ,   116.06556   )
    (   900 ,   116.04742   )
    (   950 ,   115.23499   )
    (   1000    ,   115.05154   )
    (   1050    ,   114.79189   )
    (   1100    ,   114.21699   )
    (   1150    ,   115.61914   )
    (   1200    ,   115.659 )
    (   1250    ,   114.20878   )
    (   1300    ,   114.17972   )
    (   1350    ,   113.90509   )
    (   1400    ,   113.90128   )
    (   1450    ,   112.18399   )
    (   1500    ,   110.74907   )
    (   1550    ,   102.46882   )
    (   1600    ,   108.87266   )
    (   1650    ,   106.44673   )
    (   1700    ,   107.11298   )
    (   1750    ,   109.8159    )
    (   1800    ,   109.03053   )
    (   1850    ,   109.09946   )
    (   1900    ,   109.69159   )
    (   1950    ,   111.24165   )
    (   2000    ,   110.50958   )
    (   2050    ,   110.76885   )
    (   2100    ,   111.32415   )
    (   2150    ,   110.87395   )
    (   2200    ,   111.25456   )
    (   2250    ,   110.7828    )
    (   2300    ,   110.64327   )
    (   2350    ,   109.66737   )
    (   2400    ,   110.49777   )
    (   2450    ,   111.15439   )
    (   2500    ,   111.14368   )
    (   2550    ,   112.90957   )
    (   2600    ,   112.95669   )
    (   2650    ,   113.76656   )
    (   2700    ,   115.16159   )
    (   2750    ,   113.24634   )
    (   2800    ,   114.80832   )
    (   2850    ,   114.13035   )
    (   2900    ,   114.88034   )
    (   2950    ,   114.78615   )
    (   3000    ,   115.65636   )
    (   3050    ,   113.51702   )
    (   3100    ,   111.23295   )
    (   3150    ,   112.54023   )
    (   3200    ,   113.7472    )
    (   3250    ,   114.08822   )
    (   3300    ,   114.21145   )
    (   3350    ,   114.24722   )
    (   3400    ,   114.89214   )
    (   3450    ,   114.37705   )
    (   3500    ,   115.31582   )
    (   3550    ,   116.31394   )
    (   3600    ,   116.62806   )
    (   3650    ,   115.7282    )
    (   3700    ,   116.96181   )
    (   3750    ,   116.44857   )
    (   3800    ,   116.41579   )
    (   3850    ,   115.50241   )
    (   3900    ,   114.28477   )
    (   3950    ,   115.0645    )
    (   4000    ,   116.11158   )
};
\addplot[color=green,dashdotted,thick] coordinates {
    (   50  ,   107.17507   )
    (   100 ,   139.85824   )
    (   150 ,   162.32571   )
    (   200 ,   178.35534   )
    (   250 ,   189.3063    )
    (   300 ,   194.70001   )
    (   350 ,   185.36826   )
    (   400 ,   179.75766   )
    (   450 ,   178.15863   )
    (   500 ,   168.64512   )
    (   550 ,   167.97355   )
    (   600 ,   162.42983   )
    (   650 ,   90.67583    )
    (   700 ,   93.52312    )
    (   750 ,   94.479  )
    (   800 ,   95.844  )
    (   850 ,   97.02443    )
    (   900 ,   97.58542    )
    (   950 ,   96.63466    )
    (   1000    ,   97.68407    )
    (   1050    ,   97.37601    )
    (   1100    ,   98.29556    )
    (   1150    ,   98.29418    )
    (   1200    ,   98.14613    )
    (   1250    ,   93.01807    )
    (   1300    ,   94.29428    )
    (   1350    ,   95.94171    )
    (   1400    ,   97.06634    )
    (   1450    ,   97.43693    )
    (   1500    ,   98.08194    )
    (   1550    ,   97.96506    )
    (   1600    ,   98.90475    )
    (   1650    ,   98.87555    )
    (   1700    ,   99.84997    )
    (   1750    ,   99.78158    )
    (   1800    ,   99.6702 )
    (   1850    ,   98.16645    )
    (   1900    ,   97.61793    )
    (   1950    ,   99.04007    )
    (   2000    ,   99.94081    )
    (   2050    ,   100.20518   )
    (   2100    ,   100.02155   )
    (   2150    ,   100.42495   )
    (   2200    ,   100.45884   )
    (   2250    ,   100.75571   )
    (   2300    ,   101.36146   )
    (   2350    ,   101.12904   )
    (   2400    ,   101.63117   )
    (   2450    ,   100.92247   )
    (   2500    ,   100.21237   )
    (   2550    ,   100.92301   )
    (   2600    ,   101.70579   )
    (   2650    ,   101.62393   )
    (   2700    ,   102.43058   )
    (   2750    ,   102.80602   )
    (   2800    ,   102.6858    )
    (   2850    ,   103.32783   )
    (   2900    ,   103.75158   )
    (   2950    ,   103.28856   )
    (   3000    ,   103.92001   )
    (   3050    ,   103.71492   )
    (   3100    ,   103.00567   )
    (   3150    ,   103.08028   )
    (   3200    ,   103.124 )
    (   3250    ,   103.54178   )
    (   3300    ,   103.95561   )
    (   3350    ,   104.05514   )
    (   3400    ,   103.93962   )
    (   3450    ,   103.93909   )
    (   3500    ,   104.38237   )
    (   3550    ,   104.23585   )
    (   3600    ,   104.48812   )
    (   3650    ,   104.35891   )
    (   3700    ,   104.27358   )
    (   3750    ,   103.54493   )
    (   3800    ,   103.79874   )
    (   3850    ,   104.40238   )
    (   3900    ,   104.3067    )
    (   3950    ,   104.57846   )
    (   4000    ,   104.9845    )
};
\end{axis}
\end{tikzpicture}

%% file: g_shapes_800_mk.tex
\begin{tikzpicture}
  \begin{axis}[width=2.2in, height=1.60in,
               solid,
               xlabel={\tiny $m=k$},
               title={ \small  $n=600$},
               xmin=0,xmax=4000,ymin=0,ymax=225,
               legend style={legend pos=south east, fill=white},
               clip=false]
\addplot[color=black,dotted,thick] coordinates {
    (   50  ,   28.47921    )
    (   100 ,   91.72208    )
    (   150 ,   133.65873   )
    (   200 ,   172.47946   )
    (   250 ,   99.82218    )
    (   300 ,   104.9371    )
    (   350 ,   115.74958   )
    (   400 ,   93.15554    )
    (   450 ,   97.43466    )
    (   500 ,   101.00568   )
    (   550 ,   104.63275   )
    (   600 ,   87.48263    )
    (   650 ,   88.66478    )
    (   700 ,   92.11984    )
    (   750 ,   94.46823    )
    (   800 ,   92.25569    )
    (   850 ,   91.28949    )
    (   900 ,   93.70982    )
    (   950 ,   97.30933    )
    (   1000    ,   91.79378    )
    (   1050    ,   92.08512    )
    (   1100    ,   93.46591    )
    (   1150    ,   96.41835    )
    (   1200    ,   90.50381    )
    (   1250    ,   91.93138    )
    (   1300    ,   93.52984    )
    (   1350    ,   88.7665 )
    (   1400    ,   89.69952    )
    (   1450    ,   91.05781    )
    (   1500    ,   93.4371 )
    (   1550    ,   88.57201    )
    (   1600    ,   89.75266    )
    (   1650    ,   91.54038    )
    (   1700    ,   93.6103 )
    (   1750    ,   88.86044    )
    (   1800    ,   90.59144    )
    (   1850    ,   92.29779    )
    (   1900    ,   94.3908 )
    (   1950    ,   89.68898    )
    (   2000    ,   91.3411 )
    (   2050    ,   93.21837    )
    (   2100    ,   95.31492    )
    (   2150    ,   90.80891    )
    (   2200    ,   92.42148    )
    (   2250    ,   93.84709    )
    (   2300    ,   96.18135    )
    (   2350    ,   91.52006    )
    (   2400    ,   93.21058    )
    (   2450    ,   94.86861    )
    (   2500    ,   91.14019    )
    (   2550    ,   92.34251    )
    (   2600    ,   93.84584    )
    (   2650    ,   95.48037    )
    (   2700    ,   91.95038    )
    (   2750    ,   93.21694    )
    (   2800    ,   94.56936    )
    (   2850    ,   96.05702    )
    (   2900    ,   92.73645    )
    (   2950    ,   93.83371    )
    (   3000    ,   95.10819    )
    (   3050    ,   96.54394    )
    (   3100    ,   93.24308    )
    (   3150    ,   94.25511    )
    (   3200    ,   95.65095    )
    (   3250    ,   97.12587    )
    (   3300    ,   93.72194    )
    (   3350    ,   95.07011    )
    (   3400    ,   96.09065    )
    (   3450    ,   97.57888    )
    (   3500    ,   94.27939    )
    (   3550    ,   95.35663    )
    (   3600    ,   96.80596    )
    (   3650    ,   93.90253    )
    (   3700    ,   94.86344    )
    (   3750    ,   95.78986    )
    (   3800    ,   97.09971    )
    (   3850    ,   94.37196    )
    (   3900    ,   95.26868    )
    (   3950    ,   96.41759    )
    (   4000    ,   97.57203    )
};
\addplot[color=magenta,dashed,thick] coordinates {
    (   50  ,   36.84689    )
    (   100 ,   92.63047    )
    (   150 ,   138.56173   )
    (   200 ,   145.12823   )
    (   250 ,   183.14839   )
    (   300 ,   187.63454   )
    (   350 ,   207.72511   )
    (   400 ,   192.41581   )
    (   450 ,   191.76182   )
    (   500 ,   185.05291   )
    (   550 ,   175.95726   )
    (   600 ,   134.15762   )
    (   650 ,   121.04598   )
    (   700 ,   116.03357   )
    (   750 ,   110.68746   )
    (   800 ,   90.37372    )
    (   850 ,   97.12442    )
    (   900 ,   102.90417   )
    (   950 ,   106.30975   )
    (   1000    ,   103.7595    )
    (   1050    ,   106.42632   )
    (   1100    ,   110.76191   )
    (   1150    ,   114.67532   )
    (   1200    ,   113.13302   )
    (   1250    ,   116.78107   )
    (   1300    ,   118.49741   )
    (   1350    ,   119.11085   )
    (   1400    ,   119.4985    )
    (   1450    ,   119.29585   )
    (   1500    ,   118.88748   )
    (   1550    ,   102.1887    )
    (   1600    ,   105.09004   )
    (   1650    ,   110.03871   )
    (   1700    ,   113.43994   )
    (   1750    ,   112.87866   )
    (   1800    ,   114.12858   )
    (   1850    ,   115.18757   )
    (   1900    ,   117.09926   )
    (   1950    ,   117.40604   )
    (   2000    ,   117.91219   )
    (   2050    ,   119.81536   )
    (   2100    ,   120.21929   )
    (   2150    ,   118.60153   )
    (   2200    ,   119.28684   )
    (   2250    ,   119.68212   )
    (   2300    ,   119.31036   )
    (   2350    ,   111.80656   )
    (   2400    ,   113.96752   )
    (   2450    ,   117.46666   )
    (   2500    ,   116.59448   )
    (   2550    ,   117.77667   )
    (   2600    ,   118.98185   )
    (   2650    ,   120.26108   )
    (   2700    ,   120.62927   )
    (   2750    ,   120.94145   )
    (   2800    ,   121.57119   )
    (   2850    ,   122.03444   )
    (   2900    ,   121.98225   )
    (   2950    ,   122.20655   )
    (   3000    ,   122.19473   )
    (   3050    ,   120.79498   )
    (   3100    ,   113.83772   )
    (   3150    ,   116.24315   )
    (   3200    ,   117.99931   )
    (   3250    ,   119.76065   )
    (   3300    ,   118.88915   )
    (   3350    ,   119.03205   )
    (   3400    ,   120.26735   )
    (   3450    ,   121.38264   )
    (   3500    ,   121.06402   )
    (   3550    ,   121.62221   )
    (   3600    ,   122.38228   )
    (   3650    ,   122.57058   )
    (   3700    ,   122.55656   )
    (   3750    ,   122.60269   )
    (   3800    ,   122.50125   )
    (   3850    ,   115.73049   )
    (   3900    ,   117.24603   )
    (   3950    ,   118.77597   )
    (   4000    ,   120.43955   )
};
\addplot[color=cyan,solid,thick] coordinates {
    (   50  ,   29.87661    )
    (   100 ,   88.36134    )
    (   150 ,   133.56485   )
    (   200 ,   178.06268   )
    (   250 ,   191.61199   )
    (   300 ,   202.51152   )
    (   350 ,   203.36423   )
    (   400 ,   198.71847   )
    (   450 ,   193.22627   )
    (   500 ,   187.86799   )
    (   550 ,   173.1439    )
    (   600 ,   131.25596   )
    (   650 ,   123.28933   )
    (   700 ,   127.65838   )
    (   750 ,   133.30655   )
    (   800 ,   116.85277   )
    (   850 ,   118.97858   )
    (   900 ,   123.80196   )
    (   950 ,   128.32921   )
    (   1000    ,   126.10489   )
    (   1050    ,   127.02765   )
    (   1100    ,   130.92607   )
    (   1150    ,   133.37344   )
    (   1200    ,   132.24385   )
    (   1250    ,   133.76703   )
    (   1300    ,   135.83973   )
    (   1350    ,   136.45166   )
    (   1400    ,   138.0286    )
    (   1450    ,   137.46241   )
    (   1500    ,   138.57606   )
    (   1550    ,   126.31578   )
    (   1600    ,   128.60659   )
    (   1650    ,   131.88965   )
    (   1700    ,   136.21712   )
    (   1750    ,   137.70008   )
    (   1800    ,   138.74006   )
    (   1850    ,   141.17303   )
    (   1900    ,   143.51161   )
    (   1950    ,   141.69804   )
    (   2000    ,   143.2924    )
    (   2050    ,   144.41366   )
    (   2100    ,   145.46303   )
    (   2150    ,   147.01884   )
    (   2200    ,   148.1312    )
    (   2250    ,   148.78284   )
    (   2300    ,   150.84306   )
    (   2350    ,   138.23079   )
    (   2400    ,   140.49745   )
    (   2450    ,   142.28876   )
    (   2500    ,   144.24365   )
    (   2550    ,   144.80253   )
    (   2600    ,   147.52638   )
    (   2650    ,   148.70896   )
    (   2700    ,   148.16464   )
    (   2750    ,   148.75411   )
    (   2800    ,   150.33832   )
    (   2850    ,   151.21841   )
    (   2900    ,   151.8969    )
    (   2950    ,   152.34107   )
    (   3000    ,   152.8367    )
    (   3050    ,   153.59238   )
    (   3100    ,   143.50431   )
    (   3150    ,   144.68853   )
    (   3200    ,   146.83789   )
    (   3250    ,   149.36704   )
    (   3300    ,   149.38234   )
    (   3350    ,   150.22509   )
    (   3400    ,   150.79209   )
    (   3450    ,   152.3103    )
    (   3500    ,   152.26112   )
    (   3550    ,   152.78051   )
    (   3600    ,   153.72362   )
    (   3650    ,   153.83343   )
    (   3700    ,   154.22954   )
    (   3750    ,   155.06993   )
    (   3800    ,   155.68726   )
    (   3850    ,   146.76599   )
    (   3900    ,   147.81347   )
    (   3950    ,   149.54024   )
    (   4000    ,   151.60189   )
};
\addplot[color=green,dashdotted,thick] coordinates {
    (   50  ,   28.49084    )
    (   100 ,   87.13331    )
    (   150 ,   145.07237   )
    (   200 ,   161.24535   )
    (   250 ,   190.35678   )
    (   300 ,   210.58461   )
    (   350 ,   198.08438   )
    (   400 ,   202.58101   )
    (   450 ,   203.96446   )
    (   500 ,   177.30255   )
    (   550 ,   167.8841    )
    (   600 ,   162.50278   )
    (   650 ,   122.37016   )
    (   700 ,   121.05042   )
    (   750 ,   122.61745   )
    (   800 ,   99.8063 )
    (   850 ,   102.5441    )
    (   900 ,   106.0308    )
    (   950 ,   110.47664   )
    (   1000    ,   113.66537   )
    (   1050    ,   116.63225   )
    (   1100    ,   117.68655   )
    (   1150    ,   120.66969   )
    (   1200    ,   123.00678   )
    (   1250    ,   117.21059   )
    (   1300    ,   121.84785   )
    (   1350    ,   125.5654    )
    (   1400    ,   128.12363   )
    (   1450    ,   127.21311   )
    (   1500    ,   129.17506   )
    (   1550    ,   132.46881   )
    (   1600    ,   133.94641   )
    (   1650    ,   135.22475   )
    (   1700    ,   137.44841   )
    (   1750    ,   138.0814    )
    (   1800    ,   139.69119   )
    (   1850    ,   140.78283   )
    (   1900    ,   135.59745   )
    (   1950    ,   138.68208   )
    (   2000    ,   140.49661   )
    (   2050    ,   140.99037   )
    (   2100    ,   143.17167   )
    (   2150    ,   143.88268   )
    (   2200    ,   145.11074   )
    (   2250    ,   146.39634   )
    (   2300    ,   148.11206   )
    (   2350    ,   147.85512   )
    (   2400    ,   150.54923   )
    (   2450    ,   150.97389   )
    (   2500    ,   145.91215   )
    (   2550    ,   147.60394   )
    (   2600    ,   149.44673   )
    (   2650    ,   151.2412    )
    (   2700    ,   151.53071   )
    (   2750    ,   152.75053   )
    (   2800    ,   154.29352   )
    (   2850    ,   153.6348    )
    (   2900    ,   154.85887   )
    (   2950    ,   156.2281    )
    (   3000    ,   156.35803   )
    (   3050    ,   157.52125   )
    (   3100    ,   157.82303   )
    (   3150    ,   153.6496    )
    (   3200    ,   155.11738   )
    (   3250    ,   156.57075   )
    (   3300    ,   156.94015   )
    (   3350    ,   157.95703   )
    (   3400    ,   159.12193   )
    (   3450    ,   158.52199   )
    (   3500    ,   160.0032    )
    (   3550    ,   160.42551   )
    (   3600    ,   160.71848   )
    (   3650    ,   161.40992   )
    (   3700    ,   162.25785   )
    (   3750    ,   157.89662   )
    (   3800    ,   159.45738   )
    (   3850    ,   160.98452   )
    (   3900    ,   162.06425   )
    (   3950    ,   162.15135   )
    (   4000    ,   162.77819   )
};
\end{axis}
\end{tikzpicture}

%% file: g_shapes_800_k.tex
\begin{tikzpicture}
  \begin{axis}[width=2.2in, height=1.60in,
               solid,
               xlabel={\tiny $k$},
               ylabel={\tiny GFLOPS},
               title={ \small $m=n=600$ },
               xmin=0,xmax=4000,ymin=0,ymax=225,
               legend style={legend pos=south east, fill=white},
               clip=false]
\addplot[color=black,dotted,thick] coordinates {
    (   50  ,   41.46348    )
    (   100 ,   66.86702    )
    (   150 ,   92.71413    )
    (   200 ,   69.51404    )
    (   250 ,   81.15189    )
    (   300 ,   90.80102    )
    (   350 ,   102.37086   )
    (   400 ,   89.17646    )
    (   450 ,   94.46777    )
    (   500 ,   97.66955    )
    (   550 ,   106.16591   )
    (   600 ,   87.54418    )
    (   650 ,   89.42471    )
    (   700 ,   93.60682    )
    (   750 ,   103.82359   )
    (   800 ,   95.31625    )
    (   850 ,   96.46208    )
    (   900 ,   99.60233    )
    (   950 ,   102.05421   )
    (   1000    ,   103.29497   )
    (   1050    ,   104.8878    )
    (   1100    ,   107.09629   )
    (   1150    ,   110.73172   )
    (   1200    ,   111.71368   )
    (   1250    ,   112.83311   )
    (   1300    ,   113.80598   )
    (   1350    ,   114.69503   )
    (   1400    ,   116.99766   )
    (   1450    ,   118.22275   )
    (   1500    ,   119.81637   )
    (   1550    ,   120.32669   )
    (   1600    ,   121.70067   )
    (   1650    ,   122.60363   )
    (   1700    ,   123.67343   )
    (   1750    ,   124.26168   )
    (   1800    ,   125.05435   )
    (   1850    ,   126.30481   )
    (   1900    ,   127.36242   )
    (   1950    ,   128.75637   )
    (   2000    ,   128.47578   )
    (   2050    ,   129.75622   )
    (   2100    ,   131.06756   )
    (   2150    ,   132.68444   )
    (   2200    ,   134.87637   )
    (   2250    ,   132.82601   )
    (   2300    ,   134.75977   )
    (   2350    ,   134.57687   )
    (   2400    ,   134.4313    )
    (   2450    ,   135.82095   )
    (   2500    ,   136.14724   )
    (   2550    ,   140.81691   )
    (   2600    ,   137.70081   )
    (   2650    ,   137.79166   )
    (   2700    ,   139.90161   )
    (   2750    ,   139.5498    )
    (   2800    ,   140.41423   )
    (   2850    ,   140.96238   )
    (   2900    ,   140.90005   )
    (   2950    ,   140.98643   )
    (   3000    ,   142.42913   )
    (   3050    ,   142.2952    )
    (   3100    ,   142.73565   )
    (   3150    ,   143.21958   )
    (   3200    ,   143.26042   )
    (   3250    ,   144.01382   )
    (   3300    ,   143.91939   )
    (   3350    ,   144.76595   )
    (   3400    ,   144.7568    )
    (   3450    ,   145.20331   )
    (   3500    ,   145.81241   )
    (   3550    ,   146.28475   )
    (   3600    ,   146.70445   )
    (   3650    ,   147.22548   )
    (   3700    ,   147.40034   )
    (   3750    ,   147.9595    )
    (   3800    ,   147.87173   )
    (   3850    ,   148.22648   )
    (   3900    ,   149.85907   )
    (   3950    ,   147.82671   )
    (   4000    ,   149.43653   )
};
\addplot[color=magenta,dashed,thick] coordinates {
    (   50  ,   139.49217   )
    (   100 ,   176.8655    )
    (   150 ,   187.72194   )
    (   200 ,   189.57496   )
    (   250 ,   206.63837   )
    (   300 ,   188.00346   )
    (   350 ,   199.01346   )
    (   400 ,   183.67745   )
    (   450 ,   165.97408   )
    (   500 ,   174.14734   )
    (   550 ,   167.4721    )
    (   600 ,   135.26211   )
    (   650 ,   131.0438    )
    (   700 ,   129.14964   )
    (   750 ,   126.78446   )
    (   800 ,   94.20625    )
    (   850 ,   96.58815    )
    (   900 ,   98.08319    )
    (   950 ,   102.08541   )
    (   1000    ,   102.7517    )
    (   1050    ,   102.42624   )
    (   1100    ,   104.88564   )
    (   1150    ,   107.77548   )
    (   1200    ,   108.76767   )
    (   1250    ,   107.16657   )
    (   1300    ,   108.4649    )
    (   1350    ,   109.50208   )
    (   1400    ,   109.77456   )
    (   1450    ,   110.14753   )
    (   1500    ,   111.32988   )
    (   1550    ,   107.42123   )
    (   1600    ,   108.59696   )
    (   1650    ,   110.27843   )
    (   1700    ,   111.84514   )
    (   1750    ,   112.36596   )
    (   1800    ,   113.25466   )
    (   1850    ,   114.0714    )
    (   1900    ,   115.02258   )
    (   1950    ,   115.14744   )
    (   2000    ,   115.78773   )
    (   2050    ,   116.77921   )
    (   2100    ,   117.11085   )
    (   2150    ,   116.51332   )
    (   2200    ,   116.89777   )
    (   2250    ,   117.2745    )
    (   2300    ,   119.41575   )
    (   2350    ,   116.33923   )
    (   2400    ,   116.62782   )
    (   2450    ,   117.96841   )
    (   2500    ,   117.4068    )
    (   2550    ,   119.84489   )
    (   2600    ,   120.24798   )
    (   2650    ,   119.87487   )
    (   2700    ,   120.69821   )
    (   2750    ,   120.47608   )
    (   2800    ,   121.13427   )
    (   2850    ,   121.45055   )
    (   2900    ,   121.94159   )
    (   2950    ,   121.76463   )
    (   3000    ,   121.95466   )
    (   3050    ,   122.64891   )
    (   3100    ,   123.60022   )
    (   3150    ,   120.95857   )
    (   3200    ,   122.10897   )
    (   3250    ,   122.86762   )
    (   3300    ,   122.83886   )
    (   3350    ,   123.596 )
    (   3400    ,   123.65733   )
    (   3450    ,   123.98417   )
    (   3500    ,   124.20492   )
    (   3550    ,   123.67395   )
    (   3600    ,   125.26732   )
    (   3650    ,   124.61641   )
    (   3700    ,   125.22552   )
    (   3750    ,   124.81832   )
    (   3800    ,   124.30701   )
    (   3850    ,   123.09783   )
    (   3900    ,   122.98585   )
    (   3950    ,   124.29867   )
    (   4000    ,   124.64079   )
};
\addplot[color=cyan,solid,thick] coordinates {
    (   50  ,   143.99885   )
    (   100 ,   184.65374   )
    (   150 ,   199.59047   )
    (   200 ,   203.50164   )
    (   250 ,   207.55404   )
    (   300 ,   202.48874   )
    (   350 ,   166.17222   )
    (   400 ,   171.9287    )
    (   450 ,   167.61736   )
    (   500 ,   168.60105   )
    (   550 ,   164.13412   )
    (   600 ,   132.00328   )
    (   650 ,   130.32447   )
    (   700 ,   133.27537   )
    (   750 ,   134.76941   )
    (   800 ,   117.05502   )
    (   850 ,   119.20042   )
    (   900 ,   122.14588   )
    (   950 ,   125.31859   )
    (   1000    ,   114.3901    )
    (   1050    ,   114.54341   )
    (   1100    ,   116.08243   )
    (   1150    ,   118.44418   )
    (   1200    ,   117.79152   )
    (   1250    ,   117.96376   )
    (   1300    ,   118.35503   )
    (   1350    ,   117.70127   )
    (   1400    ,   118.05934   )
    (   1450    ,   117.49404   )
    (   1500    ,   119.48499   )
    (   1550    ,   113.38914   )
    (   1600    ,   115.08308   )
    (   1650    ,   116.12148   )
    (   1700    ,   117.45795   )
    (   1750    ,   119.50473   )
    (   1800    ,   119.37724   )
    (   1850    ,   119.99562   )
    (   1900    ,   121.61122   )
    (   1950    ,   118.98388   )
    (   2000    ,   118.88791   )
    (   2050    ,   119.11935   )
    (   2100    ,   120.85862   )
    (   2150    ,   122.75629   )
    (   2200    ,   121.74003   )
    (   2250    ,   122.30137   )
    (   2300    ,   122.64995   )
    (   2350    ,   117.82889   )
    (   2400    ,   119.69548   )
    (   2450    ,   119.77452   )
    (   2500    ,   121.55008   )
    (   2550    ,   123.78996   )
    (   2600    ,   124.21115   )
    (   2650    ,   124.25216   )
    (   2700    ,   124.75757   )
    (   2750    ,   124.24559   )
    (   2800    ,   125.05834   )
    (   2850    ,   125.35082   )
    (   2900    ,   125.58513   )
    (   2950    ,   125.51405   )
    (   3000    ,   124.96031   )
    (   3050    ,   125.29928   )
    (   3100    ,   123.19861   )
    (   3150    ,   124.01551   )
    (   3200    ,   124.815 )
    (   3250    ,   125.56814   )
    (   3300    ,   125.80129   )
    (   3350    ,   125.79365   )
    (   3400    ,   126.26087   )
    (   3450    ,   125.19111   )
    (   3500    ,   127.02178   )
    (   3550    ,   126.04702   )
    (   3600    ,   126.40416   )
    (   3650    ,   126.65392   )
    (   3700    ,   125.81455   )
    (   3750    ,   125.93352   )
    (   3800    ,   126.56459   )
    (   3850    ,   124.59906   )
    (   3900    ,   125.45064   )
    (   3950    ,   125.93181   )
    (   4000    ,   126.21474   )
};
\addplot[color=green,dashdotted,thick] coordinates {
    (   50  ,   141.22299   )
    (   100 ,   180.9609    )
    (   150 ,   202.11207   )
    (   200 ,   189.88819   )
    (   250 ,   199.69491   )
    (   300 ,   210.68053   )
    (   350 ,   189.14176   )
    (   400 ,   187.263 )
    (   450 ,   190.80103   )
    (   500 ,   163.31804   )
    (   550 ,   161.56546   )
    (   600 ,   164.22702   )
    (   650 ,   147.64418   )
    (   700 ,   145.60343   )
    (   750 ,   140.38561   )
    (   800 ,   109.0146    )
    (   850 ,   111.4281    )
    (   900 ,   112.92423   )
    (   950 ,   113.39382   )
    (   1000    ,   114.09282   )
    (   1050    ,   115.3946    )
    (   1100    ,   116.40273   )
    (   1150    ,   116.69435   )
    (   1200    ,   118.25662   )
    (   1250    ,   112.83023   )
    (   1300    ,   115.14297   )
    (   1350    ,   117.80639   )
    (   1400    ,   119.23811   )
    (   1450    ,   118.49435   )
    (   1500    ,   119.2624    )
    (   1550    ,   121.40444   )
    (   1600    ,   121.38829   )
    (   1650    ,   122.33466   )
    (   1700    ,   122.85635   )
    (   1750    ,   125.28314   )
    (   1800    ,   130.0853    )
    (   1850    ,   127.45506   )
    (   1900    ,   126.04456   )
    (   1950    ,   127.23596   )
    (   2000    ,   128.65846   )
    (   2050    ,   128.04013   )
    (   2100    ,   130.61169   )
    (   2150    ,   131.52272   )
    (   2200    ,   132.17563   )
    (   2250    ,   132.02698   )
    (   2300    ,   134.85595   )
    (   2350    ,   134.39733   )
    (   2400    ,   134.96133   )
    (   2450    ,   136.90088   )
    (   2500    ,   151.49164   )
    (   2550    ,   140.261 )
    (   2600    ,   140.24809   )
    (   2650    ,   141.94448   )
    (   2700    ,   143.33059   )
    (   2750    ,   143.5941    )
    (   2800    ,   144.9096    )
    (   2850    ,   143.03618   )
    (   2900    ,   145.02293   )
    (   2950    ,   145.77773   )
    (   3000    ,   146.00405   )
    (   3050    ,   146.14848   )
    (   3100    ,   147.78793   )
    (   3150    ,   147.97068   )
    (   3200    ,   147.48043   )
    (   3250    ,   148.8677    )
    (   3300    ,   147.815 )
    (   3350    ,   148.5575    )
    (   3400    ,   149.61169   )
    (   3450    ,   149.048 )
    (   3500    ,   149.38013   )
    (   3550    ,   150.34521   )
    (   3600    ,   150.23178   )
    (   3650    ,   149.96913   )
    (   3700    ,   150.3886    )
    (   3750    ,   152.81693   )
    (   3800    ,   152.05603   )
    (   3850    ,   152.7069    )
    (   3900    ,   153.1491    )
    (   3950    ,   152.73748   )
    (   4000    ,   157.58254   )
};
\end{axis}
\end{tikzpicture}

%% file: g_shapes_800_mn.tex
\begin{tikzpicture}
  \begin{axis}[width=2.2in, height=1.60in,
               solid,
               xlabel={\tiny $m=n$},
               title={ \small  $k=600$ },
               xmin=0,xmax=4000,ymin=0,ymax=225,
               legend style={legend pos=south east, fill=white},
               clip=false]
\addplot[color=black,dotted,thick] coordinates {
    (   50  ,   25.27891    )
    (   100 ,   60.46376    )
    (   150 ,   57.81832    )
    (   200 ,   92.51995    )
    (   250 ,   76.96957    )
    (   300 ,   87.52061    )
    (   350 ,   92.23315    )
    (   400 ,   89.00649    )
    (   450 ,   90.61018    )
    (   500 ,   90.75173    )
    (   550 ,   91.24461    )
    (   600 ,   87.09475    )
    (   650 ,   84.30697    )
    (   700 ,   83.72899    )
    (   750 ,   81.20855    )
    (   800 ,   80.56307    )
    (   850 ,   78.20987    )
    (   900 ,   77.66002    )
    (   950 ,   75.8354 )
    (   1000    ,   74.43993    )
    (   1050    ,   74.67163    )
    (   1100    ,   75.30298    )
    (   1150    ,   75.26272    )
    (   1200    ,   75.83905    )
    (   1250    ,   75.43896    )
    (   1300    ,   76.63063    )
    (   1350    ,   76.68506    )
    (   1400    ,   76.75147    )
    (   1450    ,   77.49727    )
    (   1500    ,   77.7115 )
    (   1550    ,   77.68008    )
    (   1600    ,   78.11457    )
    (   1650    ,   78.21116    )
    (   1700    ,   78.54364    )
    (   1750    ,   78.92568    )
    (   1800    ,   79.13195    )
    (   1850    ,   78.25922    )
    (   1900    ,   78.89688    )
    (   1950    ,   79.0697 )
    (   2000    ,   80.05769    )
    (   2050    ,   80.59398    )
    (   2100    ,   80.78179    )
    (   2150    ,   80.23147    )
    (   2200    ,   80.6605 )
    (   2250    ,   80.9269 )
    (   2300    ,   81.32929    )
    (   2350    ,   81.18728    )
    (   2400    ,   81.23729    )
    (   2450    ,   80.6005 )
    (   2500    ,   81.34604    )
    (   2550    ,   81.58559    )
    (   2600    ,   81.86895    )
    (   2650    ,   82.1186 )
    (   2700    ,   82.2389 )
    (   2750    ,   82.04985    )
    (   2800    ,   82.14799    )
    (   2850    ,   82.19011    )
    (   2900    ,   82.14524    )
    (   2950    ,   82.59977    )
    (   3000    ,   82.68131    )
    (   3050    ,   80.43198    )
    (   3100    ,   80.83563    )
    (   3150    ,   81.12513    )
    (   3200    ,   81.25544    )
    (   3250    ,   81.5005 )
    (   3300    ,   81.56542    )
    (   3350    ,   81.5749 )
    (   3400    ,   81.64897    )
    (   3450    ,   81.77622    )
    (   3500    ,   82.08284    )
    (   3550    ,   82.12409    )
    (   3600    ,   82.2602 )
    (   3650    ,   82.17382    )
    (   3700    ,   82.36553    )
    (   3750    ,   82.48754    )
    (   3800    ,   82.67027    )
    (   3850    ,   82.90333    )
    (   3900    ,   82.89471    )
    (   3950    ,   82.63116    )
    (   4000    ,   82.90375    )
};
\addplot[color=magenta,dashed,thick] coordinates {
    (   50  ,   35.55092    )
    (   100 ,   90.98146    )
    (   150 ,   135.96467   )
    (   200 ,   166.7651    )
    (   250 ,   178.75401   )
    (   300 ,   193.97353   )
    (   350 ,   191.99875   )
    (   400 ,   184.26015   )
    (   450 ,   180.81407   )
    (   500 ,   174.71593   )
    (   550 ,   159.36031   )
    (   600 ,   145.00703   )
    (   650 ,   107.78872   )
    (   700 ,   107.18833   )
    (   750 ,   104.46526   )
    (   800 ,   97.75684    )
    (   850 ,   102.51987   )
    (   900 ,   106.25069   )
    (   950 ,   107.64463   )
    (   1000    ,   108.29667   )
    (   1050    ,   109.69012   )
    (   1100    ,   112.58113   )
    (   1150    ,   113.90944   )
    (   1200    ,   117.02451   )
    (   1250    ,   119.4682    )
    (   1300    ,   120.51778   )
    (   1350    ,   122.76591   )
    (   1400    ,   124.29456   )
    (   1450    ,   125.27581   )
    (   1500    ,   124.75825   )
    (   1550    ,   115.3135    )
    (   1600    ,   117.54373   )
    (   1650    ,   121.11285   )
    (   1700    ,   124.04685   )
    (   1750    ,   125.67316   )
    (   1800    ,   126.58349   )
    (   1850    ,   126.39904   )
    (   1900    ,   127.8408    )
    (   1950    ,   128.52892   )
    (   2000    ,   129.28507   )
    (   2050    ,   130.45204   )
    (   2100    ,   131.35857   )
    (   2150    ,   130.50478   )
    (   2200    ,   131.57492   )
    (   2250    ,   131.4668    )
    (   2300    ,   131.8414    )
    (   2350    ,   127.14462   )
    (   2400    ,   129.25341   )
    (   2450    ,   131.26962   )
    (   2500    ,   132.44496   )
    (   2550    ,   133.20973   )
    (   2600    ,   134.26669   )
    (   2650    ,   134.84504   )
    (   2700    ,   135.69455   )
    (   2750    ,   135.19343   )
    (   2800    ,   135.75574   )
    (   2850    ,   135.79246   )
    (   2900    ,   136.61498   )
    (   2950    ,   136.98664   )
    (   3000    ,   137.20457   )
    (   3050    ,   136.00482   )
    (   3100    ,   131.86577   )
    (   3150    ,   133.3623    )
    (   3200    ,   134.82721   )
    (   3250    ,   135.15904   )
    (   3300    ,   136.00545   )
    (   3350    ,   135.58217   )
    (   3400    ,   136.43851   )
    (   3450    ,   136.92766   )
    (   3500    ,   137.48317   )
    (   3550    ,   137.77045   )
    (   3600    ,   138.08327   )
    (   3650    ,   138.98379   )
    (   3700    ,   139.55922   )
    (   3750    ,   139.6035    )
    (   3800    ,   139.53493   )
    (   3850    ,   135.33441   )
    (   3900    ,   136.17003   )
    (   3950    ,   136.97956   )
    (   4000    ,   137.98644   )
};
\addplot[color=cyan,solid,thick] coordinates {
    (   50  ,   31.89589    )
    (   100 ,   93.15975    )
    (   150 ,   134.68618   )
    (   200 ,   164.70903   )
    (   250 ,   181.28378   )
    (   300 ,   195.12512   )
    (   350 ,   198.71713   )
    (   400 ,   192.65503   )
    (   450 ,   180.20076   )
    (   500 ,   174.65764   )
    (   550 ,   165.90812   )
    (   600 ,   162.81065   )
    (   650 ,   125.65822   )
    (   700 ,   129.12151   )
    (   750 ,   127.85692   )
    (   800 ,   102.00699   )
    (   850 ,   105.93882   )
    (   900 ,   107.54041   )
    (   950 ,   108.18496   )
    (   1000    ,   109.55753   )
    (   1050    ,   111.55581   )
    (   1100    ,   113.99347   )
    (   1150    ,   115.48779   )
    (   1200    ,   118.5267    )
    (   1250    ,   120.85016   )
    (   1300    ,   122.73491   )
    (   1350    ,   123.91648   )
    (   1400    ,   125.99525   )
    (   1450    ,   126.04534   )
    (   1500    ,   126.01088   )
    (   1550    ,   115.85603   )
    (   1600    ,   118.50873   )
    (   1650    ,   121.41327   )
    (   1700    ,   123.82805   )
    (   1750    ,   125.67382   )
    (   1800    ,   127.62632   )
    (   1850    ,   127.00707   )
    (   1900    ,   127.91337   )
    (   1950    ,   129.28045   )
    (   2000    ,   130.50463   )
    (   2050    ,   130.73807   )
    (   2100    ,   130.99848   )
    (   2150    ,   130.90263   )
    (   2200    ,   131.63899   )
    (   2250    ,   132.0607    )
    (   2300    ,   132.20267   )
    (   2350    ,   127.51464   )
    (   2400    ,   129.24243   )
    (   2450    ,   131.11368   )
    (   2500    ,   132.31968   )
    (   2550    ,   133.03506   )
    (   2600    ,   134.22203   )
    (   2650    ,   134.92191   )
    (   2700    ,   135.67029   )
    (   2750    ,   135.33813   )
    (   2800    ,   135.754 )
    (   2850    ,   135.9438    )
    (   2900    ,   136.70149   )
    (   2950    ,   137.03126   )
    (   3000    ,   137.1823    )
    (   3050    ,   135.88037   )
    (   3100    ,   131.91741   )
    (   3150    ,   133.35206   )
    (   3200    ,   134.73077   )
    (   3250    ,   135.25209   )
    (   3300    ,   135.70672   )
    (   3350    ,   135.34227   )
    (   3400    ,   136.31942   )
    (   3450    ,   136.66227   )
    (   3500    ,   137.37187   )
    (   3550    ,   137.66623   )
    (   3600    ,   137.80913   )
    (   3650    ,   139.31238   )
    (   3700    ,   139.5327    )
    (   3750    ,   139.68044   )
    (   3800    ,   139.57578   )
    (   3850    ,   134.95307   )
    (   3900    ,   135.88578   )
    (   3950    ,   136.52817   )
    (   4000    ,   137.68813   )
};
\addplot[color=green,dashdotted,thick] coordinates {
    (   50  ,   34.11844    )
    (   100 ,   79.10089    )
    (   150 ,   133.9711    )
    (   200 ,   164.85668   )
    (   250 ,   179.52419   )
    (   300 ,   194.60897   )
    (   350 ,   202.46567   )
    (   400 ,   209.22349   )
    (   450 ,   192.89312   )
    (   500 ,   180.68556   )
    (   550 ,   168.51499   )
    (   600 ,   160.0508    )
    (   650 ,   87.60642    )
    (   700 ,   92.16892    )
    (   750 ,   94.01305    )
    (   800 ,   95.23056    )
    (   850 ,   96.79829    )
    (   900 ,   98.36383    )
    (   950 ,   97.74828    )
    (   1000    ,   99.44196    )
    (   1050    ,   99.57959    )
    (   1100    ,   100.46059   )
    (   1150    ,   100.79218   )
    (   1200    ,   100.56267   )
    (   1250    ,   93.2948 )
    (   1300    ,   95.02503    )
    (   1350    ,   96.51133    )
    (   1400    ,   100.1179    )
    (   1450    ,   102.38764   )
    (   1500    ,   103.54269   )
    (   1550    ,   103.56593   )
    (   1600    ,   104.18916   )
    (   1650    ,   105.38864   )
    (   1700    ,   105.55891   )
    (   1750    ,   105.9027    )
    (   1800    ,   105.7379    )
    (   1850    ,   104.4362    )
    (   1900    ,   99.45603    )
    (   1950    ,   101.45409   )
    (   2000    ,   103.13313   )
    (   2050    ,   104.001 )
    (   2100    ,   105.24871   )
    (   2150    ,   104.92647   )
    (   2200    ,   105.72141   )
    (   2250    ,   106.16375   )
    (   2300    ,   106.90744   )
    (   2350    ,   106.75882   )
    (   2400    ,   107.1694    )
    (   2450    ,   106.73994   )
    (   2500    ,   102.76379   )
    (   2550    ,   103.29357   )
    (   2600    ,   104.68457   )
    (   2650    ,   106.34315   )
    (   2700    ,   107.10539   )
    (   2750    ,   107.06669   )
    (   2800    ,   107.70086   )
    (   2850    ,   108.08402   )
    (   2900    ,   108.24118   )
    (   2950    ,   108.59102   )
    (   3000    ,   108.68457   )
    (   3050    ,   107.63572   )
    (   3100    ,   107.73662   )
    (   3150    ,   104.64939   )
    (   3200    ,   106.00928   )
    (   3250    ,   106.74054   )
    (   3300    ,   107.29258   )
    (   3350    ,   107.40438   )
    (   3400    ,   107.81709   )
    (   3450    ,   108.10144   )
    (   3500    ,   108.43381   )
    (   3550    ,   108.72324   )
    (   3600    ,   108.75516   )
    (   3650    ,   108.55023   )
    (   3700    ,   108.54483   )
    (   3750    ,   105.69713   )
    (   3800    ,   106.36059   )
    (   3850    ,   106.75258   )
    (   3900    ,   108.41392   )
    (   3950    ,   108.30321   )
    (   4000    ,   108.62821   )
};
\end{axis}
\end{tikzpicture}

%% file: g_shapes_800_square.tex
\begin{tikzpicture}
  \begin{axis}[width=2.2in, height=1.60in,
               solid,
               xlabel={\tiny $m=n=k$},
               ylabel={\tiny  GFLOPS},
               title={ {\small  Square} },
               xmin=0,xmax=4000,ymin=0,ymax=225,
               legend style={at={(1.1,0.5)}, anchor=west},
               clip=false]
\addplot[color=black,dotted,thick] coordinates {
    (   50  ,   2.91019 )
    (   100 ,   24.77517    )
    (   150 ,   62.90656    )
    (   200 ,   97.65625    )
    (   250 ,   122.57161   )
    (   300 ,   135.37227   )
    (   350 ,   140.71448   )
    (   400 ,   90.93105    )
    (   450 ,   99.22617    )
    (   500 ,   102.6388    )
    (   550 ,   109.62691   )
    (   600 ,   88.08832    )
    (   650 ,   87.16796    )
    (   700 ,   89.80324    )
    (   750 ,   91.5199 )
    (   800 ,   85.80473    )
    (   850 ,   87.04872    )
    (   900 ,   89.45133    )
    (   950 ,   91.53464    )
    (   1000    ,   84.65206    )
    (   1050    ,   88.03743    )
    (   1100    ,   90.47692    )
    (   1150    ,   94.51016    )
    (   1200    ,   87.43113    )
    (   1250    ,   89.2182 )
    (   1300    ,   93.31247    )
    (   1350    ,   87.17834    )
    (   1400    ,   89.71272    )
    (   1450    ,   92.59346    )
    (   1500    ,   95.80008    )
    (   1550    ,   89.64797    )
    (   1600    ,   92.15687    )
    (   1650    ,   94.67903    )
    (   1700    ,   97.54442    )
    (   1750    ,   91.96211    )
    (   1800    ,   94.49707    )
    (   1850    ,   95.59419    )
    (   1900    ,   98.70088    )
    (   1950    ,   93.3861 )
    (   2000    ,   96.26741    )
    (   2050    ,   98.98291    )
    (   2100    ,   101.64235   )
    (   2150    ,   95.74861    )
    (   2200    ,   97.82484    )
    (   2250    ,   100.18928   )
    (   2300    ,   102.41576   )
    (   2350    ,   97.0729 )
    (   2400    ,   99.15523    )
    (   2450    ,   99.97929    )
    (   2500    ,   96.2374 )
    (   2550    ,   98.04333    )
    (   2600    ,   100.07426   )
    (   2650    ,   102.14397   )
    (   2700    ,   97.81143    )
    (   2750    ,   99.13347    )
    (   2800    ,   100.76723   )
    (   2850    ,   102.63851   )
    (   2900    ,   98.37191    )
    (   2950    ,   100.1191    )
    (   3000    ,   101.92292   )
    (   3050    ,   101.2339    )
    (   3100    ,   97.6943 )
    (   3150    ,   99.20414    )
    (   3200    ,   100.92288   )
    (   3250    ,   102.59894   )
    (   3300    ,   98.97737    )
    (   3350    ,   100.15069   )
    (   3400    ,   101.81417   )
    (   3450    ,   103.30818   )
    (   3500    ,   99.94452    )
    (   3550    ,   101.36741   )
    (   3600    ,   103.06466   )
    (   3650    ,   99.44999    )
    (   3700    ,   100.86282   )
    (   3750    ,   102.18101   )
    (   3800    ,   103.78981   )
    (   3850    ,   100.64229   )
    (   3900    ,   101.81107   )
    (   3950    ,   102.59482   )
    (   4000    ,   104.35893   )
};
\addplot[color=magenta,dashed,thick] coordinates {
    (   50  ,   3.0644  )
    (   100 ,   20.95557    )
    (   150 ,   65.39874    )
    (   200 ,   117.65311   )
    (   250 ,   146.43587   )
    (   300 ,   178.65355   )
    (   350 ,   185.84541   )
    (   400 ,   189.86844   )
    (   450 ,   197.79597   )
    (   500 ,   190.11956   )
    (   550 ,   186.64492   )
    (   600 ,   137.88833   )
    (   650 ,   100.05301   )
    (   700 ,   102.44705   )
    (   750 ,   103.41986   )
    (   800 ,   90.84387    )
    (   850 ,   94.59194    )
    (   900 ,   103.15191   )
    (   950 ,   108.17012   )
    (   1000    ,   107.27342   )
    (   1050    ,   112.0727    )
    (   1100    ,   117.31492   )
    (   1150    ,   122.6276    )
    (   1200    ,   122.53971   )
    (   1250    ,   126.9967    )
    (   1300    ,   129.15141   )
    (   1350    ,   131.25075   )
    (   1400    ,   134.01867   )
    (   1450    ,   135.25602   )
    (   1500    ,   136.37627   )
    (   1550    ,   112.76855   )
    (   1600    ,   117.65825   )
    (   1650    ,   123.55929   )
    (   1700    ,   128.43241   )
    (   1750    ,   126.4014    )
    (   1800    ,   129.58308   )
    (   1850    ,   131.31851   )
    (   1900    ,   134.43216   )
    (   1950    ,   135.06459   )
    (   2000    ,   136.78601   )
    (   2050    ,   138.63813   )
    (   2100    ,   140.19503   )
    (   2150    ,   139.08944   )
    (   2200    ,   140.63787   )
    (   2250    ,   140.67332   )
    (   2300    ,   140.92225   )
    (   2350    ,   126.98134   )
    (   2400    ,   130.46399   )
    (   2450    ,   134.26028   )
    (   2500    ,   133.62355   )
    (   2550    ,   135.87785   )
    (   2600    ,   138.37234   )
    (   2650    ,   140.86666   )
    (   2700    ,   141.26813   )
    (   2750    ,   141.05886   )
    (   2800    ,   142.53247   )
    (   2850    ,   143.47336   )
    (   2900    ,   144.27916   )
    (   2950    ,   145.00976   )
    (   3000    ,   145.28372   )
    (   3050    ,   143.85679   )
    (   3100    ,   131.57447   )
    (   3150    ,   134.54402   )
    (   3200    ,   137.15375   )
    (   3250    ,   139.56412   )
    (   3300    ,   138.20242   )
    (   3350    ,   138.99687   )
    (   3400    ,   141.45186   )
    (   3450    ,   143.08029   )
    (   3500    ,   142.91158   )
    (   3550    ,   143.81944   )
    (   3600    ,   144.85707   )
    (   3650    ,   146.26435   )
    (   3700    ,   146.98765   )
    (   3750    ,   147.25227   )
    (   3800    ,   147.32191   )
    (   3850    ,   134.828 )
    (   3900    ,   137.22  )
    (   3950    ,   139.3751    )
    (   4000    ,   141.85964   )
};
\addplot[color=cyan,solid,thick] coordinates {
    (   50  ,   3.04588 )
    (   100 ,   21.30493    )
    (   150 ,   73.40866    )
    (   200 ,   114.99127   )
    (   250 ,   150.58064   )
    (   300 ,   178.82157   )
    (   350 ,   197.01547   )
    (   400 ,   199.87695   )
    (   450 ,   198.66077   )
    (   500 ,   196.88915   )
    (   550 ,   183.38335   )
    (   600 ,   132.51534   )
    (   650 ,   121.35847   )
    (   700 ,   125.37331   )
    (   750 ,   123.85127   )
    (   800 ,   90.12137    )
    (   850 ,   95.68546    )
    (   900 ,   103.09942   )
    (   950 ,   109.34753   )
    (   1000    ,   107.57182   )
    (   1050    ,   112.9122    )
    (   1100    ,   117.84565   )
    (   1150    ,   122.85389   )
    (   1200    ,   123.45442   )
    (   1250    ,   128.11006   )
    (   1300    ,   130.80255   )
    (   1350    ,   132.71257   )
    (   1400    ,   133.95758   )
    (   1450    ,   136.11101   )
    (   1500    ,   135.77643   )
    (   1550    ,   112.16725   )
    (   1600    ,   116.89957   )
    (   1650    ,   122.93995   )
    (   1700    ,   127.49758   )
    (   1750    ,   125.43626   )
    (   1800    ,   128.75226   )
    (   1850    ,   131.48606   )
    (   1900    ,   134.8821    )
    (   1950    ,   135.00136   )
    (   2000    ,   136.50844   )
    (   2050    ,   138.61974   )
    (   2100    ,   140.53614   )
    (   2150    ,   138.98645   )
    (   2200    ,   140.08988   )
    (   2250    ,   139.83592   )
    (   2300    ,   140.05528   )
    (   2350    ,   125.83237   )
    (   2400    ,   129.28158   )
    (   2450    ,   133.15542   )
    (   2500    ,   133.041 )
    (   2550    ,   134.81172   )
    (   2600    ,   137.76866   )
    (   2650    ,   140.487 )
    (   2700    ,   140.64121   )
    (   2750    ,   140.71772   )
    (   2800    ,   142.04093   )
    (   2850    ,   143.05874   )
    (   2900    ,   143.60956   )
    (   2950    ,   144.33978   )
    (   3000    ,   144.2715    )
    (   3050    ,   142.47223   )
    (   3100    ,   130.25619   )
    (   3150    ,   133.38238   )
    (   3200    ,   136.07334   )
    (   3250    ,   138.34222   )
    (   3300    ,   137.06385   )
    (   3350    ,   138.40742   )
    (   3400    ,   140.5106    )
    (   3450    ,   142.20043   )
    (   3500    ,   142.53718   )
    (   3550    ,   143.00179   )
    (   3600    ,   144.21856   )
    (   3650    ,   145.35046   )
    (   3700    ,   145.71925   )
    (   3750    ,   146.12518   )
    (   3800    ,   145.56506   )
    (   3850    ,   133.31353   )
    (   3900    ,   135.54228   )
    (   3950    ,   137.85687   )
    (   4000    ,   140.33442   )
};
\addplot[color=green,dashdotted,thick] coordinates {
    (   50  ,   3.04518 )
    (   100 ,   24.99188    )
    (   150 ,   58.77641    )
    (   200 ,   117.0746    )
    (   250 ,   148.06006   )
    (   300 ,   181.74475   )
    (   350 ,   191.35499   )
    (   400 ,   203.50635   )
    (   450 ,   205.99109   )
    (   500 ,   190.83707   )
    (   550 ,   175.04507   )
    (   600 ,   162.81673   )
    (   650 ,   80.60508    )
    (   700 ,   88.31068    )
    (   750 ,   95.39963    )
    (   800 ,   96.5483 )
    (   850 ,   102.2778    )
    (   900 ,   107.16843   )
    (   950 ,   108.18565   )
    (   1000    ,   112.07071   )
    (   1050    ,   115.72697   )
    (   1100    ,   117.6645    )
    (   1150    ,   120.12319   )
    (   1200    ,   122.73116   )
    (   1250    ,   110.58641   )
    (   1300    ,   114.83425   )
    (   1350    ,   118.51357   )
    (   1400    ,   127.41495   )
    (   1450    ,   128.72794   )
    (   1500    ,   131.8805    )
    (   1550    ,   133.17622   )
    (   1600    ,   133.90427   )
    (   1650    ,   135.74553   )
    (   1700    ,   138.47906   )
    (   1750    ,   138.75675   )
    (   1800    ,   140.58497   )
    (   1850    ,   140.02902   )
    (   1900    ,   128.32229   )
    (   1950    ,   133.32214   )
    (   2000    ,   137.2494    )
    (   2050    ,   138.25021   )
    (   2100    ,   140.53143   )
    (   2150    ,   140.98991   )
    (   2200    ,   141.85303   )
    (   2250    ,   143.43035   )
    (   2300    ,   145.19468   )
    (   2350    ,   145.45737   )
    (   2400    ,   147.16218   )
    (   2450    ,   149.50535   )
    (   2500    ,   137.23476   )
    (   2550    ,   140.51456   )
    (   2600    ,   142.35299   )
    (   2650    ,   148.18952   )
    (   2700    ,   148.68895   )
    (   2750    ,   148.68281   )
    (   2800    ,   150.54366   )
    (   2850    ,   150.53217   )
    (   2900    ,   151.94118   )
    (   2950    ,   153.13241   )
    (   3000    ,   153.42643   )
    (   3050    ,   153.19937   )
    (   3100    ,   153.59562   )
    (   3150    ,   144.88048   )
    (   3200    ,   148.39071   )
    (   3250    ,   150.67072   )
    (   3300    ,   151.36996   )
    (   3350    ,   151.37814   )
    (   3400    ,   152.95108   )
    (   3450    ,   153.23401   )
    (   3500    ,   154.34081   )
    (   3550    ,   155.41174   )
    (   3600    ,   155.43912   )
    (   3650    ,   157.78543   )
    (   3700    ,   158.59181   )
    (   3750    ,   148.44768   )
    (   3800    ,   151.1624    )
    (   3850    ,   152.43638   )
    (   3900    ,   154.63754   )
    (   3950    ,   155.78587   )
    (   4000    ,   157.06589   )
};
\legend{
    \begin{minipage}{1.0in}{\scriptsize  MOMMS Goto}\end{minipage},
    \begin{minipage}{1.0in}{\scriptsize MOMMS $A_3 B_2 C_0$}\end{minipage},
    \begin{minipage}{1.0in}{\scriptsize MOMMS $B_3 A_2 C_0$}\end{minipage},
    \begin{minipage}{1.0in}{\scriptsize MOMMS $C_3 A_2 C_0$}\end{minipage}
}
\end{axis}
\end{tikzpicture}

%% file: 06conclusion.tex
We have developed a new family of algorithms for MMM
that effectively utilizes a cache hierarchy with multiple layers of fast memory,
using two loops at each level of the memory hierarchy.
We then demonstrated performance improvements over state-of-the-art implementations of MMM
when I/O cost to main memory is a limiting factor.
Algorithms like this are key to delaying the inevitable situation where MMM becomes memory bound.

We chose to focus on potential performance benefits of these algorithms
and demonstrated those benefits in our experiments.
However memory movements cost far more energy than flops do~\cite{shalf2010exascale}, 
so algorithms that reduce I/O costs are beneficial for the additional reason
that they reduce energy usage.

Many algorithms in libraries that implement higher-level linear algebra, such as LAPACK~\cite{LAPACK} or libflame~\cite{FLAME},
take advantage of the fact that MMM implementations in BLAS libraries are efficient when the $k$ dimension is relatively small, on the order of a couple of hundred,
as Goto's Algorithm reaches its maximal efficiency when $k$ is equal to the blocksize $k_c$.
We expect future computers to be bandwidth bound when executing MMM in such situations,
and algorithms for MMM that have larger blocksizes will be used.
To take advantage of algorithms that use larger blocksizes, 
LAPACK and FLAME can use larger blocksizes, however this is currently disadvantageous because the larger their blocksizes are,
the more time is spent during inefficient unblocked subproblems.
According to this line of thought, LAPACK and FLAME would benefit from algorithms that do not suffer from this weakness.
One possibility is to use recursive algorithms, as advocated in Peise and Bientinesi~\cite{peise2017algorithm}.


Hard drives and other similarly slow storage devices can be thought of as another layer of the memory hierarchy.
Because of this, we believe the methodology in this paper can be used to instantiate
out-of-core algorithms for MMM.
A major difference between such out-of-core algorithms and the ones in this paper targeting LRU caches
is that out-of-core algorithms may require explicit transfers to disk and explicit overlapping of I/O and computation.

Future work is to generalize the MOMMS family of algorithms to other dense linear algebra operations,
much in the way that Goto's algorithm~\cite{GotoBLAS} was generalized to the rest of the level-3 BLAS~\cite{goto2008high}.
The key point is that most suboperations during the other level-3 BLAS operations (that operate on structured matrices) are regular, unstructured MMM operations~\cite{poorman_journal}.